 \documentclass[final,5p,times,twocolumn,authoryear]{elsarticle}

\usepackage{amssymb}
\usepackage{lipsum}
\usepackage{amsthm}
\usepackage{url}
\usepackage{hyperref}
\usepackage[nopatch]{microtype}
\usepackage{booktabs}
\usepackage{multirow}
\usepackage{footnote}
\usepackage{threeparttable}

\newcommand{\ion}[2]{#1\textsc{#2}}

\journal{New Astronomy}

\begin{document}

\begin{frontmatter}



\title{Unveiling Compact Planetary Nebulae: Broad-Band Survey Analysis and LAMOST Confirmation}


\author[LAP,USP]{L. A. Guti\'errez-Soto}
\ead{gsotoangel@fcaglp.unlp.edu.ar}
\author[UNC,CONICET]{M. Bel\'en Mari}
\author[UNC,CONICET]{W. A. Weidmann}
\author[LAP,UNLP]{F. R. Faifer}

\affiliation[LAP]{organization={Instituto de Astrofísica de La Plata (CCT La Plata - CONICET - UNLP)}, addressline={B1900FWA}, city={La Plata}, country={Argentina}}
\affiliation[USP]{organization={Departamento de Astronomia, IAG, Universidade de São Paulo}, addressline={Rua do Matão, 1226, 05509-900}, city={São Paulo}, country={Brazil}}
\affiliation[UNC]{organization={Observatorio Astronómico de Córdoba, Universidad Nacional de Córdoba}, addressline={Laprida 854}, city={Córdoba}, country={Argentina}}
\affiliation[CONICET]{organization={Consejo de Investigaciones Científicas y Técnicas de la República Argentina}, city={Buenos Aires}, country={Argentina}}
\affiliation[UNLP]{organization={Facultad de Cs. Astronómicas y Geofísicas, UNLP}, addressline={Paseo del Bosque S/N, B1900FWA}, city={La Plata}, country={Argentina}}

\begin{abstract}
Planetary nebulae (PNe) are pivotal for advancing our knowledge of stellar evolution and galactic chemical enrichment. Recent progress in surveys and data analysis has revolutionized PN research, leading to the discovery of new objects and deeper insights into their properties. We have devised a novel photometric selection method, integrating GAIA and Pan-STARRS photometry, to identify compact PN candidates. This approach utilizes color-color diagrams, specifically $(G - g)$ versus $(G_{BP} - G_{RP})$ and $(G - r)$ versus $(G_{BP} - G_{RP})$, as primary criteria for candidate selection. The subsequent verification step involves confirming these candidates through LAMOST spectroscopic data. By cross-referencing a comprehensive dataset of PNe, GAIA, Pan-STARRS, and LAMOST DR7 spectra, we explore the potential of our approach and the crucial role played by these surveys in the field of PN research.
The LAMOST spectra provide compelling evidence supporting our selection criteria, especially for compact PNe characterized by strong emission lines and low continuum. This characteristic spectral profile in LAMOST data underscores its effectiveness in confirming compact PNe, enabling clear differentiation based on distinctive spectral features.
Applying these criteria to a catalog of emission line objects, we have selected a PN candidate. 
Detailed analysis of its LAMOST spectrum unveiled classical Balmer emission lines and high-ionization lines (\ion{He}{ii}, [\ion{Ar}{v}], [\ion{Ar}{iii}], and [\ion{Ne}{iii}]), characteristic of high-ionization PNe, with an absence of low-excitation lines. Utilizing the 1D-photoionization code {\sc cloudy}, our modeling revealed crucial parameters, including an ionizing source with an effective temperature of 180$\times10^{3}$ K, luminosity around 3,400 L$_{\odot}$, and gas abundances encompassing various elements. Comparing the PN's evolution track, the progenitor star was estimated to have a mass of 2M$_{\odot}$.
Our findings show the greatest promise for cleanly separating compact PNe from other objects and provide a robust framework for further exploration of these surveys in the context of planetary nebulae.

\end{abstract}



\begin{keyword}
planetary nebulae: general \sep ISM: lines and bands \sep surveys 



\end{keyword}

\end{frontmatter}




\section{Introduction}
\label{sec:intro}

Planetary nebulae (PNe) mark the ultimate phase in the life cycle of low-to-intermediate mass stars (0.8–8.0 solar masses, $M_\odot$), 
during which the expelled material from the asymptotic giant branch (AGB) and post-AGB stages becomes ionized by the intense ultraviolet (UV) 
radiation emitted by the hot descendant star.
As the emission nebula expands, the glowing shells of ionized gas scatter 
into the interstellar medium (ISM) leaving behind just the dying star core, which becomes a white dwarf 
(WD).

The number of these kinds of objects discovered in the Galaxy is relatively low (\(\sim3\,500\); \citealp{Parker:2016}).
However, the current number of identified PNe in the Milky Way is far away (representing
only about 15-30\%) from the estimated total of Galactic PNe (\citealp{Frew:2008}; \citealp{Jacoby:2010}),
showing that a small fraction of PNe has been cataloged until now \citep{Frew:2017}. In this context, there 
is a significant count of yet to be found PNe.
The models of \citet{Moe:2006}, for example, predict that there are 46\,000 $\pm$ 22\,000 PNe
for the general case, but the number is reduced to $\sim$6\,600 \citep{Marco:2005} 
if close binaries (e.g. common-envelope phase) are required. The finding of new PNe becomes increasingly difficult, 
probably due to their weak signal, either because they are more distant and/or because they are 
obscured by interstellar dust \citep{Kwitter:2022}.
Moreover, PNe are, perhaps, one of the most difficult objects to classify, since they can be mimicked by many other 
astronomical objects such symbiotic stars (SySt), reflection nebulae, galaxies H II regions, 
and even candidate globular clusters \citep{2010PASA...27..129F, Parker:2022a}.

It is worth noting that planetary nebulae have a relatively short lifespan of only 5\,000-25\,000   
years \citep{Badenes:2015} compared to the overall stellar life cycle. However, \citet{Fragkou:2022} 
shows an apparent age for a PN of around 70\,000 years, which is a very uncommon result. 
So why is it essential to discover a new planetary nebula? The answer lies in the vital 
clues that PNe provides to the understanding of late-stage stellar evolution and 
Galactic chemical enrichment. Their strong emission lines allow the determination of 
gas abundances, expansion and radial velocities, as well as temperatures of central stars of PNe (CSPNe). 
The PN phase enriches the ISM with nitrogen, carbon, helium
and dust, important components for the formation of future generations of stars. 
PNe yield information on the nuclear burning, dredge up, and mass loss of stellar 
progenitors (see e.g. \citealp{Kwitter:2022}) and thus for the development of 
accurate models of the chemical enrichment rates of the Galaxy.

Over the years the study of nebulae has been hampered by some problems.
For instance, one of the most important issues is the previous lack of
 precise distances to most Galactic PNe. Fortunately, this problem
tends to be solved through accurate Gaia CSPN distances.
However, despite these advancements, most of the known galactic PNe are 
still too distant and faint for Gaia DR3 and correct CSPN identification 
remains a problem \citep{Parker:2022}; so these kind of problems 
could be addressed by increasing the number of PNe in the Galaxy, 
because better study and statistics could be done. Consequently, 
several previous attempts have utilized diagnostic diagrams incorporating various 
emission-line intensities to distinguish PNe from other resolved emission sources 
such as H~{\sc ii} regions, supernova remnants (SNRs), and other objects. One of 
the widely employed methods relies on the H{$\alpha$}/[\ion{N}{ii}] 
and H{$\alpha$}/[\ion{S}{ii}] emission-line ratios \citep{Sabbadin:1977, Fesen:1985, Riesgo:2006}, 
where empirical selection criteria based on observed emission line ratios between PNe, 
supernova remnants, and H{\sc ii} regions have been developed. 
Furthermore, \citet{Viironen:2009a, Viironen:2009b} utilized 
two color-color diagrams based on IPHAS and 2MASS photometry 
to discriminate PNe from stellar and other emission 
line sources within the Galactic context. Additionally, in 
a similar vein, \cite{2013A&A...552A..74W} employed 
the VVV survey to differentiate between PNe and other stellar objects.

The most applied methodology to search for compact PNe candidates consists 
of trying to identify them using narrowband and broadband photometry and 
then revealing their true nature through spectroscopic follow-up. 
This is always required with such samples because 
the detection success rate is low  due to the fact that other compact emitters usually dominate 
the photometric diagram.

Recently, new colour criteria to identify compact PNe were developed by \citet{Akras:2019b} using 
the classification tree model in both 2MASS and AllWISE photometric data. \citet{Gutierrez-Soto:2020}
searched for compact PNe in the Javalambre and Southern Photometric Local
Universe Survey data (J-PLUS; \citealp{Cenarro:2019} and S-PLUS; \citealp{Mendes:2019}, respectively, using a combination
of narrow- and broad-band photometry. Other new colour-colour diagnostic diagrams were proposed
by \citet{Vejar:2019} to identify new PNe using broad-band filters. In this work, we propose new color criteria based on 
the GAIA and Pan-STARRS surveys that highlight compact PNe with strong emission lines. After applying this 
criterion on the sample of emission lines of \citet{Skoda:2020}, which was constructed using LAMOST DR4 data, 
we were able to quickly identify objects with prominent emission lines. 

This paper not only introduces pioneering color criteria based braod-band data for selecting compact planetary nebulae but also represents a significant leap in their identification and classification. 
Additionally, it presents the first discovery of a previously unknown planetary nebula using these color criteria. 
Supported by its LAMOST spectra, which exhibit prominent nebular emission lines, including high-ionization lines, 
we provide evidence confirming the nature of this object. The absence of red-shifted emission lines in 
its optical spectrum eliminates the possibility of its extragalactic nature. This article is organized as follows: 
Section~\ref{sec:metho} describes the observations related to GAIA, Pan-STARRS, and LAMOST. 
It also presents the color-color diagrams that we have developed for selecting compact planetary 
nebulae using broad-band photometry. We also explore the feasibility of the criteria by analyzing 
the photometry and LAMOST spectra of known PNe, and the discovery of the new planetary nebula along 
with its respective optical spectra. Section~\ref{sec:model} outlines the {\sc cloudy} model of the 
new planetary nebula and provides a comparison of the best-fit model with the post-AGB stellar evolutionary track. 
Finally, the main results and conclusions are discussed in Section~\ref{sec:conclu}.

\section{PN Candidate Identification: Broad-Band Photometry and LAMOST Confirmation}
\label{sec:metho}


Our research relies on three key data sources:
The third GAIA data release, DR3, provides precise astrometry and photometry data for approximately 1.8 billion sources \citep{Vallenari:2023}. It includes filters such as $G$, $G_{BP}$, and $G_{RP}$, enhancing its utility for various research applications, including the identification of new variable stars and central stars of planetary nebulae. Particularly, the combination of the $G_{BP}$ and $G_{RP}$ filters enables the selection of sources with temperatures adequate for ionizing any surrounding nebula.
Pan-STARRS1  (PS1; \citealp{Kaiser:2010}) is a wide-ranging optical imaging survey known for its ability to detect objects about 1 magnitude fainter than the Sloan Digital Sky Survey (SDSS) in the $z$-band \citep{York:2000}. It covers 75\% of the night sky and combines photometry and astrometry data from multiple epochs. We specifically utilized the $g$ and $r$ filters of Pan-STARRS to detect emission lines in compact planetary nebulae. Notable emission lines detected in the $g$-band include [\ion{O}{iii}] ($\lambda$5007, $\lambda$4959), \ion{H}{i} lines (H$\beta$ $\lambda$4861, H$\gamma$ $\lambda$4340, H$\delta$ $\lambda$4102), \ion{He}{ii} ($\lambda$4686), \ion{He}{i} ($\lambda$4471), [\ion{S}{ii}] ($\lambda$4074, $\lambda$4070, $\lambda$4078), and \ion{N}{i} ($\lambda$5200, $\lambda$5198). In the $r$-band, detected lines encompass \ion{H}{i} (H$\alpha$ $\lambda$6563), [\ion{N}{ii}] ($\lambda$6584, $\lambda$6548), \ion{He}{i} ($\lambda$5876), [\ion{S}{ii}] ($\lambda$6720, $\lambda$6731, $\lambda$6716), \ion{S}{iii} ($\lambda$6312), [\ion{Cl}{iii}] ($\lambda$5538), and [\ion{O}{i}] ($\lambda$6300) \citep{Vejar:2019}. 
LAMOST is a 4-meter telescope designed for spectroscopic sky surveys \citep{Cui:2012, Zhao:2012}. It employs 4,000 fibers to capture 4,000 spectra in one exposure. The low-resolution LAMOST extensive spectral database, with a resolving power of about 1\,800, is a valuable resource for identifying various types of stars, including cataclysmic variables (CVs). LAMOST plays a crucial role in confirming the true nature of objects that cannot be determined through photometry alone.

\subsection{Color Criteria for Compact Planetary Nebulae Using Broad-Band Photometry}
\label{sec:color}

In the optical, SDSS colors were employed by \citet{Kniazev:2014} to differentiate between 
PNe and other point sources in the outskirts of M31.
Many other surveys cover an extensive area of the sky, 
observing many astronomical objects with broad-band filters.
This is the case for GAIA and Pan-STARRS1, which cover a large area of the sky, meaning that many emission line objects are likely to have been observed by such broadband surveys rather than the narrow-band ones \citep{Parker:2005, Drew:2005, Drew:2014}.  
Therefore, in an attempt to find new emission line objects, even compact PNe, we propose to
create color-color diagrams based on these surveys, to identify this class of objects.  

\begin{figure*}
\centering
\begin{tabular}{ll}
\includegraphics[width=0.5\linewidth, trim=40 10 50 10, clip]{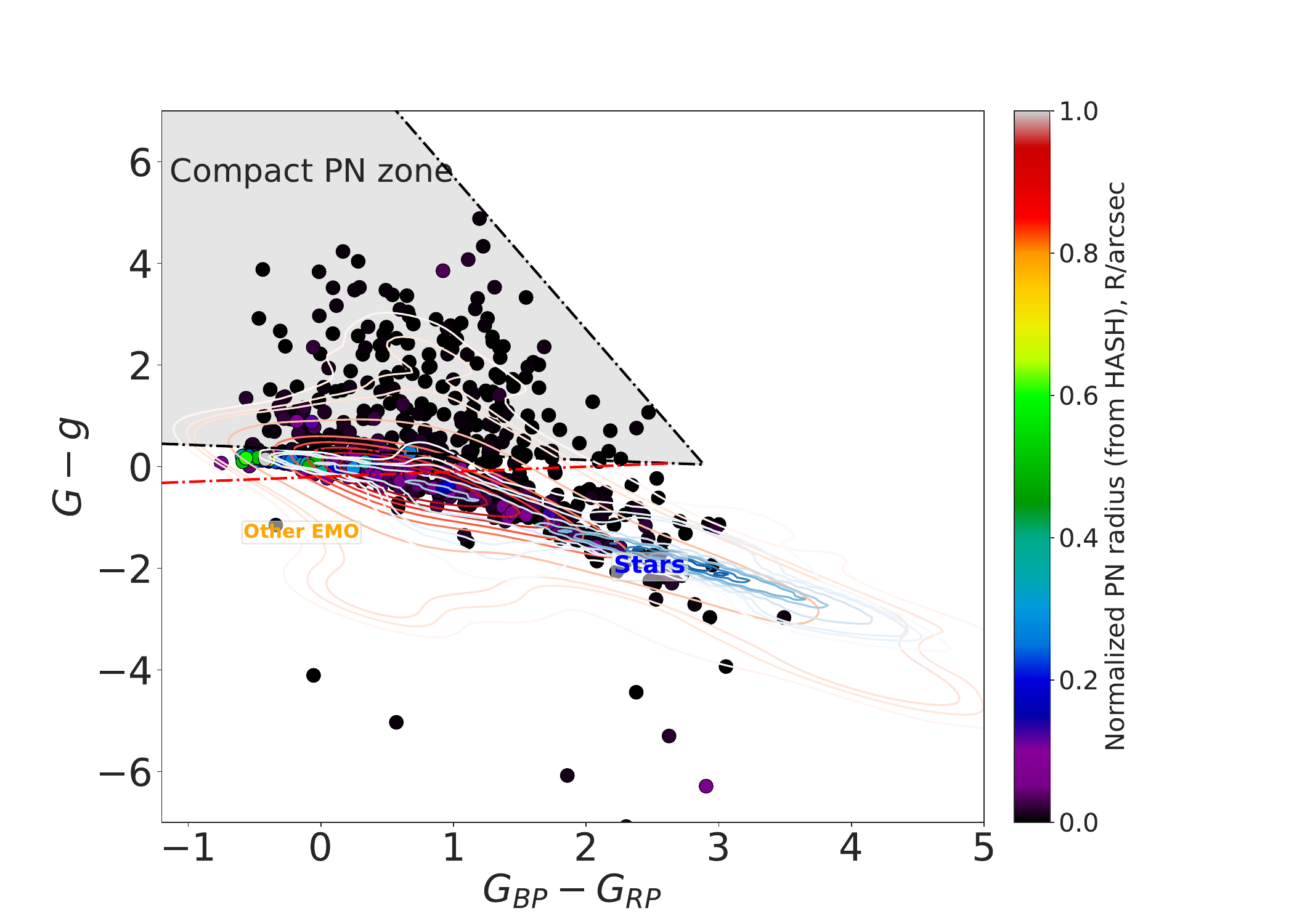} &
\includegraphics[width=0.5\linewidth, trim=40 10 50 10, clip]{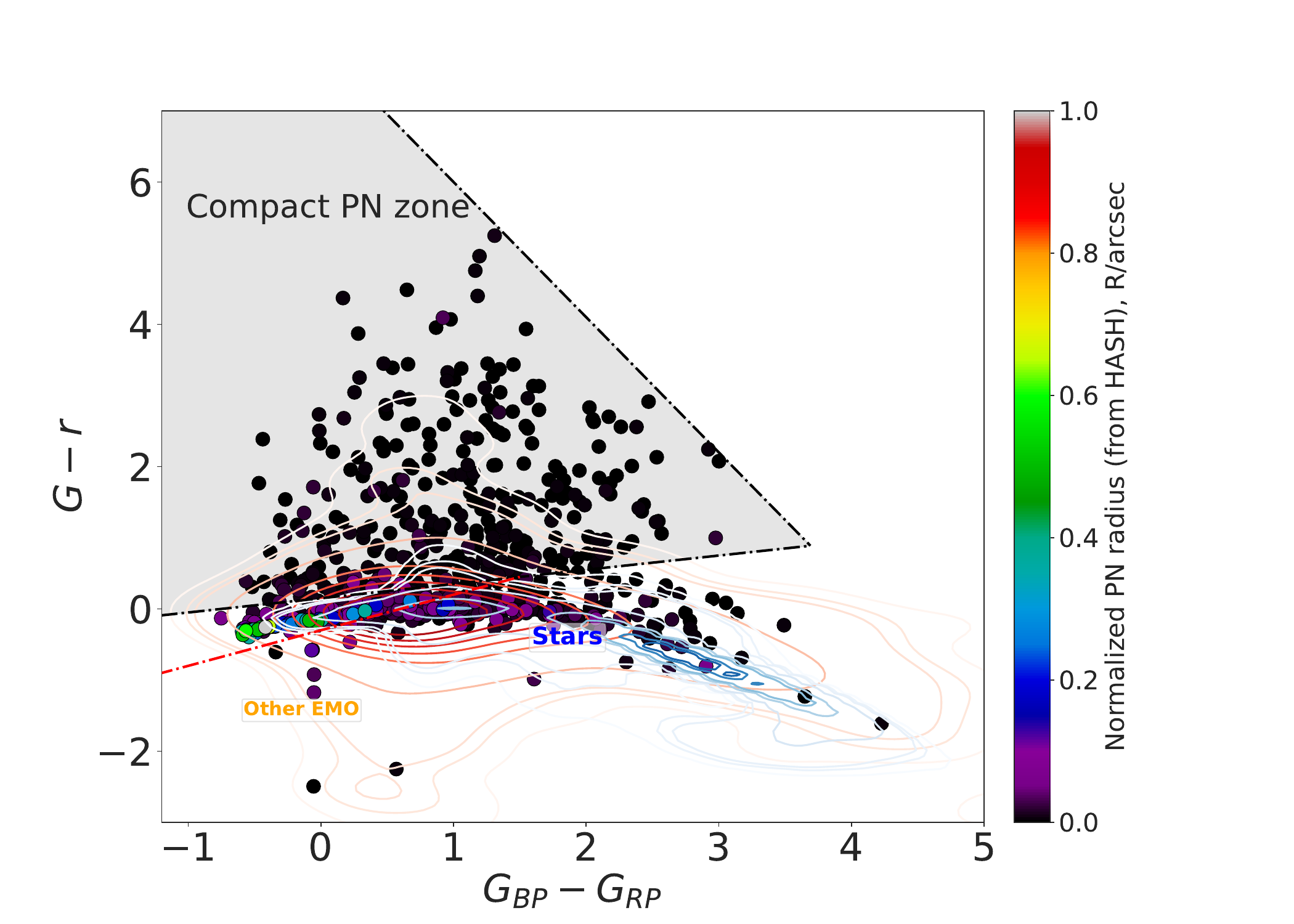}
\end{tabular}
\caption{Diagram \((G - g)\) versus \((G_{BP} - G_{RP})\) and \((G - r)\) versus \((G_{BP} - G_{RP})\) combine GAIA and Pan-STARRS1 photometry, showcasing various emission line objects: red circles represent true PNe from the central stars catalog by  \citet{Chornay:2021}, with their normalized radii indicated by the colorbar, derived from HASH. Orange contours mark the positions of SySt, CVs, SNRs, YSOs, and AeBe stars in the diagrams, while blue contours represent stars from the catalog by \citet{Smart:2021}. The black dotted lines delineate the chosen compact PN zones in both diagrams, characterized by minimal contamination from other emission line objects. The region between the red dotted lines and the black ones defines the area where the most extended PNe are located. Further details are available in the main text.} 
\label{fig:gaia-ps}
\end{figure*}

We initiated the construction of potential color-color diagrams aimed at distinguishing PNe from other emission line objects and stars using GAIA data. However, our initial attempts at separation were not successful. This was primarily due to the fact that PNe, normal stars, and other emission line stars such as CVs, all occupied the same region within the diagram. To address this challenge, we adopted a combined approach, integrating data from two surveys, Pan-STARRS1 and GAIA DR3. This combined approach led to the identification of distinct color-color diagrams that isolate compact PNe exhibiting strong emissions lines, as illustrated in Figure~\ref{fig:gaia-ps}. Notably, these diagrams include only the true PNe sourced from the catalog of \citet{Chornay:2021}, encompassing over 2\,000 sources in Gaia EDR3 identified as probable CSPNe or compact nebula detections. To isolate genuine PNe, we cross-matched these sources with GAIA DR3 and Pan-STARRS1, while simultaneously filtering out saturated sources by applying the condition $r \ge 13.5$ in the Pan-STARRS1 band. This process yielded a total of 882 PNe. Additionally, our analysis encompassed cataclysmic variables (CVs) \citep{Downes:2006}, young stellar objects (YSOs) \citep{Rigliaco:2012}, supernova remnants (SNRs) \citep{Green:2019}, and symbiotic stars (SySt) from \citet{Akras:2019a}\footnote{Note that each catalog of objects was cross-matched with GAIA and Pan-STARRS1 using \texttt{TOPCAT}}. 
Using the \((G - g)\) versus \((G_{BP} - G_{RP})\) and \((G - r)\) versus \((G_{BP} - G_{RP})\) color-color diagrams, it becomes possible to differentiate compact PNe with excess emission in the $g$- and $r$-broad band filters.

In Figure~\ref{fig:gaia-ps}, PNe are depicted with scaled bar colors representing their respective radii obtained from HASH, which indicates the maximum radius. The left panel of the figure displays the \((G - g)\) versus \((G_{BP} - G_{RP})\) color-color diagram. As indicated by the colorbar in the plot, the most compact PNe exhibit \((G - g)\) values greater than 0, reaching values up to 5.
In contrast, the most extended PNe have \((G - g)\) colors around 0, gradually decreasing from approximately 0.8 in the \((G_{BP} - G_{RP})\) color. This results in negative values in \((G - g)\). Overlaid on the plot, orange contours delineate the positions of various emission line objects, including CVs, SySt, YSOs, AeBe stars, and SNRs. These contours transition to white along the periphery, indicating a decrease in the number of objects. Similarly, blue contours represent normal stars sourced from the GAIA catalogue \citep{Smart:2021}. Like the orange contours, their color transitions to white as the number of objects decreases. This observed behavior in extended PNe in the plot is also reflected in these emission line objects and normal stars. 

In the \((G - r)\) versus \((G_{BP} - G_{RP})\) color-color diagram displayed in the right panel of the figure, while many PNe exhibit (\(G - r\)) values close to 0, others occupy a more diverse range within this color space, spanning from 0 to 5.5, and extending from -1 to 5 in the \((G_{BP} - G_{RP})\) color. Again, the most compact PNe are characterized by high (\(G - r\)) values, while the extended PNe tend to have (\(G - r\)) values around 0. Additionally, the most extended PNe are associated with \((G_{BP} - G_{RP})\) color values of less than approximately 0. The other emitters and normal stars are predominantly concentrated in the region with (\(G - r\)) values close to 0.

Although numerous PNe overlap with other emitters in both diagrams, some PNe are distinctly positioned in the upper regions of these diagrams, showing comparatively lower contamination from other sources. This observation implies that objects within this specific regions are more likely to be genuine compact PNe. To isolate this area and minimize contamination, we visually demarcated it using straight lines, delineated by dashed and pointed black lines. We also identified regions in both color-color diagrams that encompass the most extended PNe. These regions are bounded by the black dashed line and the red dashed and pointed line, as depicted in Figure~\ref{fig:gaia-ps}.

\subsection{Validating the PN Color Selection Criteria}
\label{sec:applyieng}

To validate our color criterion for selecting compact PN candidates in the photometric surveys mentioned earlier and to leverage the LAMOST spectroscopic survey for PN confirmation, we identified PNe listed in HASH with LAMOST DR7 spectra, GAIA DR3, and Pan-STARRS1 photometry, using a tolerance radius of 2 arcseconds. This resulted in a total of 25 true PNe and 19 probable and likely PNe. After removing the saturated sources, we obtained a list of 20 true PNe and 13 probable and likely PNe.

Figure~\ref{fig:gaia-ps-apply} illustrates the distribution of the identified true PNe as red circles and the probable/likely PNe as green open circles within two color-color diagnostic diagrams. In the \((G - g)\) versus \((G_{BP} - G_{RP})\) color-color diagrams, 9 out of the true PNe are situated within the compact PN region, and almost all of them are exceptionally compact objects displaying small diameters of less than 10 arcseconds. For visual references and additional details, refer to the inset colored images in Figure~\ref{fig:spectra-image-trurPN-better} and the accompanying information in Table~\ref{tab:TruePN-inf} of~\ref{sec:spectra-tPNe}. As expected, these compact PNe exhibit intense emission lines as evident in the spectra provided in Figure~\ref{fig:spectra-image-trurPN-better}. Among them, the PN Abell 30, with a larger radius of 127.0 arcseconds, shows intense [\ion{O}{iii}] emission lines in Figure~\ref{fig:spectra-image-trurPN-justonecompact}. Given its spectral characteristics, other emission lines within this PN could also be impacting the $g$ filter. This particular PN occupies the very extended PN zone within the \((G - r)\) versus \((G_{BP} - G_{RP})\) color-color diagram. 

Similarly, 9 PNe are located within the compact PN zone of the \((G - r)\) versus \((G_{BP} - G_{RP})\) color-color diagram, among which 8 PNe overlap with those located in the compact zone of the \((G - g)\) versus \((G_{BP} - G_{RP})\) color-color diagram. One of these 9 PNe, namely M 2-2, has a radius 0f 13 arcsecond and is located in the extended PN zone in the \((G - g)\) versus \((G_{BP} - G_{RP})\) color-color diagram. Refer to the spectra and colored image in Figure~\ref{fig:spectra-image-trurPN-justonecompact} for further details.

On the other hand, 7 common PNe are situated in the zone selected for the most extended PNe in both diagrams. These PNe are notably extended, with most of them having diameters exceeding 50 arcseconds (one exception is TS 1, which has a diameter of 9.2 arcseconds), as can be observed in the colored images provided in the inset of Figure~\ref{fig:spectra-image-trurPN-medium} and the detailed information in Table~\ref{tab:TruePN-inf}. The spectra of these objects (shown in Figure~\ref{fig:spectra-image-trurPN-medium} of ~\ref{sec:spectra-tPNe}) indicate that their H$\alpha$ emissions, while present, do not appear as intense when compared to the continuum. In fact, one of these PNe do not show H$\alpha$ in its LAMOST spectra (AMU 1). Finally, three PNe are located outside both the compact PN and extended PN zones in the two diagrams. As is shown in Figure~\ref{fig:spectra-image-trurPN-outside}, their spectra show the emission lines; however, it does not seem to be so intense as to significantly affect the $g$- and $r$-band, which is dominated by the continuum of the central stars. All four objects have diameters greater than 10 arcseconds (for more detailed information, see Table~\ref{tab:TruePN-inf}).

In the case of the probable and likely PNe, the majority of them are situated in the regions of very extended PNe. From their LAMOST spectra, only three objects exhibit emission lines such as [\ion{O}{iii}] 4958.9 \AA, [\ion{O}{iii}] 5006.8 \AA, and H$\alpha$. It is important to note that these emission lines are very weak, with the continuum from the central stars becoming a significant contributor. These PNe are Dr 27, Fr 2-23, and Abell 28. For the remaining objects, their LAMOST spectra resemble typical spectra of white dwarf (WD) stars, as depicted in Figure~\ref{fig:spectra-image-LikelyPN-medium} of~\ref{sec:spectra-lpPNe}. According to the HASH catalog, all these objects are categorized as very extended PNe, with the smallest having a diameter of 60 arcseconds and the largest with a radius of 3600.0 arcseconds (refer to Table~\ref{tab:LikelyPN-inf} of~\ref{sec:spectra-lpPNe} for further details). Of these objects, only one, Kn 27, falls outside both zones in the two diagrams. Its spectrum clearly displays H$\alpha$, but this information alone is inconclusive in revealing its true nature (see Figure~\ref{fig:spectra-image-LikelyPN-outside}).


\subsection{The first case of PN identification using LAMOST}
\label{sec:find}

All indications strongly support the reliable selection of compact PNe exhibiting strong emission lines using these color criteria. These intense emission lines, including [\ion{O}{iii}], H-Balmer, [\ion{N}{ii}], among others, in conjunction with a relatively low continuum, significantly influence the $g$- and $r$-band photometry. Leveraging these distinct features, we take advantage of these characteristics in our search for these types of PNe in broad-band surveys. 

Subsequently, we applied these criteria to the emission line catalog from \citet{Skoda:2020}
to further confirm the selection of PNe. This catalog was created by identifying 
emission line objects in LAMOST DR4 using an active learning approach. 
They divided their final sample of emission line objects into three subgroups: 
one comprising objects with coincidences in \texttt{SIMBAD}, a second subgroup consisting of sources listed by \citet{Hou:2016}, and a third list encompassing objects that were neither found in \texttt{SIMBAD} nor listed by \citet{Hou:2016}. In order to assess the possibility of discovering new  PNe utilizing the aforementioned color criteria,  these criteria were directly applied to the aforementioned third list, which comprised approximately 1\,000 objects that had not previously been reported in the literature. 


\begin{figure*}
\centering
\begin{tabular}{ll}
  \includegraphics[width=0.5\linewidth, trim=20 10 50 10, clip]{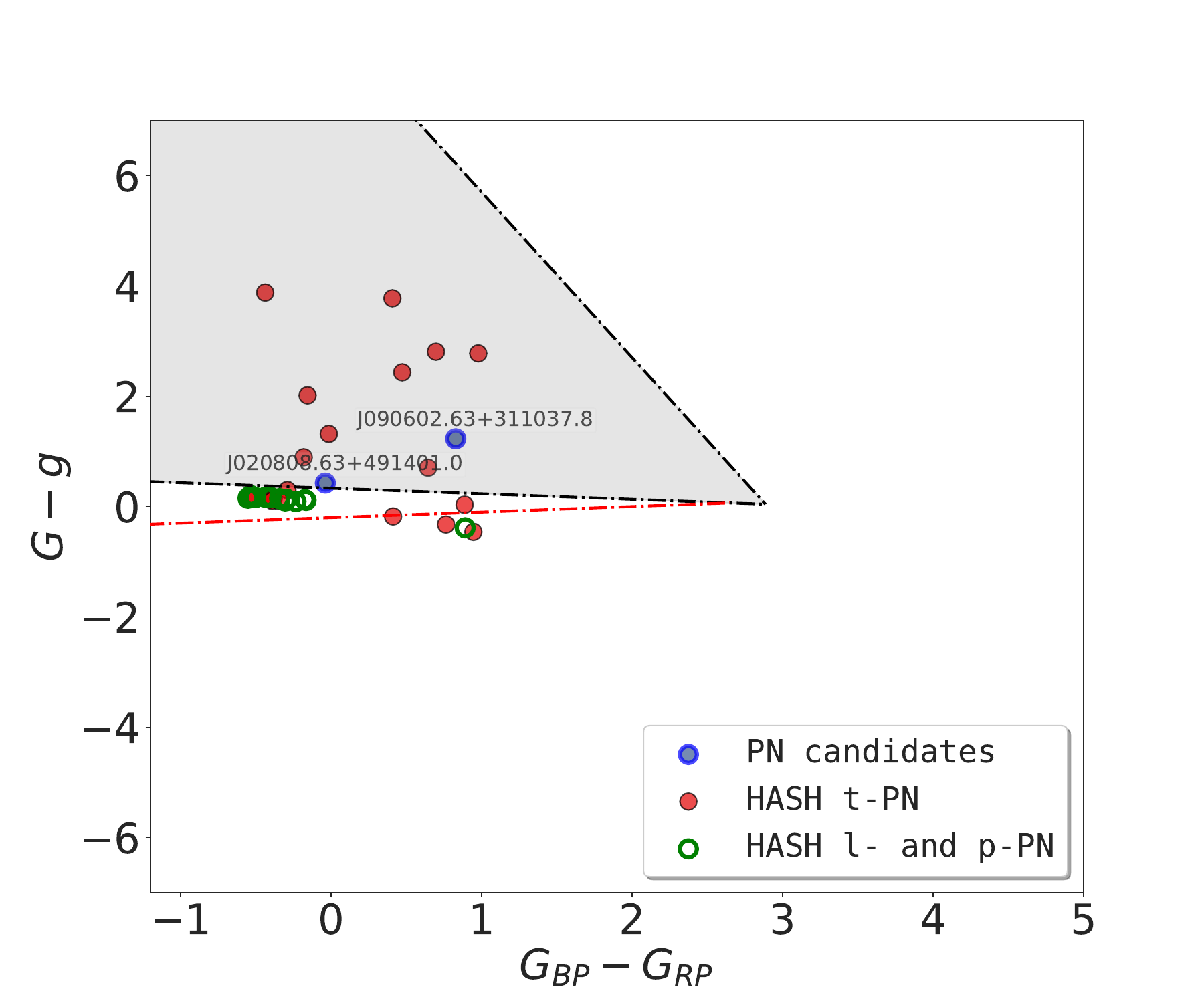} &
  \includegraphics[width=0.5\linewidth, trim=20 10 50 10, clip]{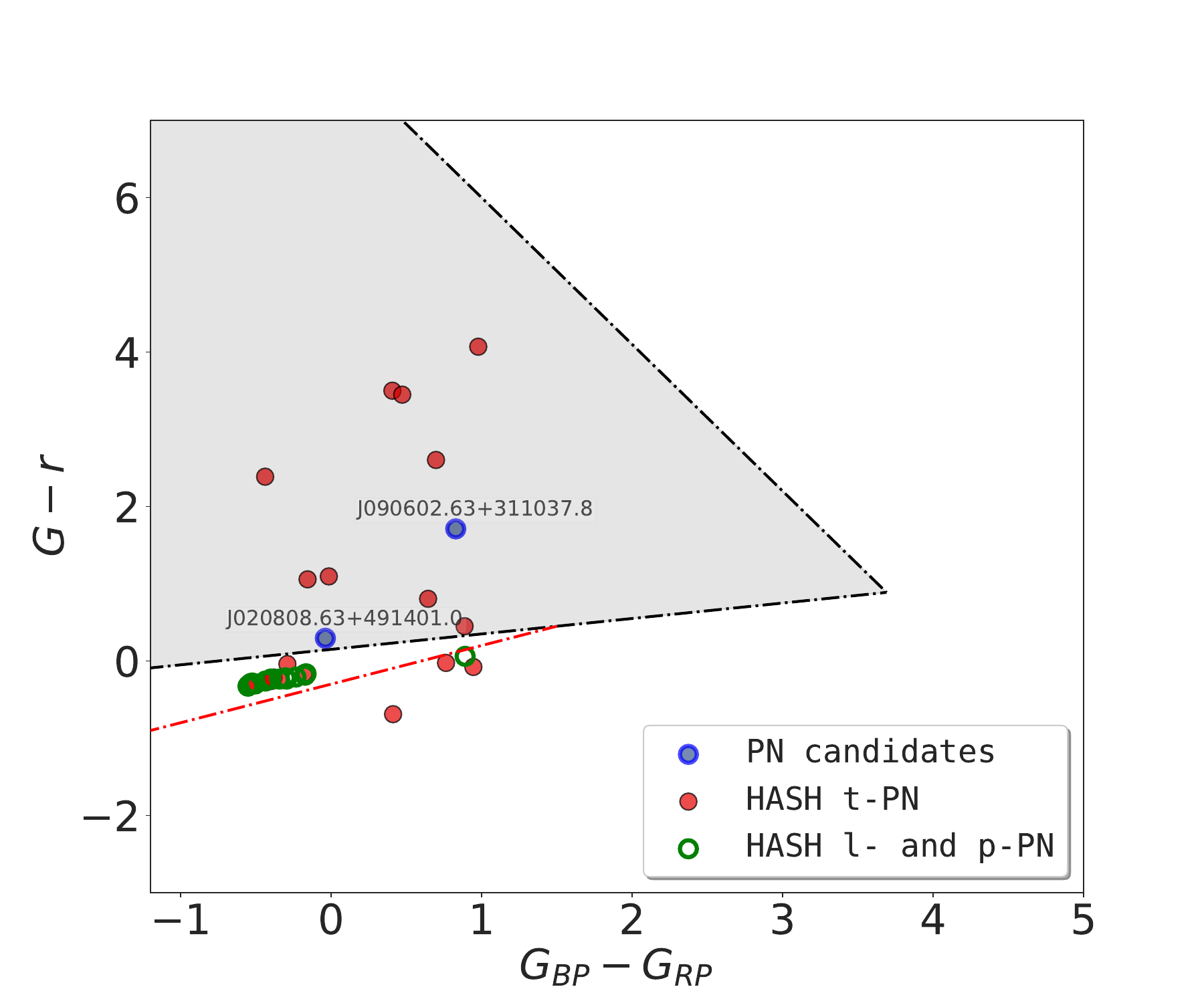}
  \end{tabular}
  \caption{As depicted in Figure \ref{fig:gaia-ps}, blue circles represent objects classified as PN candidates from the list of emission line sources by \citet{Skoda:2020}. The labels within the figure indicate the LAMOST IDs of these objects. Red circles represent true PNe from HASH, while green open circles represent the probable and likely PNe from HASH, all of which have associated LAMOST spectra. Saturated objects have been removed from the plot. 
    } 
  \label{fig:gaia-ps-apply}
\end{figure*}

\begin{figure*}
\centering
  \includegraphics[width=\linewidth]{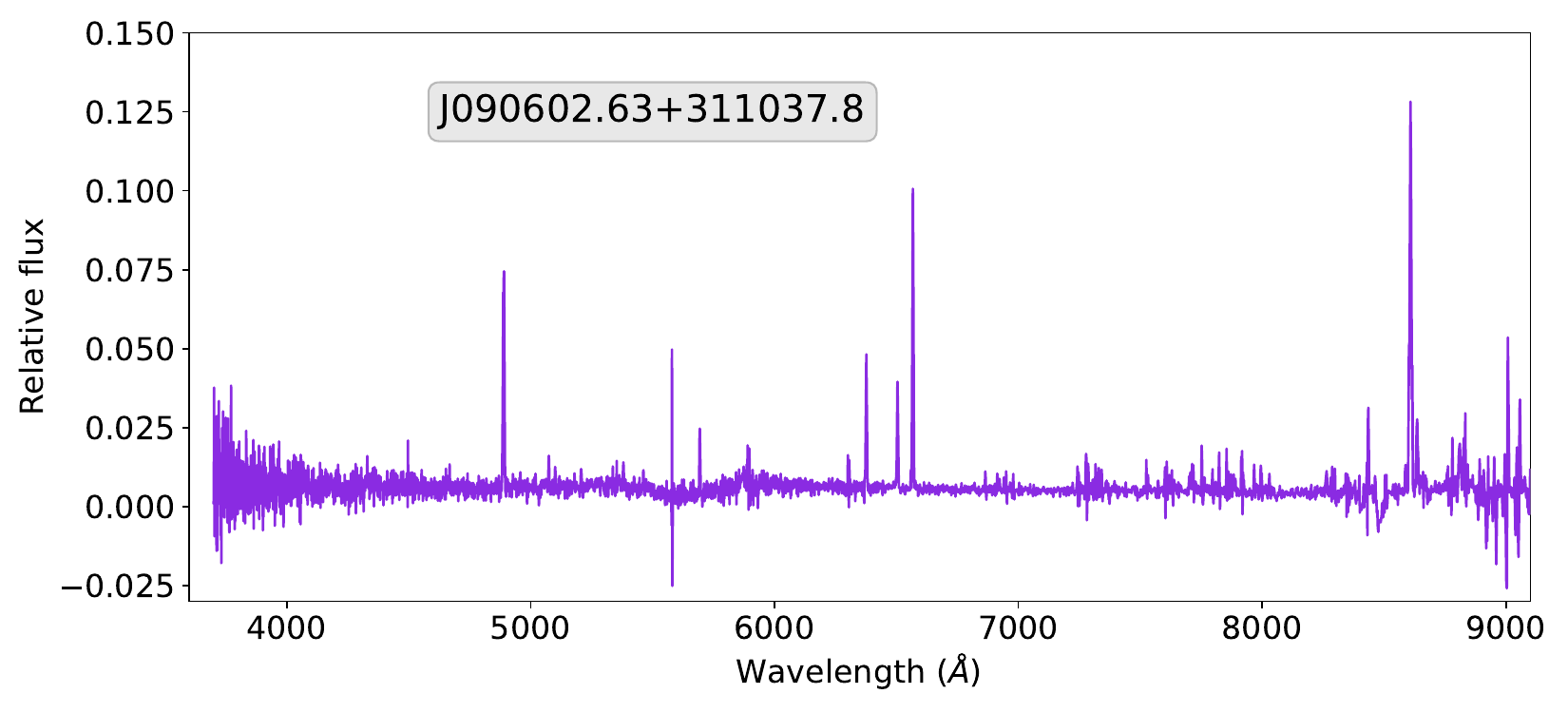} 
    \caption{The LAMOST low-resolution spectra of the candidate (J090602.63+311037.8) considered for PN status show evidence of emission lines. However, these spectra lack the complete set of characteristic emission lines typically associated with PNe. This absence led to their exclusion as likely PN category.}
  \label{fig:other-candidates}
\end{figure*}

Two objects simultaneously met the two criteria, represented by the blue circles in Figure~\ref{fig:gaia-ps-apply}. 
We retrieved the spectra of these objects from the LAMOST database. 
The two  objects display strong  emission lines.  Only one of these two objects displays the typical emission lines of PNe,
such as [\ion{O}{iii}],
\ion{He}{ii} together with the Balmer ones.
Based on this observation, we consider this particular object to be the most probable 
candidate for a planetary nebula. The other object does not display all 
characteristic emission lines and were therefore not considered as likely PNe candidates.
For the spectra of the latter object, refer to Figure~\ref{fig:other-candidates}.

\begin{figure*}[hbt!]
\centering
  \includegraphics[width=\linewidth]{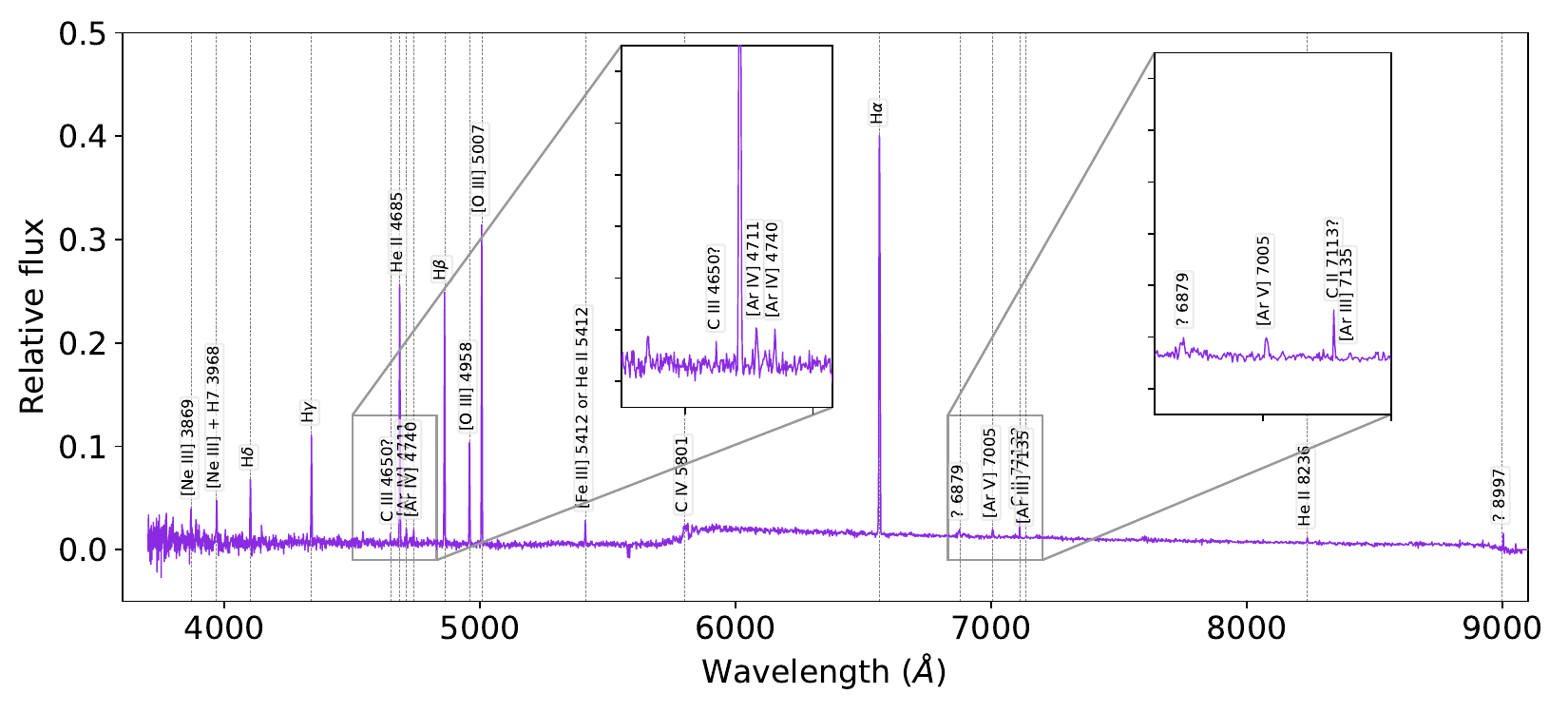}
  \caption{Low-resolution spectrum from LAMOST of J020808.63+491401.0.
    The most prominent emission lines detected in the spectra are given by the dashed vertical
    lines.} 
  \label{fig:spectra}
\end{figure*}

Figure~\ref{fig:spectra} shows the spectrum of the new PN finding in the list of emission
line sources of \citet{Skoda:2020}. There is an offset between the blue and red arms of
the spectrum. This probably occurs due to the fact that the blue- and red-arm spectra were
processed separately with the 2D Pipeline and joined together after the ﬂux calibration
\citep{Xiang:2015}. No scaling or shifting was made in cases where the blue- and
red-arm spectra did not have the same ﬂux level in the overlapping wavelength region,
as it was unclear whether the misalignment was caused by poor ﬂat-ﬁelding/flux-calibration
or sky subtraction, or a combination of both \citep{Chen:2016}. 
This PN seems a very high ionization object because \ion{He}{ii}$\lambda$4685 emission line 
looks as strong as the H$\beta$ line.
Also, the presence of [\ion{Ar}{v}],
[\ion{Ar}{iii}] and [\ion{Ne}{iii}]
are indicators of being a high
ionization PN. No lines of ions in low stages of ionization were detected,
 such as 
[\ion{N}{ii}], [\ion{O}{ii}] and [\ion{S}{ii}]. 
A preliminary analysis was performed in the next
section in an attempt to derive the physical parameters of the planetary nebula comparing
its low-resolution spectra from LAMOST with modelled {\sc cloudy} 1D-spectra.

The spectrum shows a stellar feature: the emission of \ion{C}{iv} at 5\,801-12 \AA, which 
is characteristic of very hot stars such as [WR] and early-O(H). Unfortunately, 
this information is insufficient to confirm whether the star is H-rich or H-poor.

It is important to mention that the LAMOST spectra are not calibrated in absolute flux but they are
calibrate in relative flux (see, \citealp[]{Song:2012} for more details about the procedure of 
flux calibration), then any analysis will have to be done in terms of its ratio lines. 

\begin{figure*}[hbt!]
  \centering
  \begin{tabular}{l l}
\includegraphics[width=0.5\linewidth]{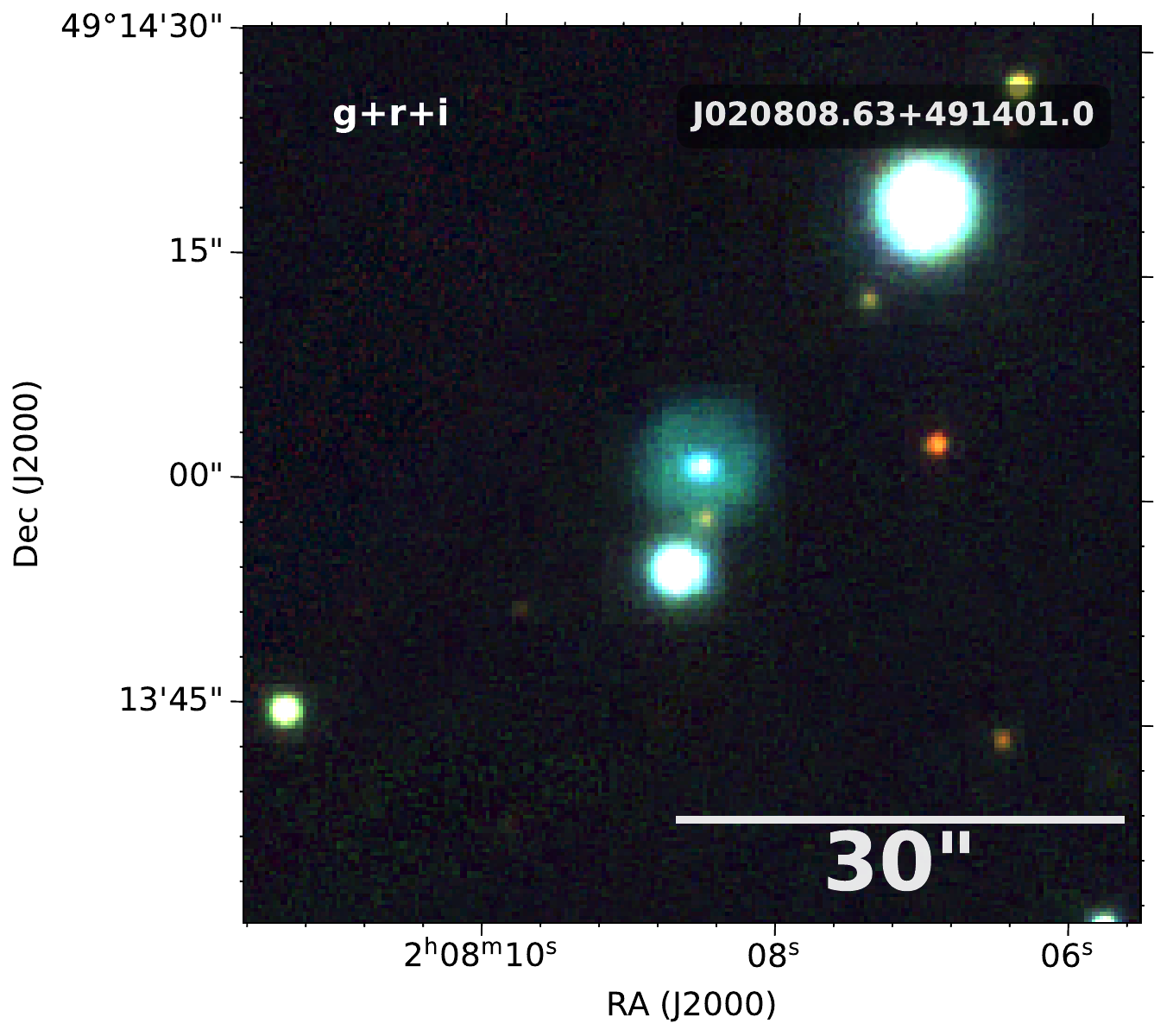}
\includegraphics[width=0.5\linewidth]{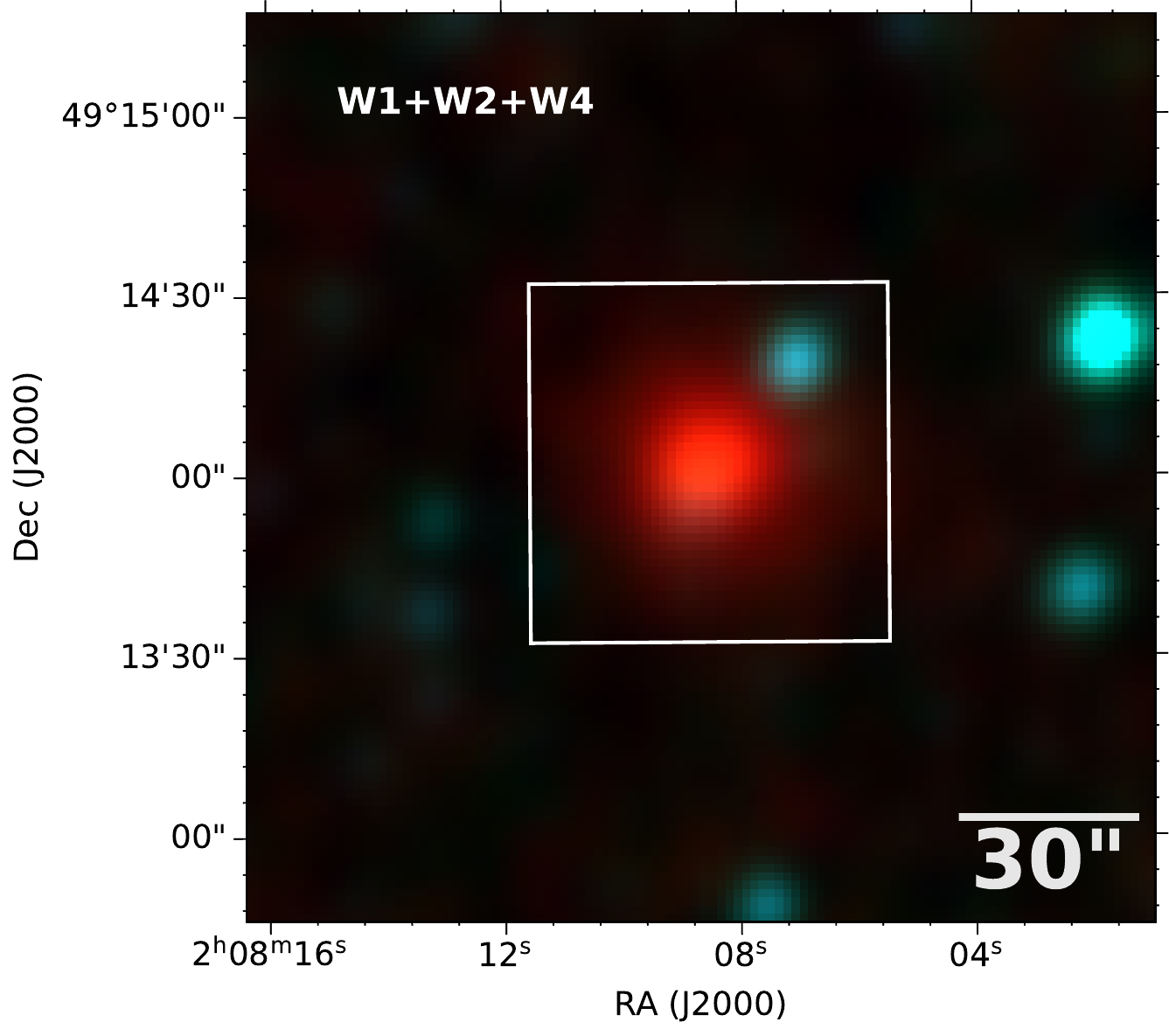}
\end{tabular}  
  \caption{Pan-STARRS optical (\textit{left}) and WISE IR (\textit{right}) coloured
    images of the new PN. To create the optical colored image were used the $g$, $r$, and $i$
    images for the blue, green, and red channels, respectively. In the same way, W1, W2, and W4
    filters were used to create the IR image. Additionally, a box (highlighted in white) is 
    overlaid on the IR image, representing the same size as the optical image.} 
  \label{fig:image}
\end{figure*}

The optical and IR images of the PN with LAMOST
identification J020808.63+491401.0 are showed in Figure~\ref{fig:image}.
It exhibits the PanSTARRS coloured
images\footnote{These RGB images were made by implementing
the python package \texttt{aplpy} \citep[][]{aplpy:2019}}.
%
%
The image shows clearly a nebular component surrounding 
a central star. Figure~\ref{fig:image} also shows the
WISE RGB image, with the filters W1, W2, and W4 in
the blue, green and red channels, respectively.
%
The WISE image shows that the object is W4 bright.  
At 22$\mu$ (W4-filter) it appears as an almost
circular (but slightly elongated in the north-south direction)
diffuse halo (of angular diameter of $\simeq$ 50$^{\prime\prime}$) surrounding
a core of bright emission centered around J020808.63+491401.0.

According to the GAIA coordinates (R.A, DEC.) = (02:08:08.61, +49:14:01.30)
in the J2000 system, 
and the geometrical distance
from \citet{Bailer:2021} of 2.31 kpc inferred from the third GAIA data release,
it is found that J020808.63+491401.0 is located $|\mathrm{z}| \sim$ 0.47 kpc bellow
Galactic plane. 

\subsection{Comparing with other high-ionization PNe}
\label{sec:comp}

In order to check the high-excitation nature of J020808.63$+$491401.0,
we compare its spectra and images, by eye and no further
detailed analysis, with three well-known high excitation PNe: NGC~2242,
NGC~4361 and PRTM~1. This step in our procedure was only made
to see if visually the spectrum and the images of the new possible PN
 resembles the already confirmed PNe and therefore continue with
the analysis of this object. All these three PNe have effective temperatures,
$\mathrm{T_{eff} \sim 105\,925}$, $125\,893$ and $79\,433$~K
and luminosity, $\mathrm{L \sim 599}$, $3\,467$ and $2\,344~$L$_{\odot}$,
respectively \citep{Weidmann:2020}.

\begin{figure*}[hbt!]
\centering
\includegraphics[width=\linewidth]{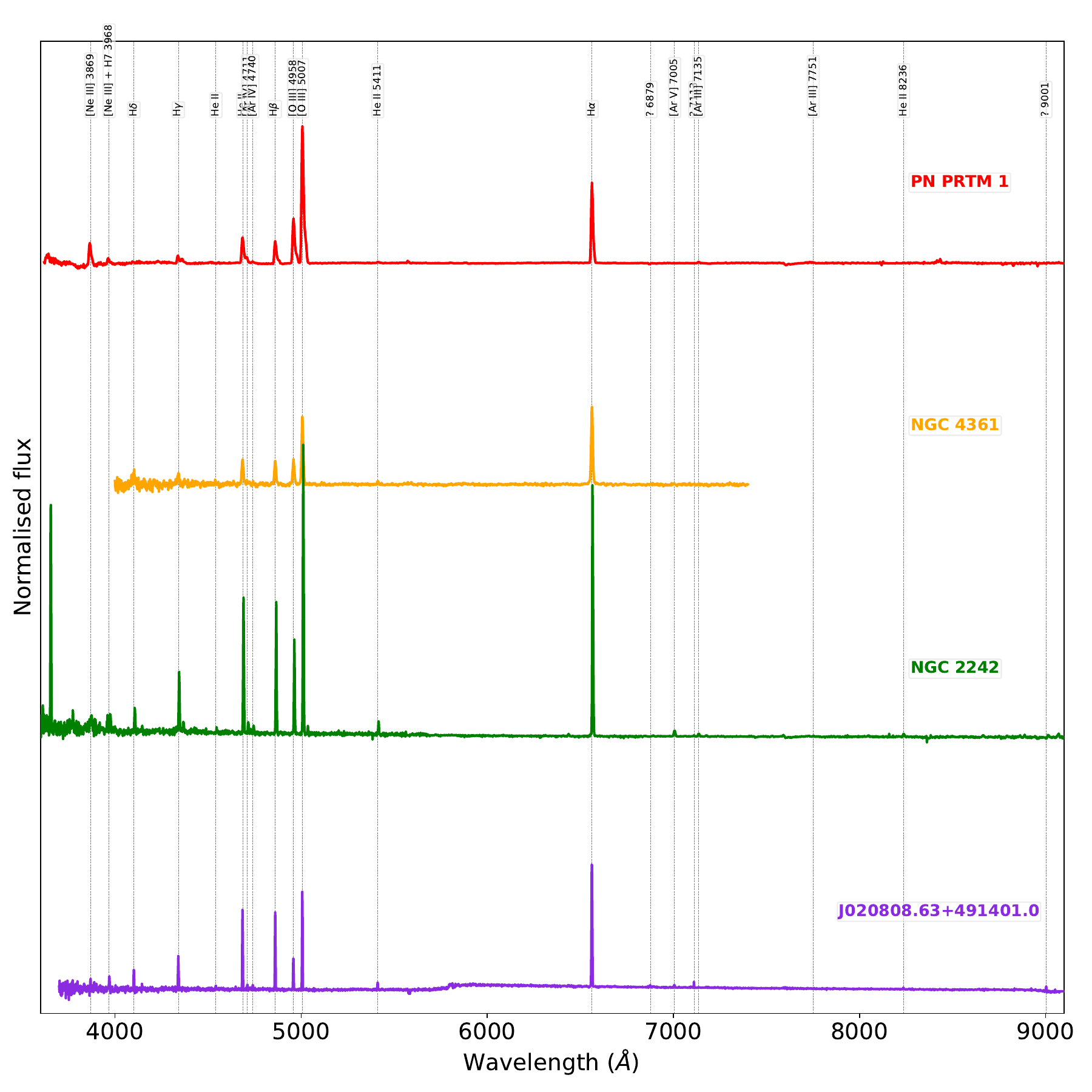}
\caption{Spectra of 3 known PNe and the new PN. From upper to lower the ID of the
  sources are PN PRTM 1, NGC~4361, NGC~2242, and the recent discovery
  J020808.63+491401.0. The spectra have all been scaled and normalized
  for display purposes. The spectra of the known PNe were got from HASH database.} 
  \label{fig:compare-spectra}
\end{figure*}

Figure~\ref{fig:compare-spectra} shows the individual spectra of the three known 
PNe together with that of J020808.63+491401.0. In all of them, the shape of
the continuum is quite similar, in which the emission lines stand out.
The spectra of J020808.63+491401.0 exhibit almost the same emission lines,
the typical hydrogen Balmer ones and some high-ionization lines produced by
transitions generated
by high-energetic photons. In all spectra, low-ionization lines are not present.

\begin{figure}[hbt!]
  \centering
\includegraphics[width=0.5\columnwidth]{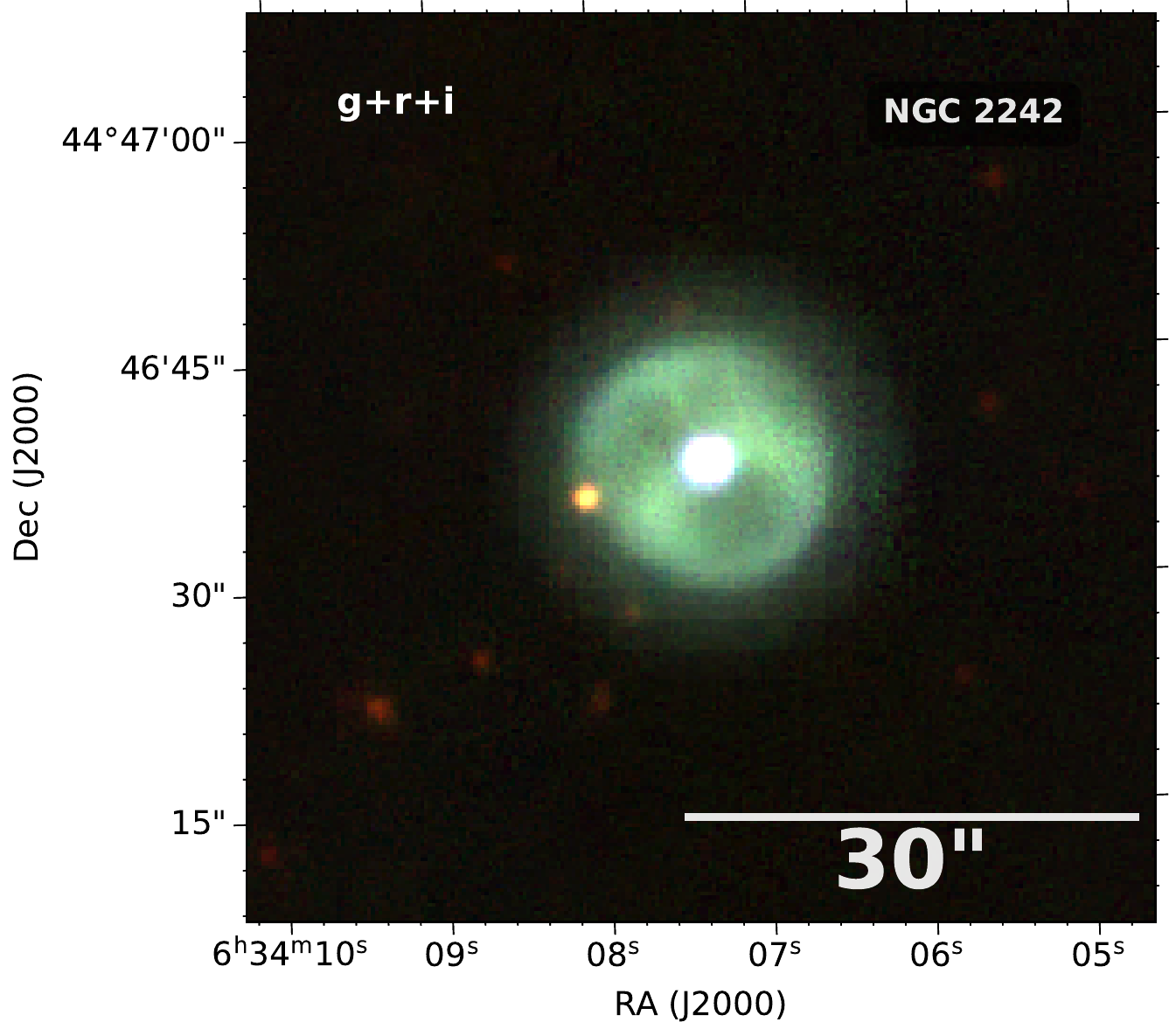}
\includegraphics[width=0.48\columnwidth]{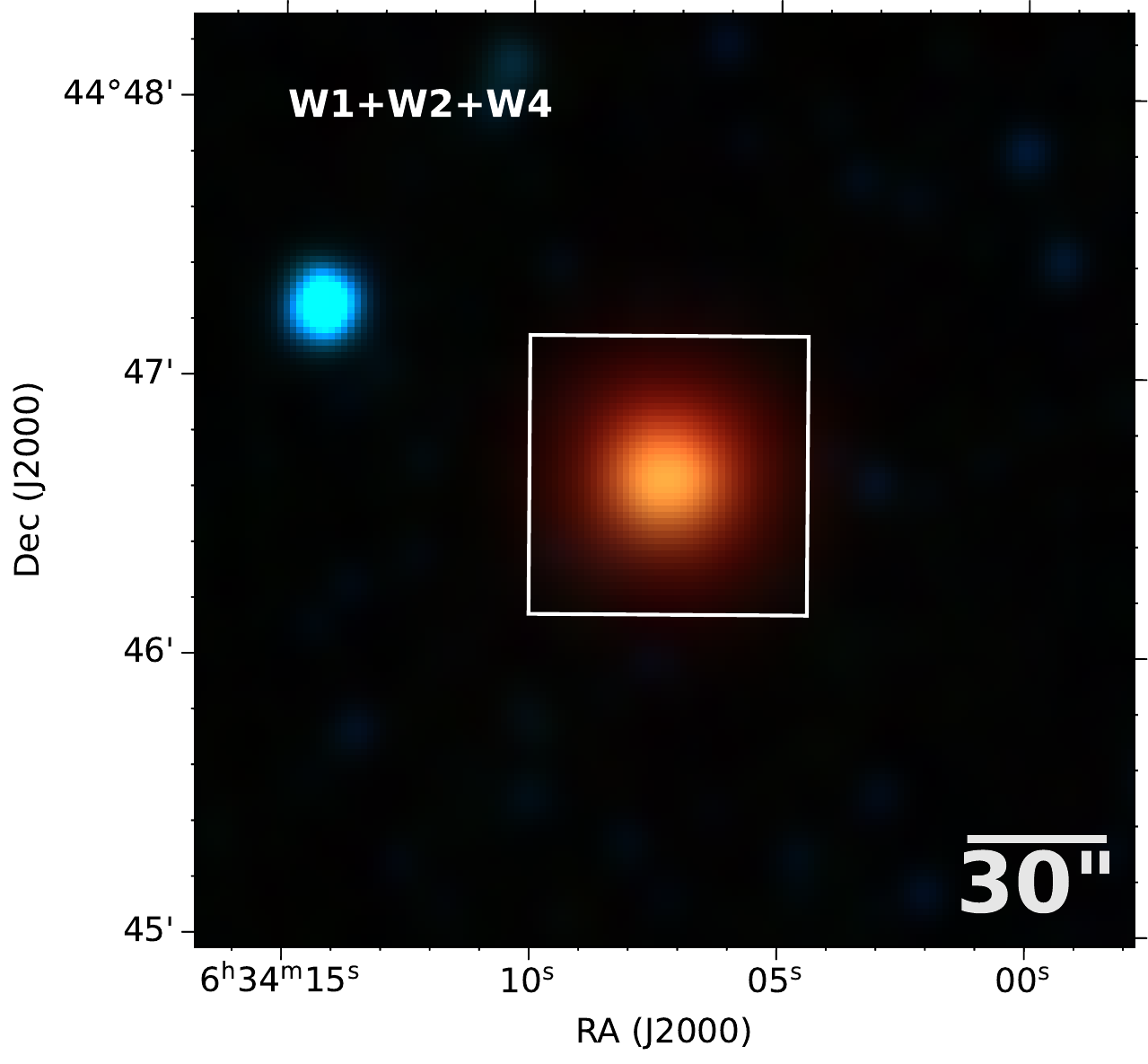}\\
\includegraphics[width=0.5\columnwidth]{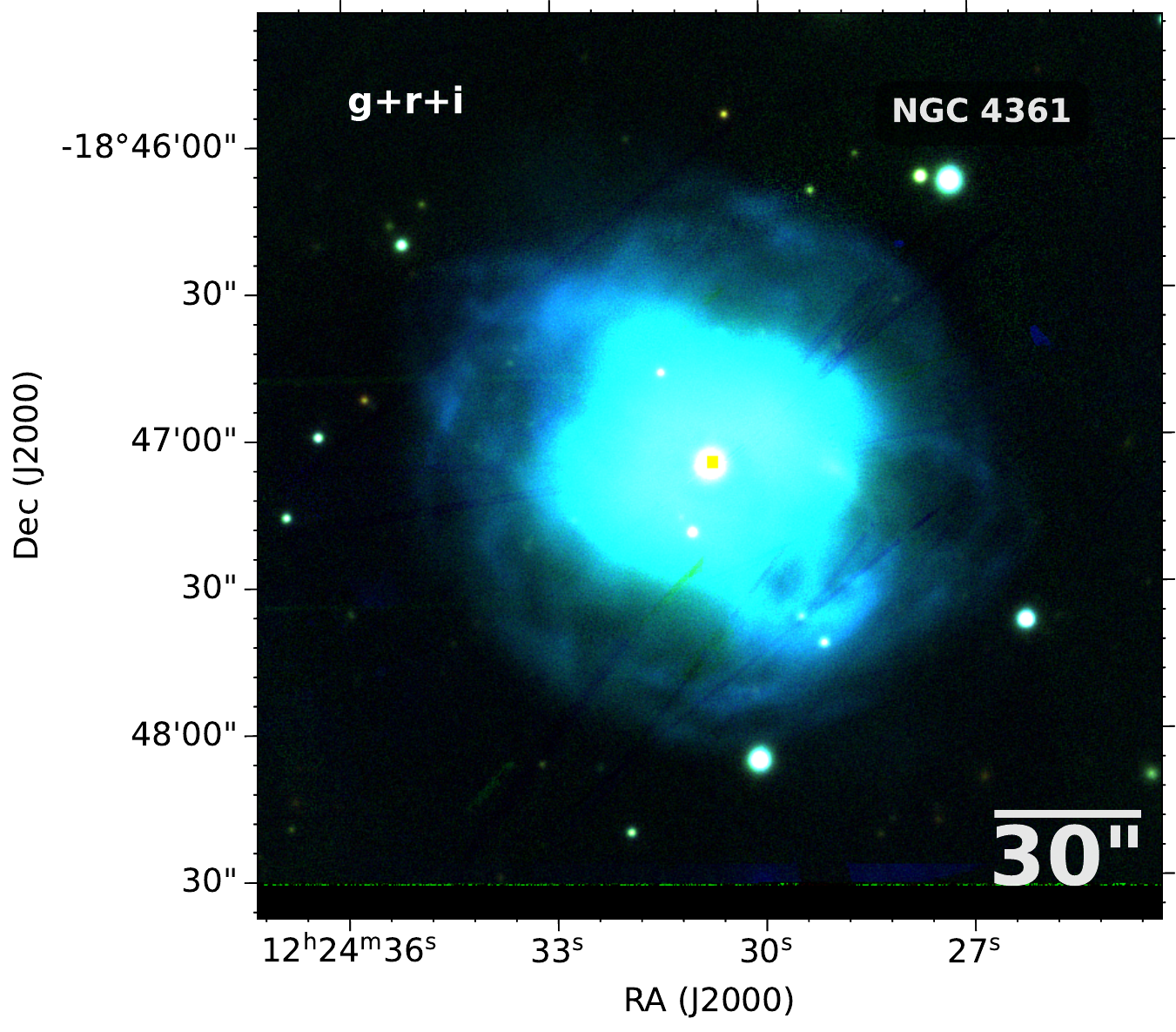}
\includegraphics[width=0.48\columnwidth]{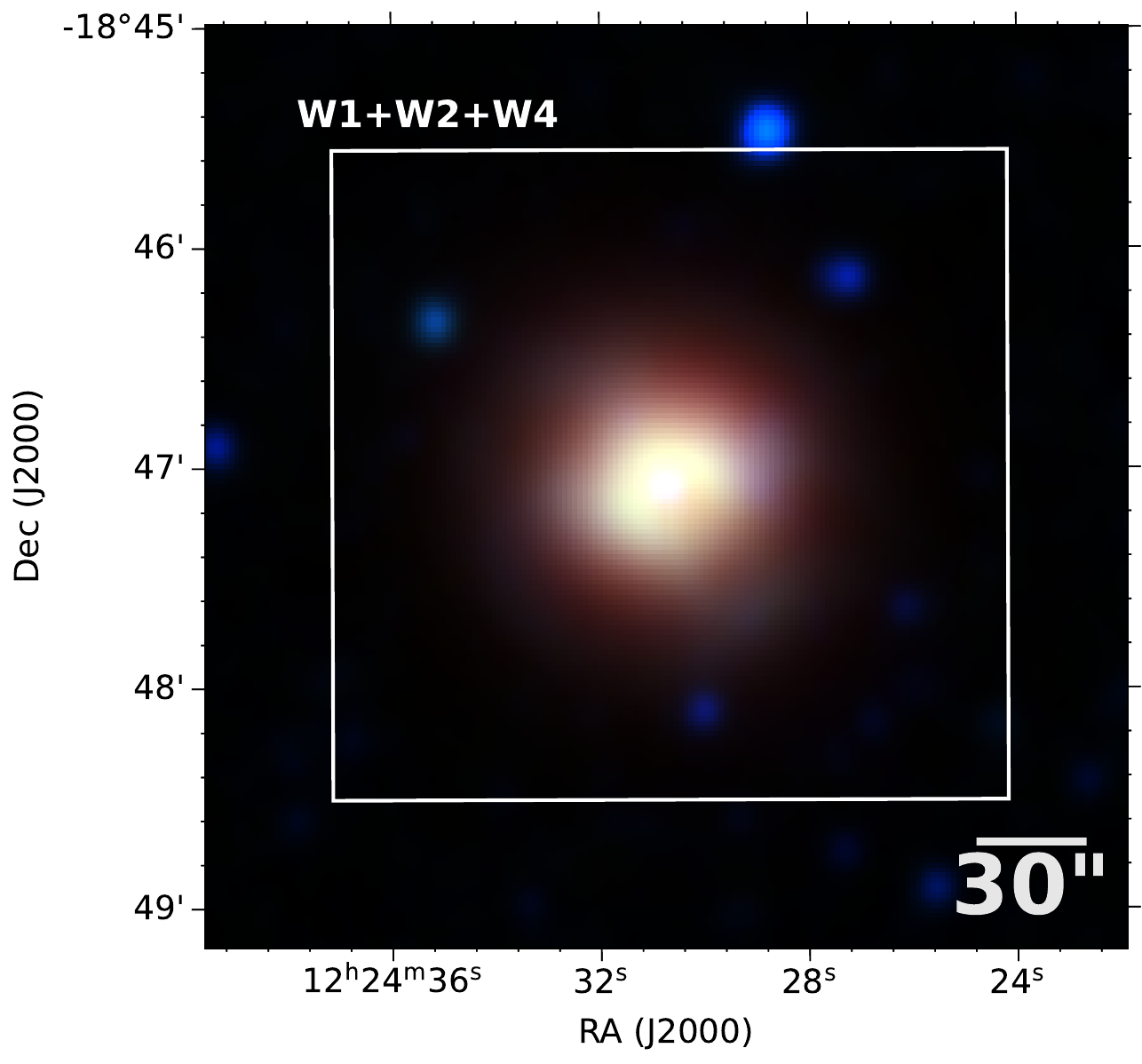}\\
\includegraphics[width=0.515\columnwidth, trim=280 10 330 10, clip]{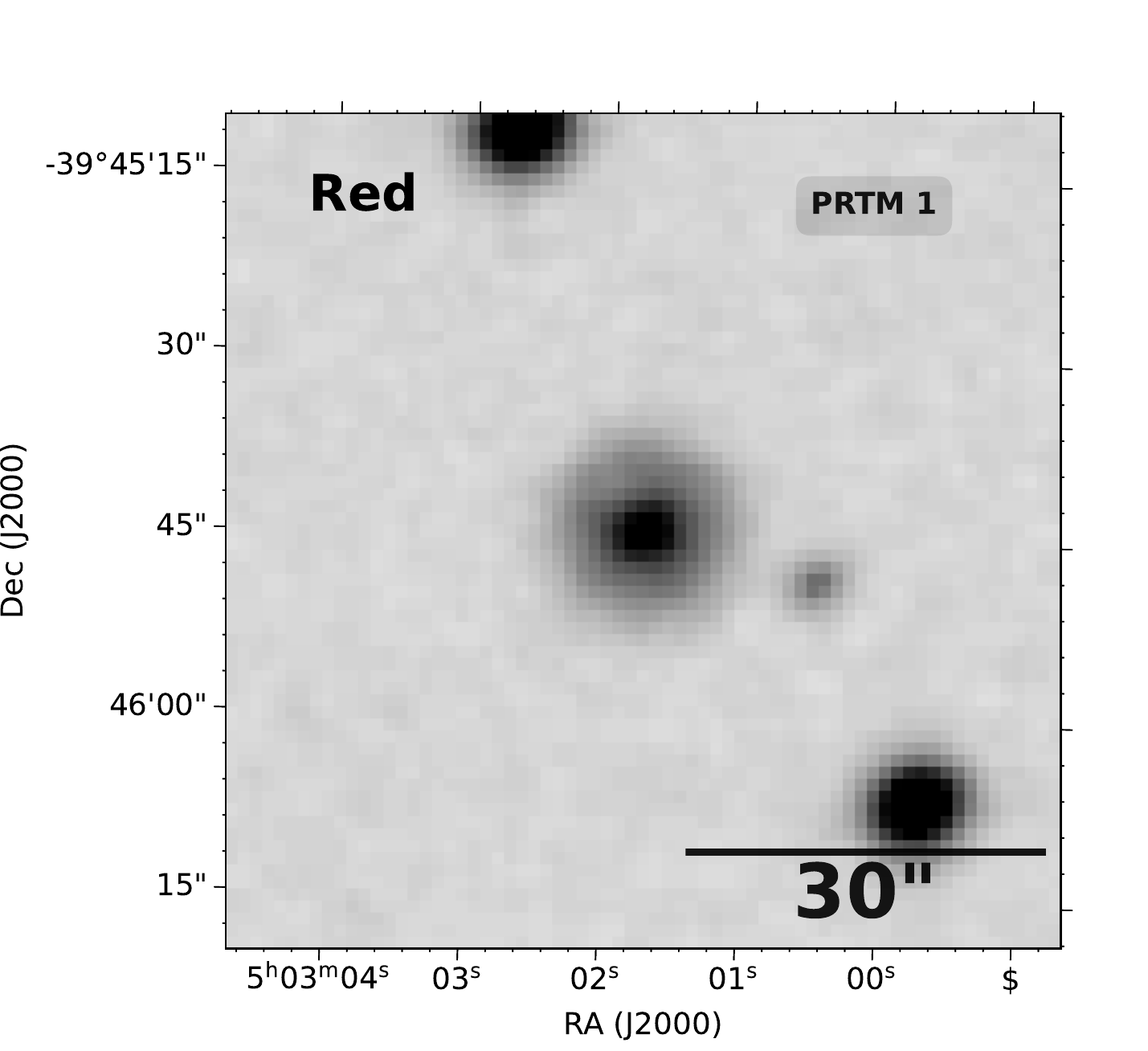}
\includegraphics[width=0.45\columnwidth, trim=58 0 0 0]{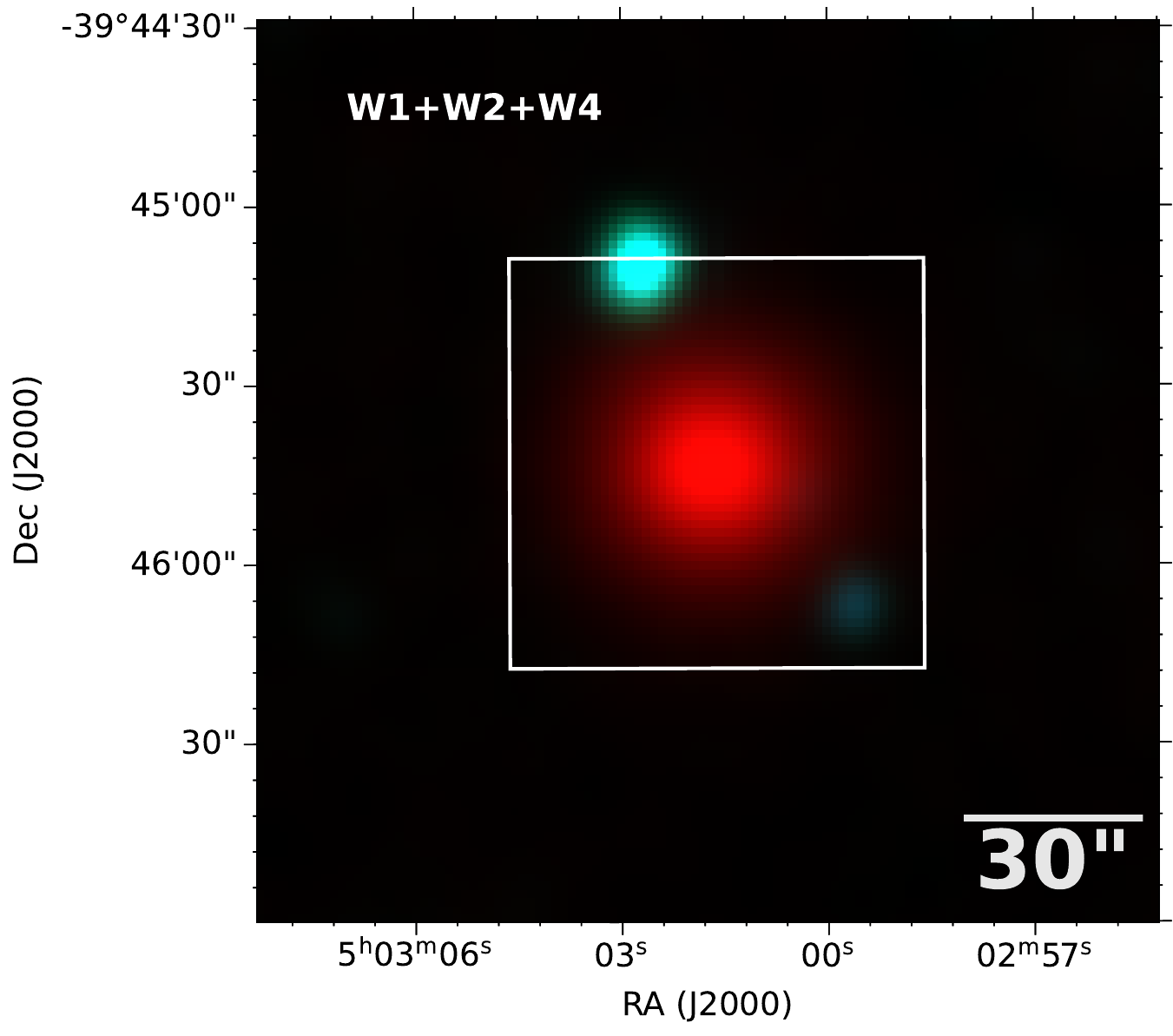}\\

  \caption{Optical (\textit{right}) and IR (\textit{left}) images of the 
    high-ionization PNe: NGC~2242 (\textit{upper}), NGC~4361(\textit{medium}), and
    PN PRTM~1(\textit{bottom}).
    The optical coloured images were constructed by combining the
    $g$, $r$ and $i$ Pan-STARRS filters in the green, blue and red channels, respectively.
    The infrared RGB images were constructed by using the the W1 (3.368 $\mu$\,m), W2
    (4.618 $\mu$\,m), and W4 (22.194 $\mu$\,m) WISE filters. As shown in Figure~\ref{fig:image}, 
    the white box overlaid on the IR images represents the size of the optical images.} 
  \label{fig:images-known}
\end{figure}

Figure~\ref{fig:images-known} displays the optical (\textit{left}) and IR (\textit{right})
images of the three known PNe. The optical images of two of the known
PNe are morphologically similar to those of J020808.63+491401.0, which presents a round 
(or an enveloping) shape. The case of NGC~2242 (\textit{upper left panel}) is a
round double shell high excitation PN with bright CSPN. While  PRTM~1
(\textit{bottom left panel}) is a bright CSPN in a round PN. J020808.63+491401.0
shares the same image features as these two PNe. It exhibits a round shell with a bright central star
(see left panel of Figure~\ref{fig:image}). NGC~4361 (\textit{middle left panel} in 
Figure~\ref{fig:images-known}) is a bit different than the LAMOST object. 
It is a bright slightly oval high excitation PN
with bright CSPN and complex internal structures. The IR image of NGC~2242 (\textit{upper})
and PRTM 1 (\textit{bottom}) is also quite similar to the IR image of J020808.63+491401.0
(right panel of Figure~\ref{fig:image}), on which a more external shell looks like
emitting in IR. Also, an intense emission is perceptively in the W4 filter. As in the optical
image, the IR image of NGC~4361 (\textit{middle}) is quite different from those
of J020808.63+491401.0. Figure~\ref{fig:images-known} shows, apparently, 
all these PNe have a similar shape. 
 In addition, in all of them, the central star is clearly perceptible, much like the LAMOST PN, 
 potentially indicating the high-temperature nature of their central stars.

In conclusion, the spectrum of J020808.63+491401.0 is very similar to the spectra
of the three PNe presented here, indicating that it is probably a true PN.
Furthermore, this similarity could indicate that it is a planetary nebula with a
high degree of ionization. Similarly, comparing optical and IR images
with these confirmed PNe emphasizes that the emitter could be a real PN.

\section{Investigating New PN through LAMOST Spectral Modeling}
\label{sec:model}

\begin{table}[hbt!]
	\begin{tabular}{lc} 
    \hline
              Parameter & Value \\
               \hline
              \textit{Observed spectrum} & \\
              RA (J2000) & 02:08:08.61 \\
              Dec (J2000) & 49:14:01.30 \\
              EC  & 6.6 \\
              V$_r$(Helio) [km s$^{-1}$] &   $-$40 \\
              V$_{exp}$(H$\beta$) [km s$^{-1}$] & $<$100 \\
        \hline
        \textit{Best-fit model} & \\
		$\log(\mathrm{Temperature~of~central~ionizing~source}) [\mathrm{K}]$  & 05.26  \\
		$\log(\mathrm{Luminosity~of~central~ionizing~source}) [\mathrm{erg~s^{-1}}]$  & 37.12  \\
		$\log(\mathrm{H-density) (cm^{-3})} $ & 03.78  \\
                 $\log(\mathrm{R_{in}) (cm)}$ & 16.60\\
                $\log(\mathrm{R_{out}) (cm)}$ & 17.05 \\
                Distance [kpc] & 02.31  \\
                $\log(\mathrm{He/H})$ & $-$0.92 \\
                $\log(\mathrm{C/H})$ & $-$4.15 \\
                $\log(\mathrm{N/H})$ & $-$4.72 \\
                $\log(\mathrm{O/H})$ &  $-$3.83\\ 
                $\log(\mathrm{Ne/H})$ & $-$4.58 \\
                $\log(\mathrm{Si/H})$ & $-$5.00\\ 
                $\log(\mathrm{S/H})$ &  $-$6.00\\ 
                $\log(\mathrm{Ar/H})$ & $-$6.24\\
                \hline
                E(B-V) & 00.03 \\
                 Total $\chi^2$ & 07.91  \\
                 \hline
         \textit{Output parameters from the model} & \\
         T([\ion{O}{iii}]) [K] & 21 600 \\
         $\mathrm{<n_{e}> [cm^{-3}]}$ & 7 400 \\
         \hline
	\end{tabular}
 \caption{Best-fit {\sc cloudy} model parameters for LAMOST J020808.63+491401.0, along with other derived physical properties obtained from observed spectra. The excitation class (EC) was determined using the expression from \cite{1990ApJ...357..140D}, and the expansion velocity (V$_{exp}$) was calculated following \cite{1984A&A...136..200S}. The heliocentric radial velocity (V$_r$(Helio)) was computed using nebular lines in the blue arm at wavelengths 4\,686, 4\,861, 4\,959, and 5\,007\AA. Additionally, the model output parameters include the estimated temperature of the nebula derived from the [\ion{O}{iii}] emission lines and the average electronic density.}
 \label{tab:parameter-best-fit}
\end{table}

Given the limited number of observational constraints, we have adopted a very
simple model of a planetary nebula, which consists of a homogeneous
gaseous sphere surrounding a hot star that radiates as a black body.
Although the LAMOST spectra are not in physical flux unity,
they are in calibrating relative flux, which makes it possible to compare 
them with other spectra in terms of normalized flux. Therefore, we compare the LAMOST 
spectra of the observed PN with the models, using the normalized spectra to H$\beta$.

Our modeling process utilized version c22.01 of the {\sc cloudy} photo-ionization 
code \citep{Ferland:2017}. {\sc cloudy} is a comprehensive tool employed to deduce 
the physical characteristics and elemental abundances of PNe by accounting for a 
wide array of ionization processes. It excels in simulating the physical conditions 
of non-equilibrium gas clouds exposed to an external radiation field, resolving 
thermal, statistical, and chemical equilibrium equations self-consistently within 
a non-local thermodynamic equilibrium (NLTE) framework. The {\sc cloudy} code 
incorporates a diverse species database of 625 entities, spanning atoms, ions, 
and molecules, along with multiple databases for modeling spectral lines. 
Furthermore, we make reference to prior research endeavors employing {\sc cloudy} 
for similar applications, such as simulating broadband photometry 
for the Large Synoptic Survey Telescope (LSST) (\citet{Vejar:2019} 
and formulating new color criteria for the identification of PNe in 
surveys like J-PLUS and S-PLUS \citet{Gutierrez-Soto:2020}.

In this vein, we utilized {\sc pyCloudy} \citep{Morisset:2013} a
Python package\footnote{\url{https://sites.google.com/site/pycloudy/home}} which
provides a set of tools for interfacing with the photoionization code {\sc cloudy} (\url{www.nublado.org}).
{\sc pyCloudy} was employed to generate input parameters for the model and process the resulting output files.

We assume a spherically symmetric gaseous material surrounding a 
central ionizing source, with dimensions determined by the 
inner ($R_{\mathrm{in}}$) and outer ($R_{\mathrm{out}}$) radii. 
The central ionizing radiation is assumed to be a black body with a temperature 
of T$_{\mathrm{BB}}$ (K) and luminosity L
(erg~s$^{-1}$). 
We assume a uniform filling factor of unity and use {\sc pyCloudy} to iteratively 
generate input files for the models. 
Many models were generated in an effort to find the best fit.

In agreement with the LAMOST spectra and similarities with these high-ionization 
PNe (see Figure \ref{fig:compare-spectra}), we considered a range of effective 
temperatures between 100$\times10^3$~K and 200$\times10^3$~K in steps of 10$\times10^3$~K, 
%
luminosities ranging from 400 $L_{\odot}$ to 10\,100 $L_{\odot}$ in steps of 300 $L_{\odot}$, 
and hydrogen densities ranging from 500 to 6\,000 cm$^{-3}$ in steps of 500 cm$^{-3}$.
We note that the spectra of LAMOST J020808.63+491401.0 are quite similar to the spectrum of NGC~2242. 
Therefore, we first adopted the abundances mainly from NGC~2242 and adjusted the values of He and Ar 
to obtain a better match with the observed spectra. We also used the inner and outer radii that showed a 
good fit to the data. 
We used the distance (geometrical) estimates of the object based on the parallax of GAIA of 2.31 kpc, 
as determined by \citet{Bailer:2021}.
After producing the models, we considered  that the observed spectra could be affected by
dust interstellar extinction. 
Therefore, we applied the reddening curve of R(V) = 3.1 to each modelled 1D-spectrum, using color excesses, E(B-V), 
ranging from 0.0 to 0.1 in steps of 0.01, and also for 0.2. 
This was done by implementing the Python package \texttt{dust\_extinction}\footnote{\url{https://dust-extinction.readthedocs.io/en/stable/##}}. 
This produced more than 40\,000 models. 

\subsection{The best-fit model}
\label{sec:best-fit}

The authors \citet{Helton:2010, Mondal:2018, Pavana:2019, Mondal:2020, Pandey:2022a, Pandey:2022b}
used the $\chi^2$ method to compare observed spectroscopic data,
mainly from novas sources, with modelled {\sc cloudy} spectra to find the best
fit models. Following their approach, we selected the final best-fit models 
through multiple iterations that involved testing various input parameters, 
as previously demonstrated, using $\chi^2$ minimization. To accomplish this, 
we compared the fluxes of the observed lines with the lines generated by the models,
enabling us to automatically assess the goodness of fit by calculating the $\chi^{2}$ values. 
We determined the minimum $\chi^2$ value for the new PN, which corresponded to the best-fit model, 
employing the following equation:

\begin{equation}
\label{eq:chi-red}
   \chi^{2}_{\mathrm{min}} = \sum^{8}_{i = 1} \frac{(M^{\mathrm{best}}_i - O_i)^2}{\sigma^{2}_i}
\end{equation}

Here, $M_i$, $O_i$, and $\sigma_i^2$ represent the modelled line flux ratios, 
observed line flux ratios, and uncertainty in the observed line flux ratios, respectively.
The number 8 in the summation symbol represents the number of parameters used to calculate the $\chi^{2}$ value. 
In this case, we used the 8 most intense flux lines of the blue arm of the spectra, which are
 [\ion{Ne}{iii}] 3967.5 \AA,
  H$\delta$, 
  H$\gamma$, 
  \ion{He}{ii} 4685.9 \AA, 
  H$\beta$, 
  [\ion{O}{iii} 4958.9 \AA,
  [\ion{O}{iii}] 5006.8 \AA~and
  \ion{He}{ii} 5411.7 \AA

For this exercise, we are not utilizing the lines in the red-arm of the spectrum of this object, 
where the most intense line is H$\alpha$. The reasons for this choice are explained in Section~\ref{sec:find}. 
We measured the flux of each line by fitting a 1D-Gaussian to the 10 \AA~region centered on the 
line and then integrating over the best-fit Gaussian within a range of approximately 5$\sigma$. 
Being $\sigma$ the standard deviation of the individual line fit Gaussian. 
 We used the  \texttt{estimate\_line\_parameters} function from the \texttt{Astropy specutils.fitting} 
 package\footnote{\url{https://specutils.readthedocs.io/en/stable/index.html}} to obtain the parameters that best fit the 1D Gaussian model.

The uncertainty in the flux ratios at 1$\sigma$ of the observed lines was computed following the equation presented by \citet{Tresse:1999}:

\begin{equation}
\label{eq:err-sigma}
  \sigma_{F} = \sigma_{c} D \sqrt{2N_{\mathrm{pix}} + \frac{EW}{D}}
\end{equation}

 Where $\sigma_{c}$ refers to the mean standard deviation per pixel of
the continuum on each side of the line, $\mathrm{N_{pix}}$ is the number of pixels
under the line, which is equated to the FWHM of the standard Gaussian profile 
following \citet{Mayya:2023}, $D$ denotes the spectral dispersion ($\sim 1.0002$ \AA~per pixel for LAMOST). 
Finally, EW is the equivalent width of the line. 


\begin{table*}
	\centering
	\caption{Observed and best-fit {\sc cloudy} model line fluxes were 
 obtained for the emission lines used to identify the model that best 
 reproduces the spectra of LAMOST J020808.63+491401.0. 
 The flux measurements of the emission lines were normalized using the H{$\beta$}  
 flux line as a reference. The provided wavelength values correspond to laboratory measurements.}
	\label{tab:abundances}
	\begin{tabular}{lcccc} 
                \hline
		\hline
		Line                & $\lambda$(\AA) & Observed Flux  & Model  Flux & $\chi^{2}$  \\
		\hline
		[\ion{Ne}{iii}] + H7 3968  & 3967.46 & 0.17 $\pm$ 0.05 & 0.15 &  0.118 \\
	    H{$\delta$}              & 4101.74 & 0.24 $\pm$ 0.05 & 0.22&  0.067 \\
		H{$\gamma$}              & 4340.71 & 0.43 $\pm$ 0.04 & 0.50 &  2.170\\
        \ion{He}{ii}              & 4685.99 & 1.05 $\pm$ 0.06 & 0.99 &  0.977\\
        H{$\beta$}               & 4861.33 & 1.00 $\pm$ 0.06 & 1.00 &  0.000\\
        $[$\ion{O}{iii}$]$            & 4958.91 & 0.39 $\pm$ 0.03 & 0.41 &  0.529\\
        $[$\ion{O}{iii}$]$            & 5006.84 & 1.24 $\pm$ 0.06 & 1.24 &  0.000\\
        \ion{He}{ii}               & 5411.52 & 0.08 $\pm$ 0.01 & 0.09 &  4.044\\
        \hline
	\end{tabular}
\end{table*}

The value of $\mathrm{\chi^{2}{\min}}$ indicates the {\sc cloudy} generated spectra model that
matches the observed spectra most closely. Although our model reproduces a wide range of 
observable effects, there are certain limits to its phenomenology.
To identify the best fit, we evaluated approximately 40\,000 generated models and selected those 
that matched the spectra of J020808.63+491401.0 very well, based on the minimum 
value of $\chi^{2}$ calculated from the expected flux ratio lines of all the models. 
The parameters of the best-fit model
are presented in Table~\ref{tab:parameter-best-fit}. The model
achieved a $\chi^{2}$ value of approximately 7.9, representing the best fit.
The temperature of the central ionizing source was determined to be approximately 180$\times$10$^{3}$ K, 
which is a very high temperature as expected, given that a high ionizing source is necessary 
to produce high-excitation emission lines. The luminosity of the central ionizing star in the 
best-fit model was estimated to be approximately 3\,400 L$_{\odot}$. The color excess of the 
model that matched the observed one most closely was 0.03. This finding suggests minimal reddening effects attributed to interstellar dust on the light passing through, affirming negligible impact.

\begin{figure*}[hbt!]
\centering
\includegraphics[width=\linewidth, trim=10 30 10 10, clip]{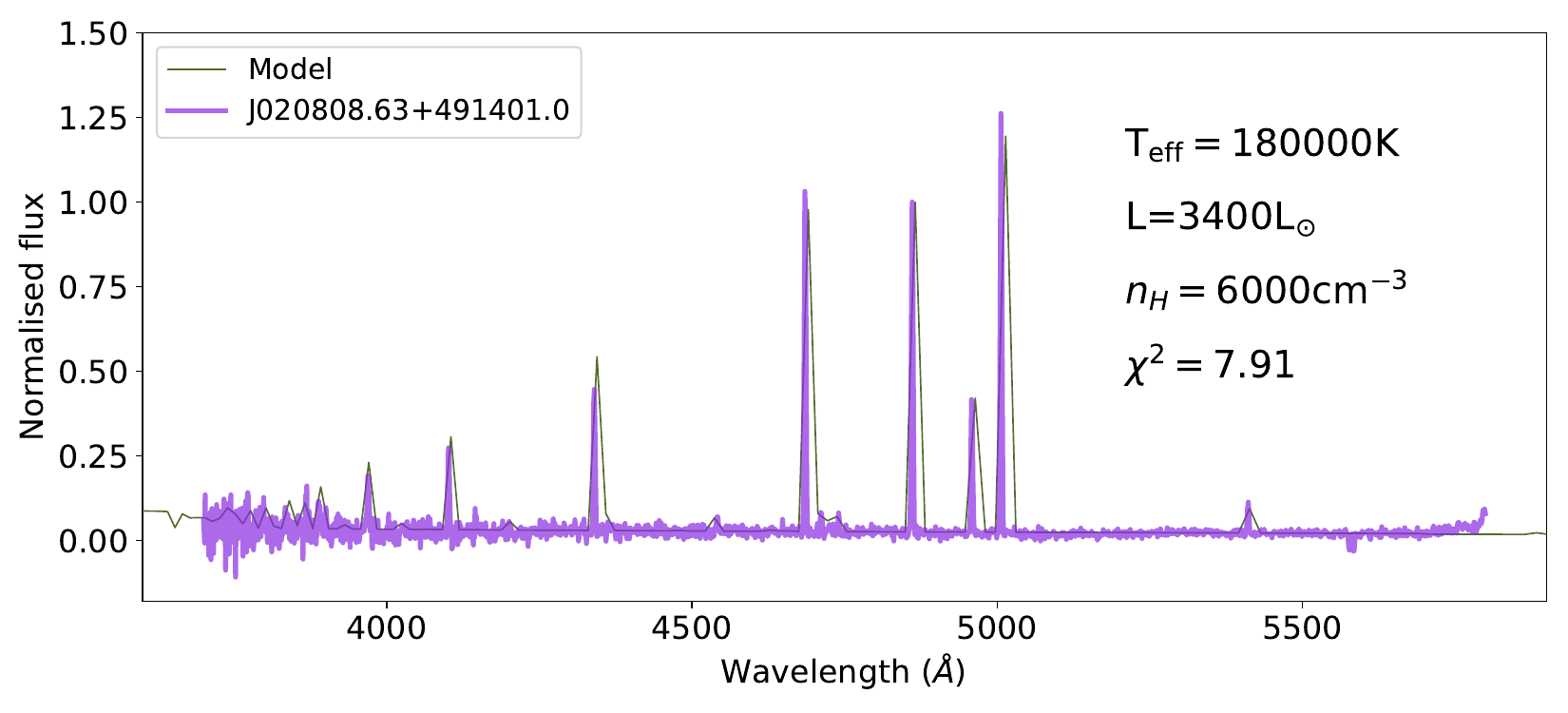}
\caption{The best {\sc Cloudy} fitted model, characterized by a $\chi^2$ value of approximately 8, represents the minimum value obtained through the comparison of the observed spectra with each model. Only the blue arms of the observed spectra are displayed. The spectra were normalized to H{$\beta$}.} 
  \label{fig:spectra-obs-model}
\end{figure*}

Table~\ref{tab:abundances} shows the relative flux to H{$\beta$} of the observed emission lines, 
as well as the {\sc cloudy} modelled values for the 8 emission lines and their 
corresponding $\chi^{2}$ values. The observed and modelled emission lines are in very good agreement, 
as evidenced by their low values of $\chi^{2}$. For instance, H{$\delta$} and [\ion{O}{iii}] 5006.8 \AA~have $\chi^{2}$ values 
of approximately 0.07 and 0.00, respectively. For comparison, we have overlaid the {\sc cloudy} 
model that represents the best-fitted spectra on the observed 
optical spectra in Figure~\ref{fig:spectra-obs-model}. 
As can be seen in the figure, all of the observed emission lines are well-reproduced by the model.

\subsection{Comparison with post-AGB stellar evolutionary tracks}
\label{sec:tracks}

\begin{figure}[hbt!]
\centering
  \includegraphics[width=\linewidth]{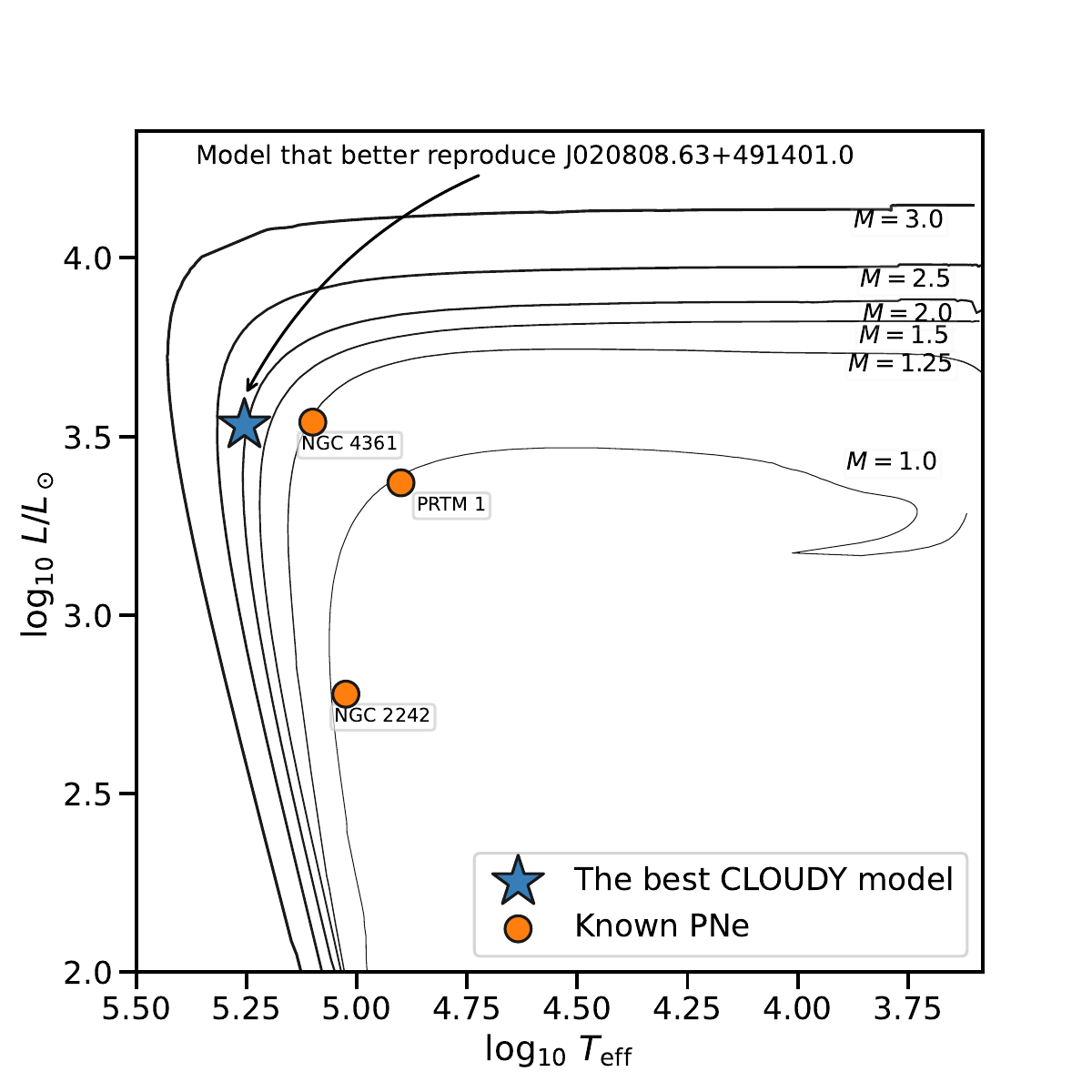}
  \caption{HR diagram of 
  H-rich  
  PN central stars, showing the observed luminosity
    and effective temperature of {\sc cloudy} model that better match the observed spectra
    (big blue stars), as well as post-AGB evolutionary tracks from \citet{Miller:2016} (shown as 
    solid lines and labeled with the initial stellar mass in solar masses). 
    The known PNe that have been selected to be similar to the new PNe 
    discussed in Section~\ref{sec:comp} are represented as filled orange circles. 
    The temperatures and luminosities of these known PNe were obtained from 
    the catalog of central stars compiled by \citet{Weidmann:2020}.} 
 \label{fig:track-evolutive}
\end{figure}

Figure~\ref{fig:track-evolutive} displays the position of the post-AGB evolutionary 
models that best match the observed PNe in the luminosity versus effective temperature diagram, 
indicated by the blue stars. The diagram also shows the position of three known PNe (represented 
by orange circles) that have been used for comparison with the new ones. The solid lines 
represent post-AGB evolutionary tracks for stars with initial masses of 1.00, 1.25, 1.50, 2.00, 2.50, 
and 3.00 M$_{\odot}$ and final masses of 0.532, 0.566, 0.583, 0.583, 0.616, and 0.706 M$_{\odot}$, 
respectively, assuming approximately solar metallicity \citep{Miller:2016}. 
The evolution proceeds from right to left on the diagram, followed by top to bottom.
 
Comparing the model central ionizing black body 
($\mathrm{T_{eff}} \sim 180\times$10$^{3}$ K and 
L $\sim$ 3\,400 L$_{\odot}$) 
with the post-AGB tracks would imply an initial mass of the progenitor star around 2 M$_{\odot}$. 
This stellar initial mass corresponds to a final mass of 0.5826 M$_{\odot}$ 
in the theoretical tracks of \citet{Miller:2016}, meaning that around 1.42 M$_{\odot}$ should have 
been lost to mass loss, which mostly occurs during the AGB stage.


\section{Discussions and conclusions}
\label{sec:conclu}

We present color criteria based on broad-band 
photometry from GAIA and Pan-STARRS for the selection of compact PN candidates. 
We have observed that the \((G - g)\) versus \((G_{BP} - G_{RP})\) and \((G - r)\) versus \((G_{BP} - G_{RP})\) color-color 
diagrams allows us to distinguish very compact PNe. This distinction is particularly significant for PNe 
with strong emission lines and low continuum, which can notably impact the $g$- and $r$-bands.
While this method maintains a conservative approach and is not exhaustive, 
it aids in identifying candidates highly likely to be authentic PNe.
Additionally, we have identified a distinct region in both color-color 
diagrams that corresponds to extremely extended PNe, characterized by their very low surface brightness. 
To assess the potential of utilizing broad-band photometry and the associated color criteria, in combination with 
the LAMOST survey for identifying compact PN candidates and subsequent spectroscopic confirmation, we have analyzed 
 HASH PNe with GAIA and Pan-STARRS photometry, as well as LAMOST spectra. Our analysis has shown that the most compact 
PNe, whose LAMOST spectra display pronounced emission lines and very weak continuum, are concentrated within the 
previously defined compact PNe region of the color-color diagram. In contrast, the most extended PNe, as indicated 
by their LAMOST spectra dominated by the central stars, with the continuum playing a significant role, is found 
within the extended region of the diagram.

In the case of  probable/likely HASH PNe, most were found in the region of very extended PNe. 
Analysis of LAMOST spectra revealed that only three objects exhibited weak emission lines, while the rest showed spectra resembling white dwarf stars. 
These objects varied significantly in size, with the largest measuring a radius of 3600.0 arcseconds and the smallest at 60.0 arcseconds. 
Only one object, Kn 27, fell outside both categories in the two color-color diagrams, with its spectrum displaying emission lines.
To heighten the likelihood of PN candidates being genuine PNe, it becomes imperative for them to meet both color criteria.

To further assess the applicability of the approach introduced here, 
we applied the selection criteria to the catalog of emission line 
objects presented by \citet{Skoda:2020}. This endeavor resulted in the discovery of a new PN. 
This catalog was created using an active deep-learning method
to identify objects with emission lines in the LAMOST spectra database.
The initial step of our selection process involved utilizing the
\((G - g)\) versus \((G_{BP} - G_{RP})\) and \((G - r)\) versus \((G_{BP} - G_{RP})\) color-color diagrams, 
to scrutinize the list of objects from \citet{Skoda:2020}.
From this examination, we identified two candidates that simultaneously met both color criteria. Subsequently, after analyzing their LAMOST spectra, we singled out the object with the LAMOST ID J020808.63+491401.0 as a genuine PN.  
 This object holds GAIA galactic coordinates, specifically 
%
$\alpha = 02^\mathrm{h}~08^\mathrm{m}~08.^{\!\!{\mathrm{s}}}61$, 
$\delta = +49^{\circ}~14^{'}~01.^{\!\!{''}}30$, J2000.
The distance of the object was estimated to be 2.31 kpc placing
it is at 0.47 kpc from the Galactic plane.
We propose the IAU nomenclature: PN~G135.6$-$11.7.

We got the low-resolution spectra of J020808.63+491401.0 
from the LAMOST spectroscopic database. Its spectrum
shows the classical Balmer-H emission lines together with
high-ionization lines and lack of low-excitation lines,
indicating the presence of a hot central star ionizing the nebular gas.
This spectral characteristic in conjunction with optical and infrared 
images that look similar to that high-ionization known as PNe reinforces 
the fact that the object is a real PN.

We use the 1D-photoionization code {\sc cloudy} to model
the observed LAMOST spectrum of the new PN finding, which is
the first attempt to derivate the physical parameters of the object.
Our model foresees the presence of a hot central ionizing source
with a temperature of 180$\times$10$^{3}$ K, and with a
rough luminosity of 3\,400 L$_{\odot}$. Our results are 
in agreement with previous studies of PN central stars, 
which have found that the majority of them have masses 
in the range of 0.53-0.68 M$_{\odot}$ and have evolved 
from stars with initial masses of 1-3 M$_{\odot}$ \citep{Weidmann:2020}. 
However, the relatively high effective temperature and luminosity of 
the central star of this new PN suggests that it may be younger 
than most PNe in the catalog of \citet{Weidmann:2020}. 

The estimated initial mass of the progenitor star, combined with the estimated final 
mass from the post-AGB tracks implies that the star lost about 1.42 M$_{\odot}$ 
of mass during its evolution, which is consistent with the expected mass loss during 
the AGB phase.

The estimated elemental abundances
from the best-fitted {\sc cloudy} model are:
He/H $\approx$ 0.12, C/H $\approx 7.08\times10^{-5}$, N/H $\approx 1.91\times10^{-5}$,
O/H $\approx 1.48 \times10^{-4}$, Ne/H $\approx 2.63\times10^{-5}$,
Si/H $\approx 1.00 \times10^{-5}$, S/H $\approx 1.00\times10^{-6}$,
Ar/H $\approx 5.75 \times10^{-7}$ for the ionized gas. By considering these
physical parameters, it was plotted along the evolutionary tracks of
\citet{Miller:2016}, finding that is the product of a relatively low mass
(around 2 M$_{\odot}$) progenitor star. The best-fitted {\sc cloudy} model 
is available in the GitHub repository:
\url{https://github.com/AngelGSoto/PNe-LAMOST/tree/main/parameters-best-model}.

This study presents a compact method for 
discovering new compact PNe using data from GAIA, Pan-STARRS, 
taking advantage of the extensive sky mapping provided by these surveys, and LAMOST.
Our results highlight the potential of this approach, alongside the 
inherent challenges, 
such as potential contamination. This paves the way for a more 
comprehensive exploration of these three surveys to uncover additional PNe. 
As such, our intention is to continue 
the exploration of these surveys to uncover more PNe. This work 
represents a significant advancement in the identification and 
characterization of objects in the field of PNe, and we are 
committed to furthering our discoveries through future investigations.

\section*{Acknowledgements}
LAG-S acknowledges funding for this work
from CONICET and FAPESP grants 2019/26412-0.
The Pan-STARRS1 (PS1) Surveys and the PS1 public science
archive have been made possible through contributions by the
Institute for Astronomy, the University of Hawaii, the Pan-
STARRS Project Office, the Max Planck Society and its
participating institutes, the Max Planck Institute for Astronomy,
Heidelberg, and the Max Planck Institute for Extraterrestrial
Physics, Garching, the Johns Hopkins University, Durham
University, the University of Edinburgh, the Queen’s University
Belfast, the Harvard-Smithsonian Center for Astrophysics, the
Las Cumbres Observatory Global Telescope Network Incorpo-
rated, the National Central University of Taiwan, the Space
Telescope Science Institute, the National Aeronautics and Space
Administration under grant No. NNX08AR22G issued through
the Planetary Science Division of the NASA Science Mission
Directorate, National Science Foundation grant No. AST-1238877,
the University of Maryland, Eotvos Lorand University
(ELTE), the Los Alamos National Laboratory, and the Gordon
and Betty Moore Foundation.
This work presents results from the European Space Agency
(ESA) space mission Gaia. Gaia data are being processed by
the Gaia Data Processing and Analysis Consortium (DPAC).
Funding for the DPAC is provided by national institutions, in
particular the institutions participating in the Gaia MultiLat-
eral Agreement (MLA). The Gaia mission website is \url{https://www.cosmos.esa.int/gaia}. 
The Gaia archive website is \url{https://archives.esac.esa.int/gaia}.
Guoshoujing Telescope (the Large Sky Area Multi-Object Fiber Spectroscopic
Telescope LAMOST) is a National Major Scientific Project built by the Chinese
Academy of Sciences. Funding for the project has been provided by the National
Development and Reform Commission. LAMOST is operated and managed by the
National Astronomical Observatories, Chinese Academy of Sciences.

Scientific software and databases used in this work include 
\texttt{TOPCAT}\footnote{\url{http://www.star.bristol.ac.uk/~mbt/topcat/}}
\citep{Taylor:2005}, simbad and vizier from Strasbourg Astronomical
Data Center (CDS)\footnote{\url{https://cds.u-strasbg.fr/}} 
and the following  python packages: numpy, astropy,
specutils, APLpy, matplotlib, seaborn.

\appendix

\section{True PNe with Gaia, Pan-STARRS, and LAMOST Data}
\label{sec:spectra-tPNe}

\begin{figure*}
\centering
\begin{tabular}{ll}
  \includegraphics[width=0.5\linewidth, trim=1 50 10 10, clip]{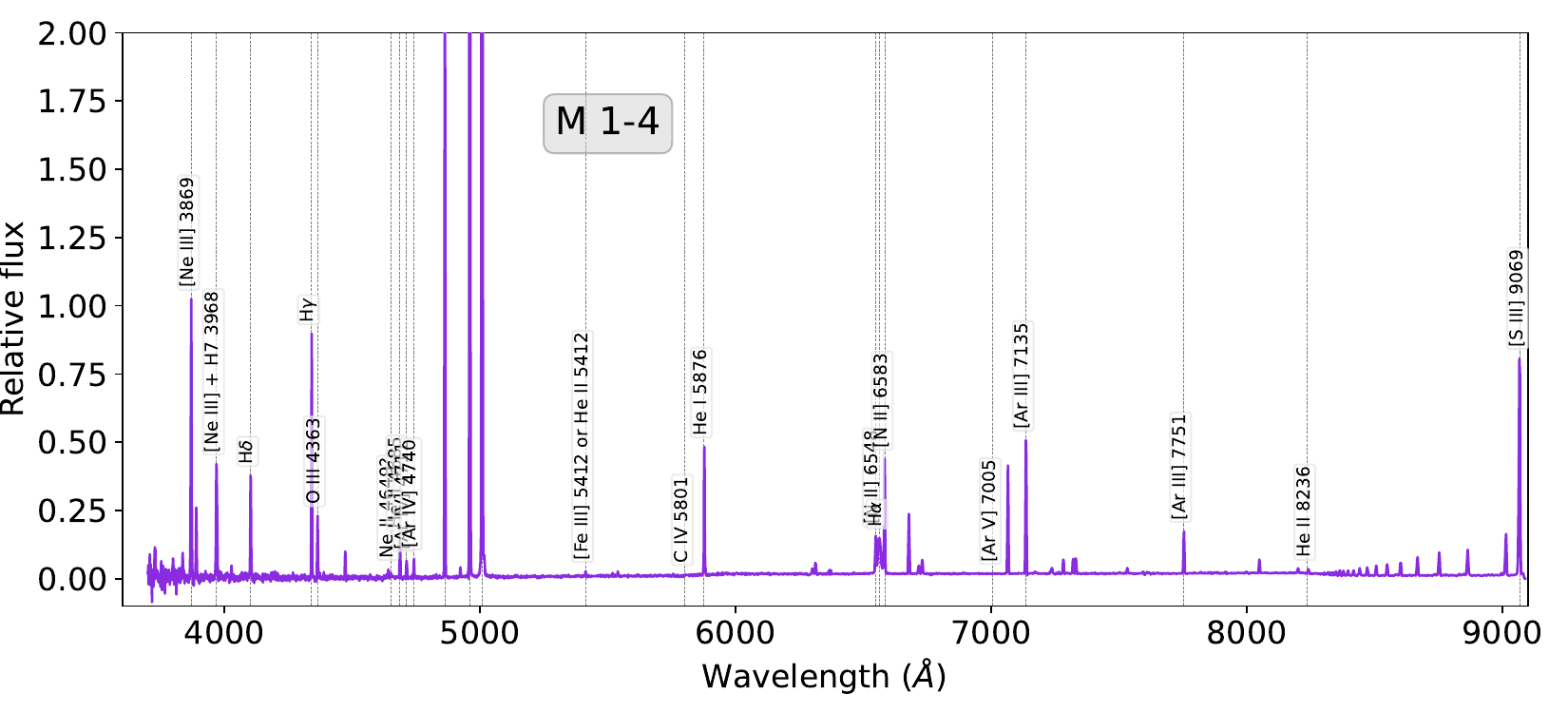} \llap{\shortstack{%
      \includegraphics[width=0.15\linewidth, clip]{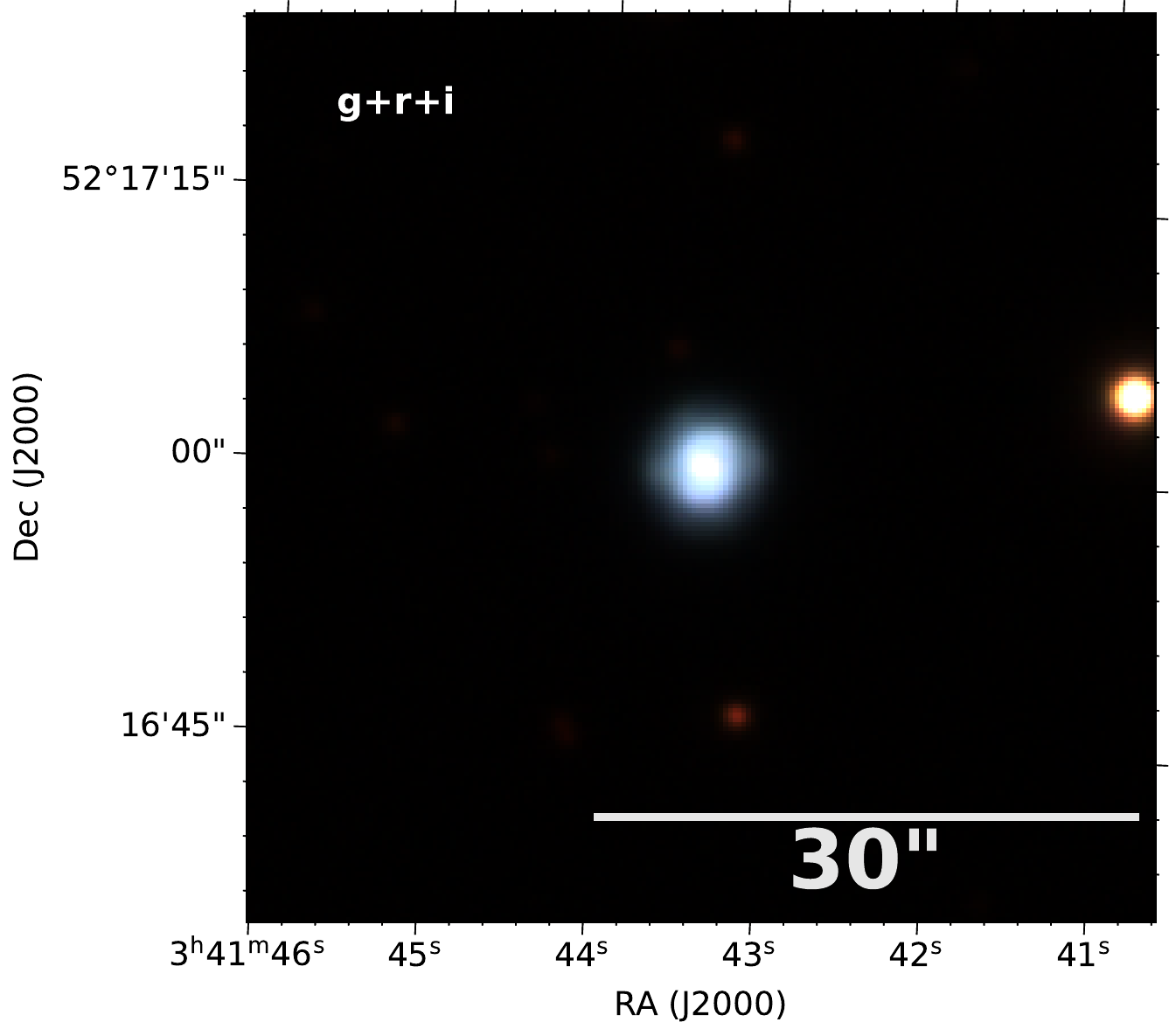}\\
      \rule{0ex}{0.85cm}%
      }
    \rule{0.25cm}{0ex}}  &
  \includegraphics[width=0.5\linewidth, trim=10 50 10 10, clip]{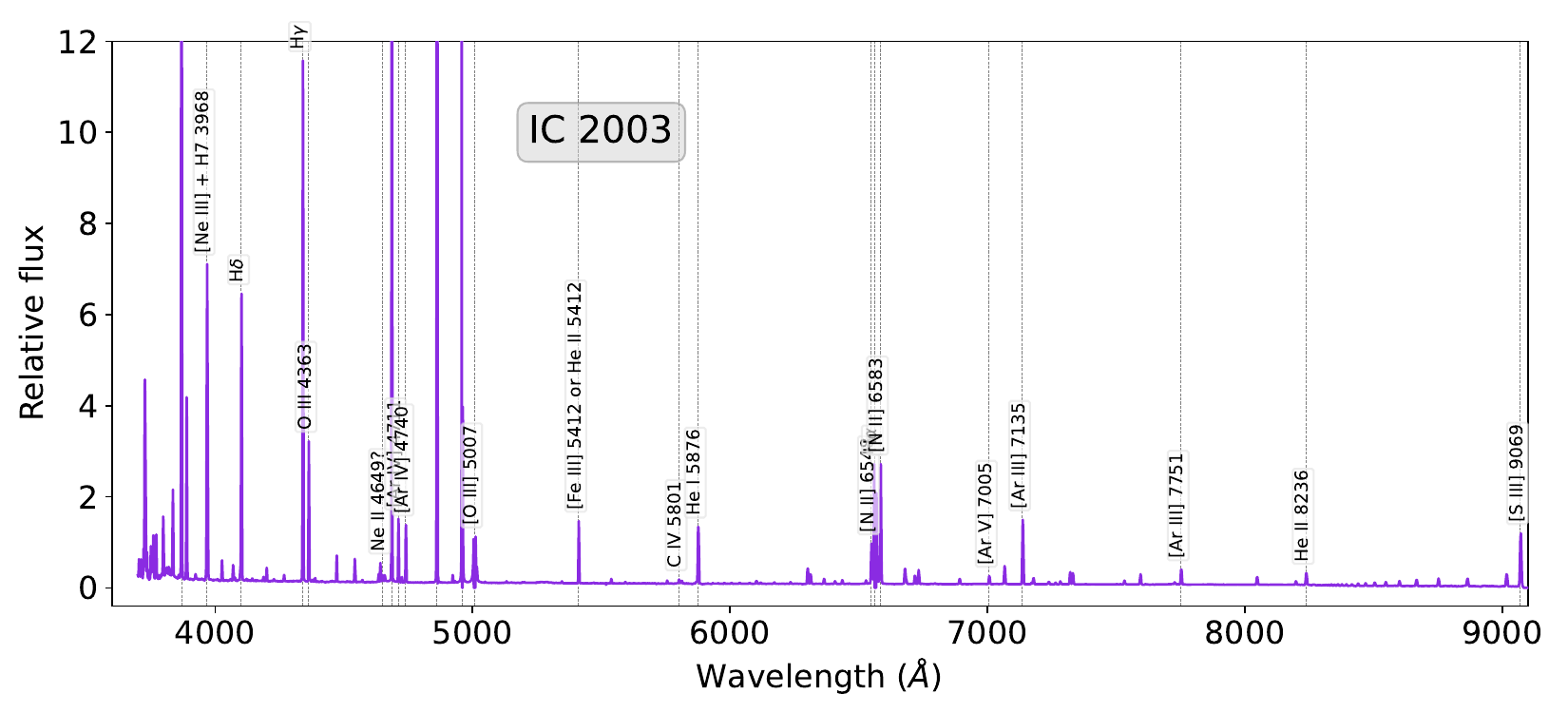} \llap{\shortstack{%
      \includegraphics[width=0.15\linewidth, clip]{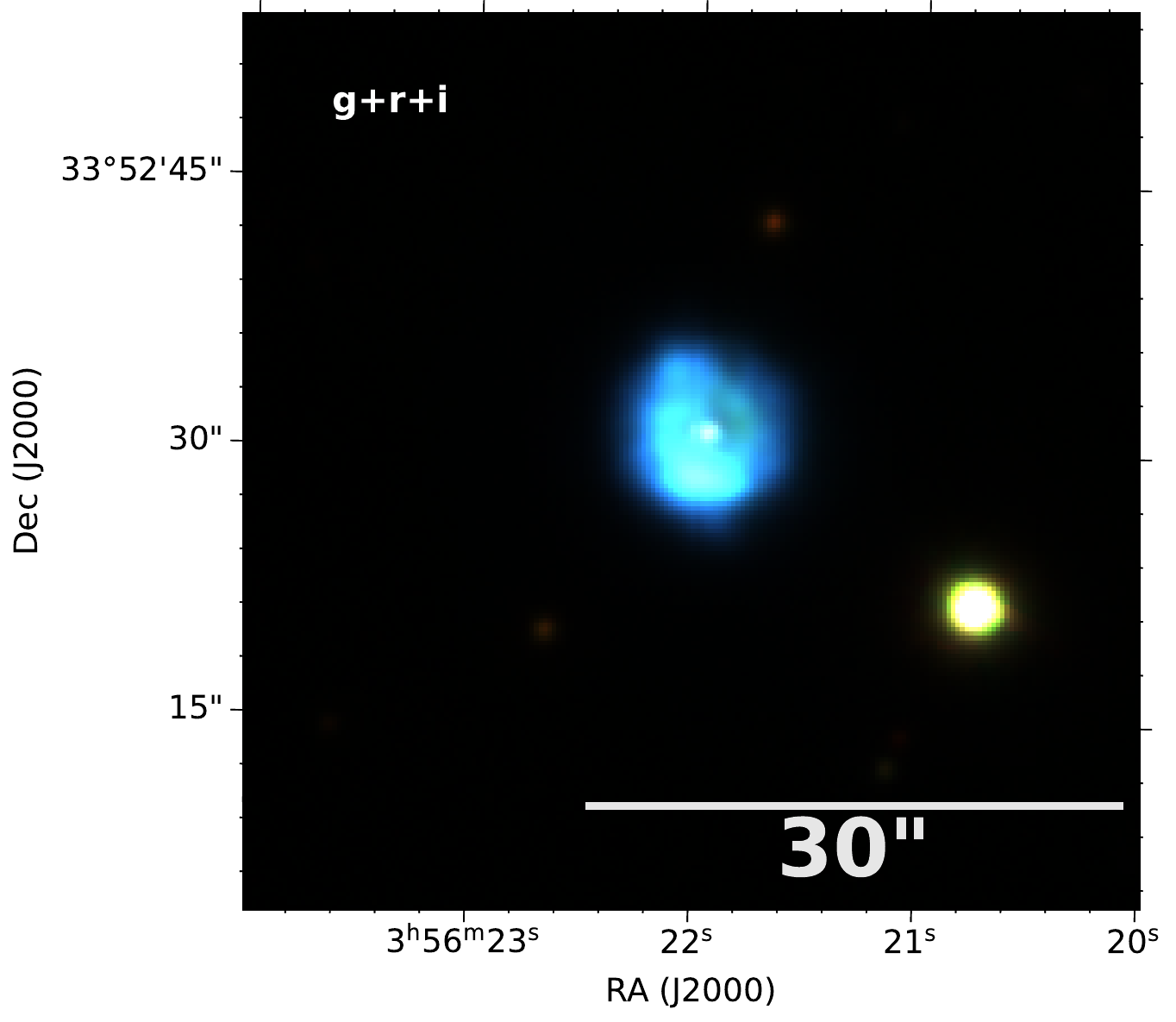} \\
      \rule{0ex}{0.85cm}%
      }
    \rule{0.2cm}{0ex}} \\
  
  \includegraphics[width=0.5\linewidth, trim=1 50 10 10, clip]{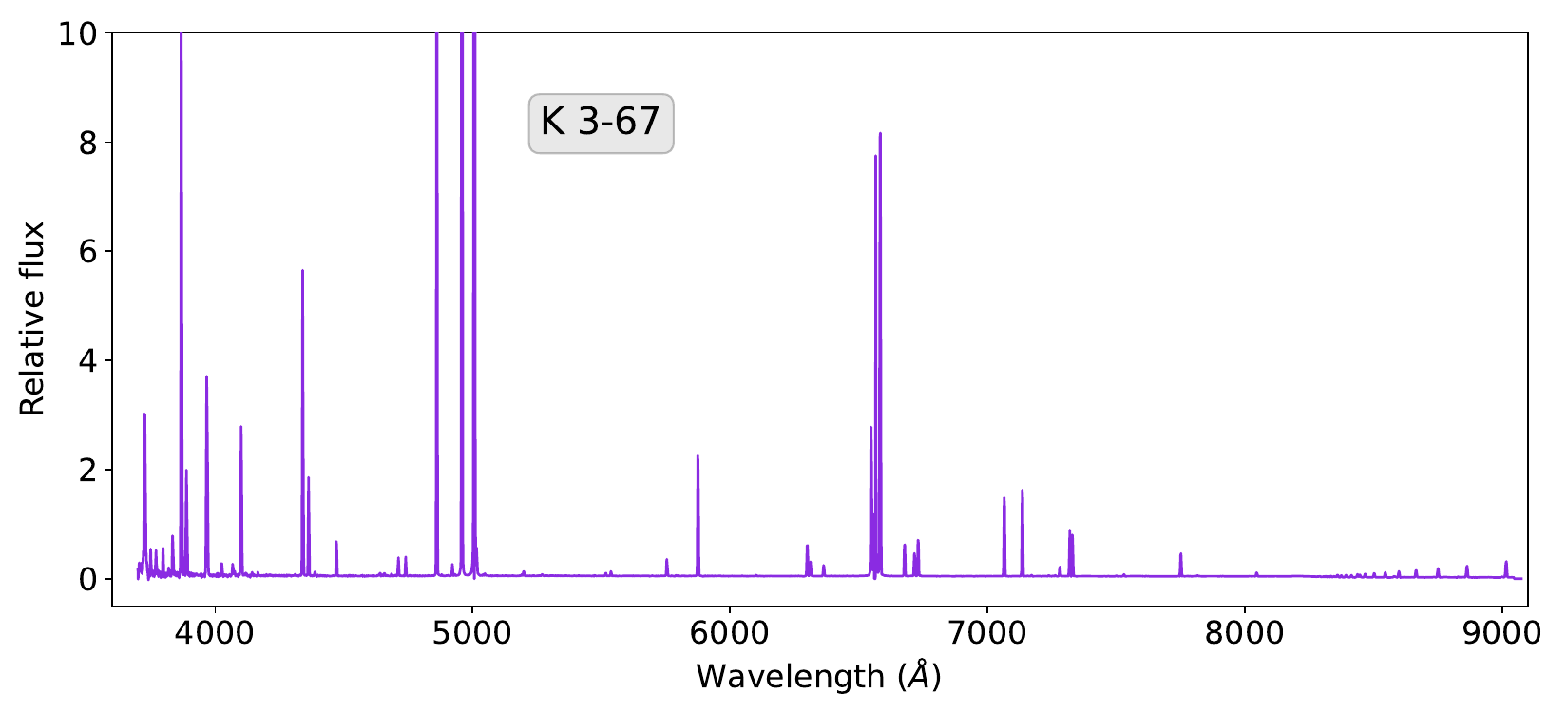} \llap{\shortstack{%
      \includegraphics[width=0.15\linewidth,  clip]{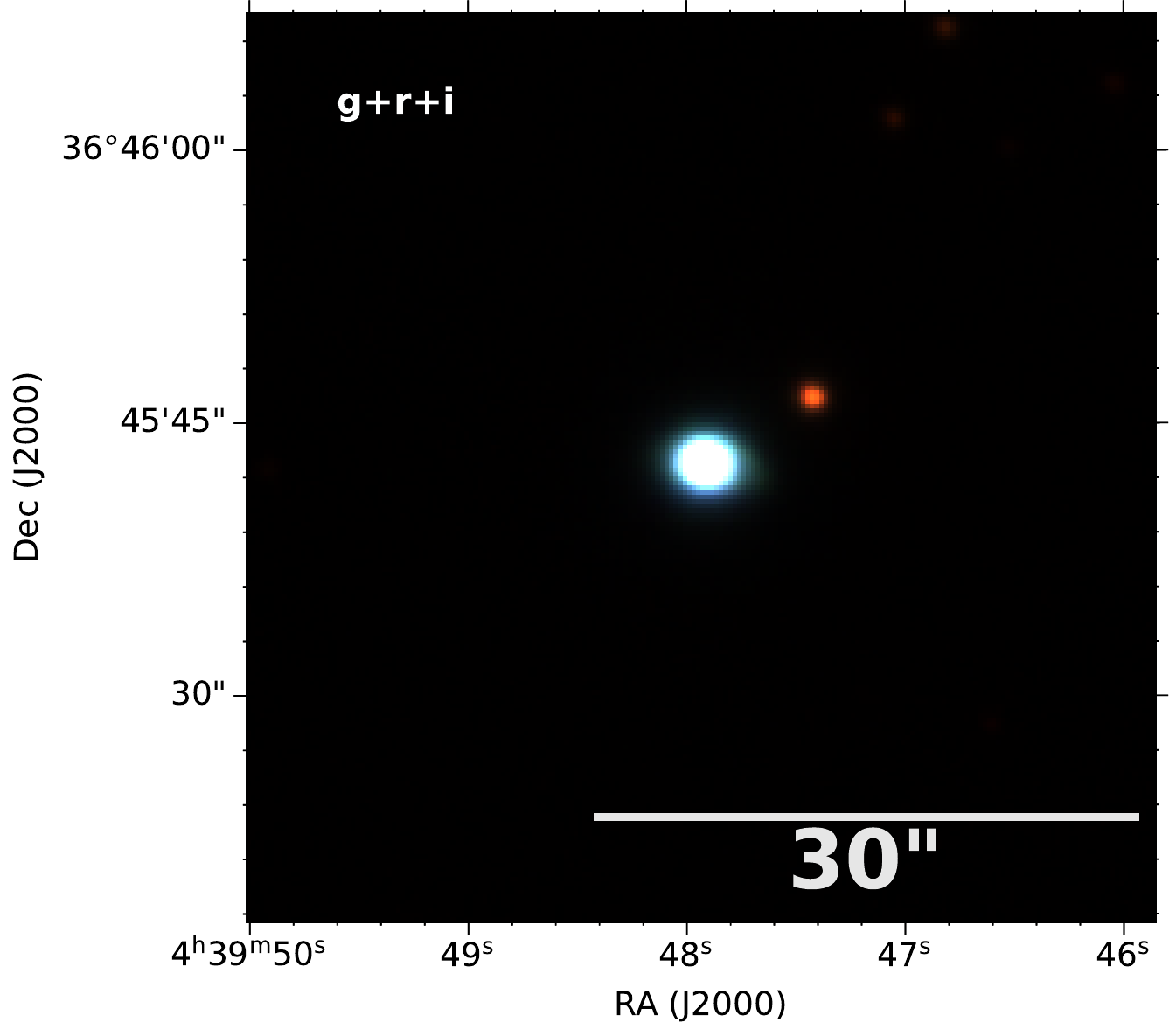}\\
      \rule{0ex}{0.85cm}%
      }
    \rule{0.25cm}{0ex}} &
  \includegraphics[width=0.5\linewidth, trim=10 50 10 10, clip]{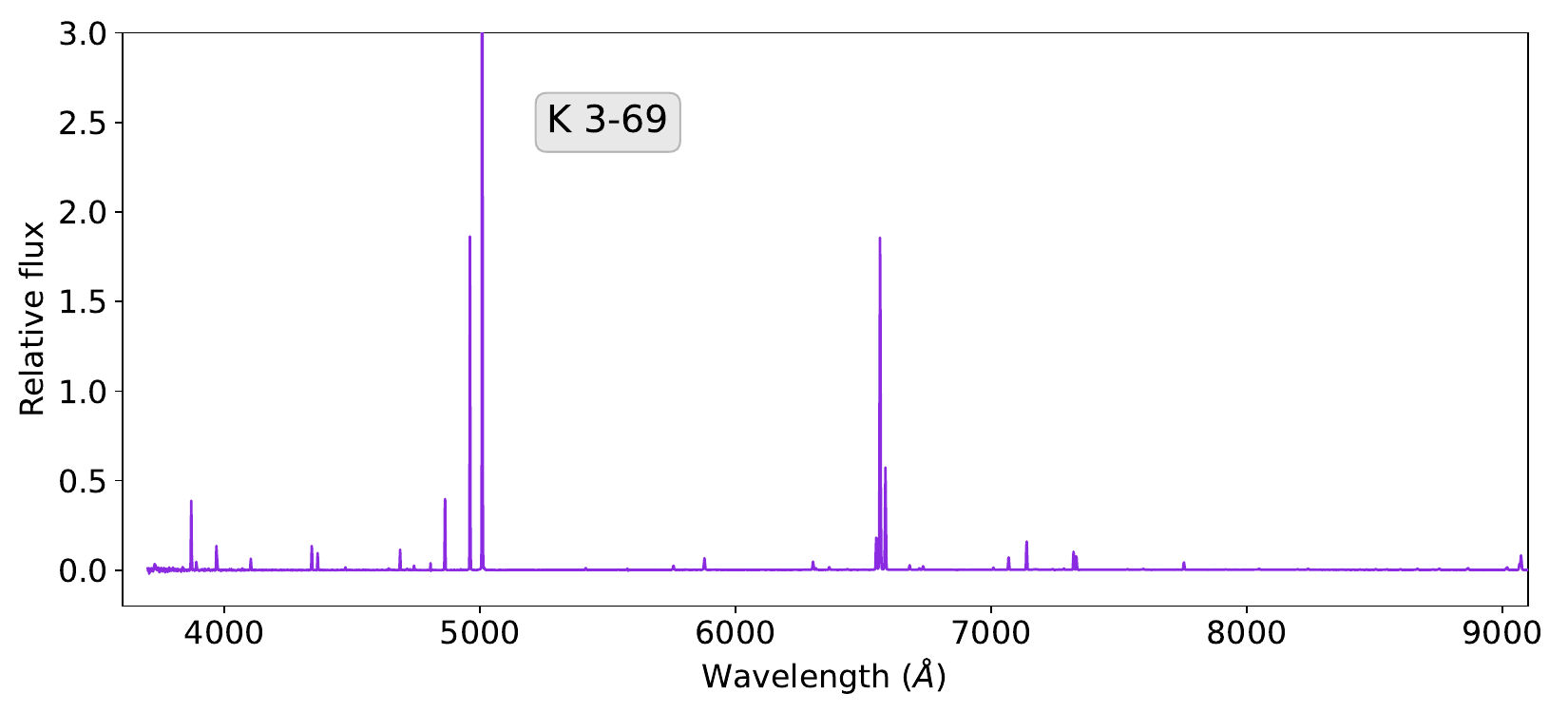} \llap{\shortstack{%
      \includegraphics[width=0.15\linewidth, clip]{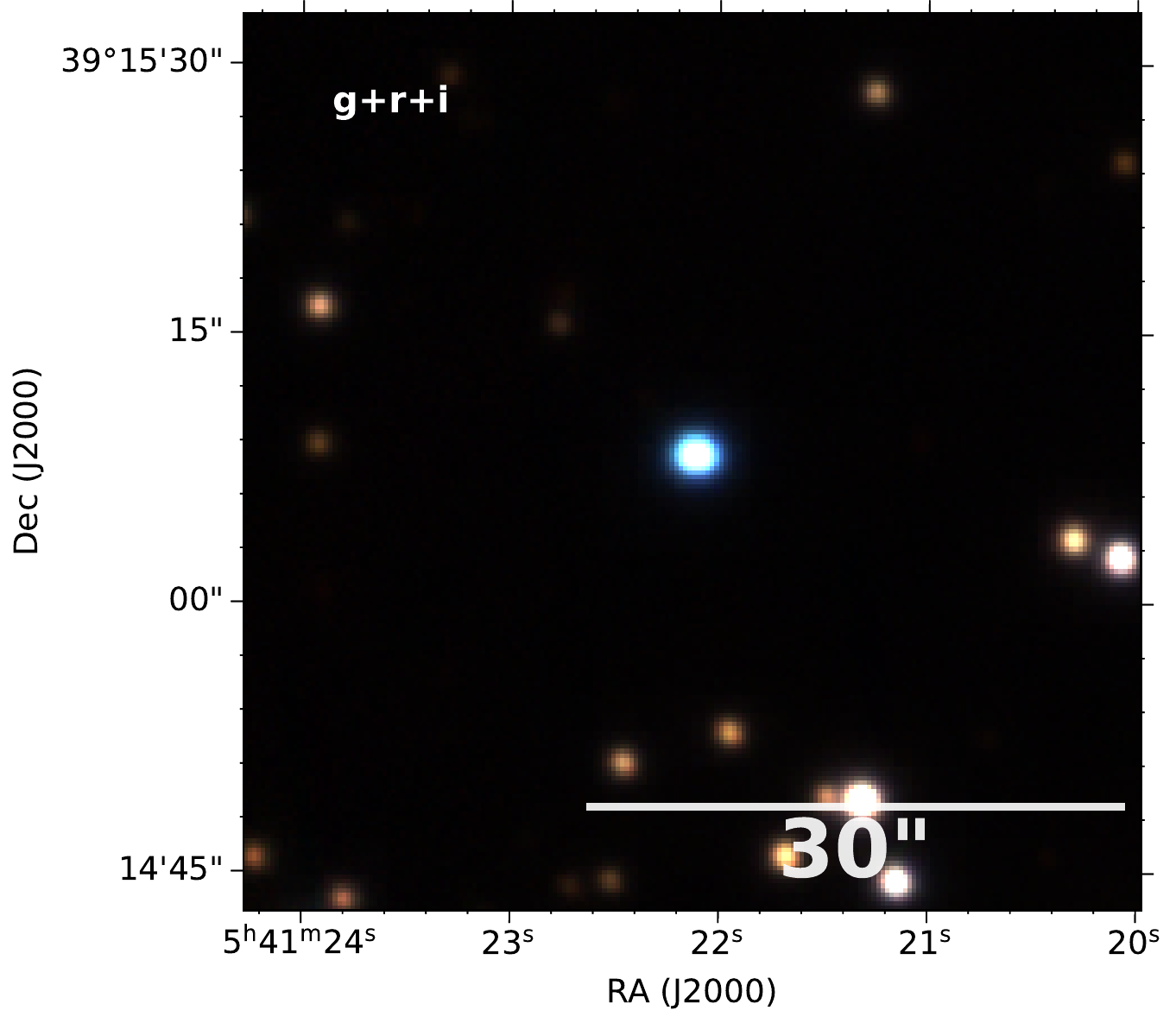}\\
      \rule{0ex}{0.85cm}%
      }
    \rule{0.2cm}{0ex}} \\
  
  \includegraphics[width=0.5\linewidth, trim=1 50 10 10, clip]{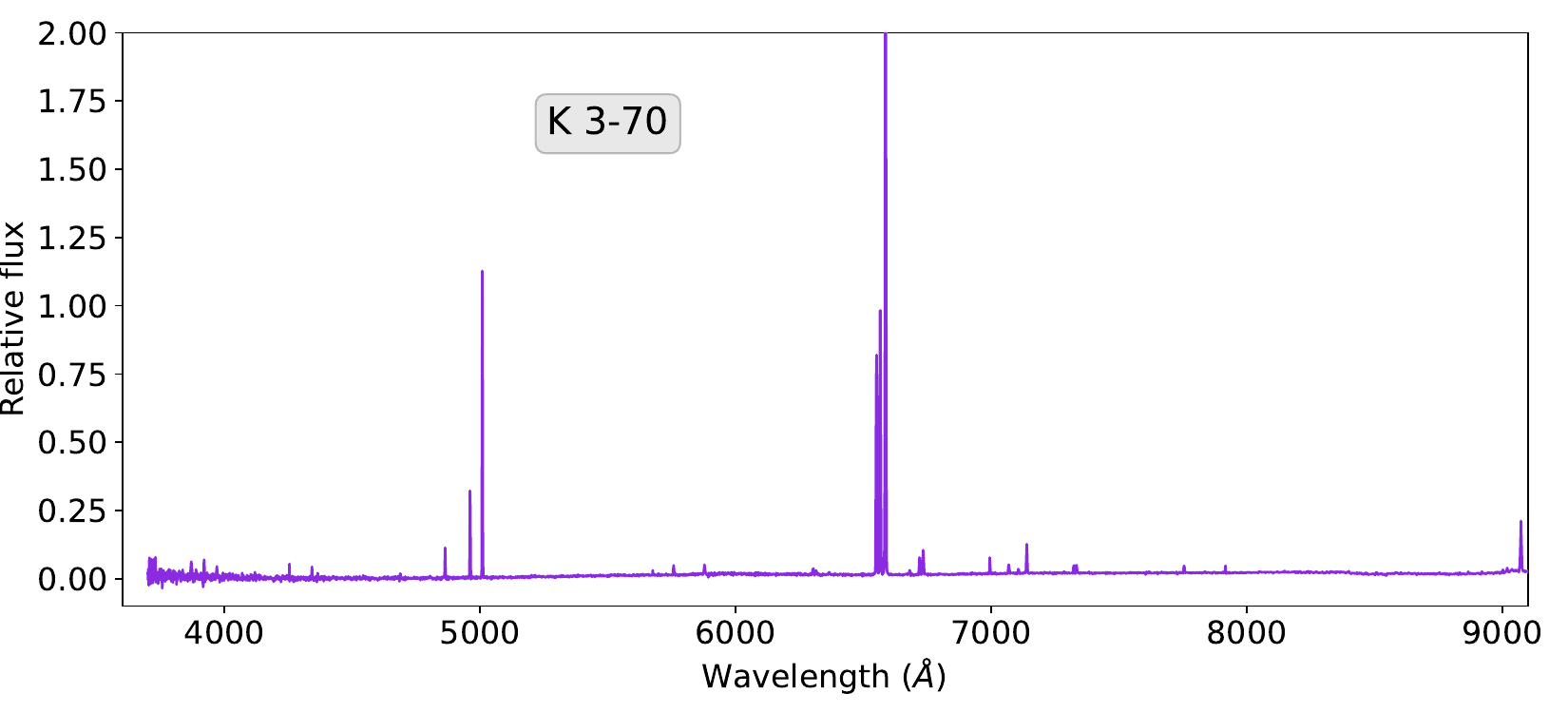} \llap{\shortstack{%
      \includegraphics[width=0.15\linewidth, clip]{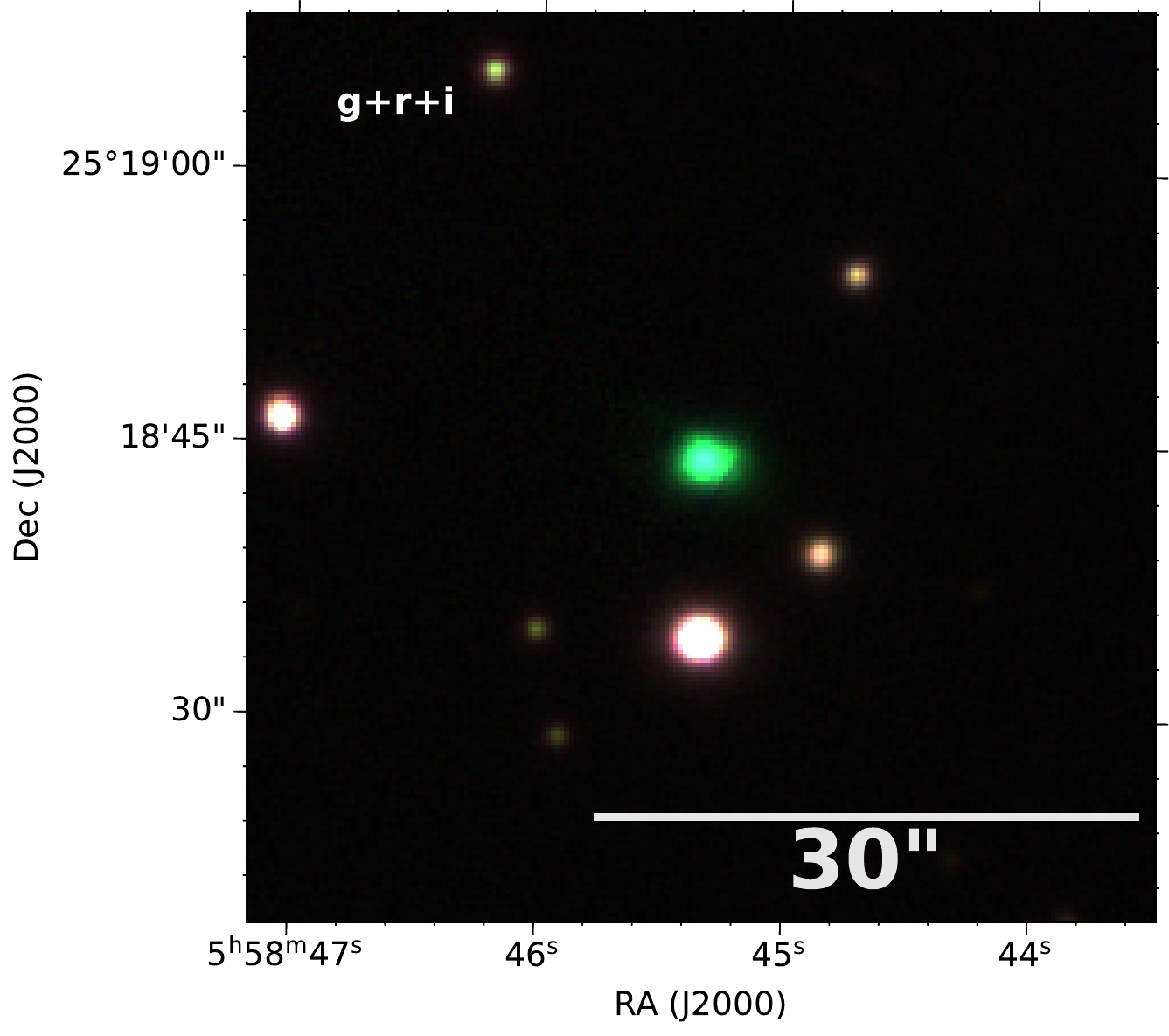}\\
      \rule{0ex}{0.75cm}%
      }
    \rule{0.25cm}{0ex}} & 
  \includegraphics[width=0.5\linewidth, trim=10 50 10 10, clip]{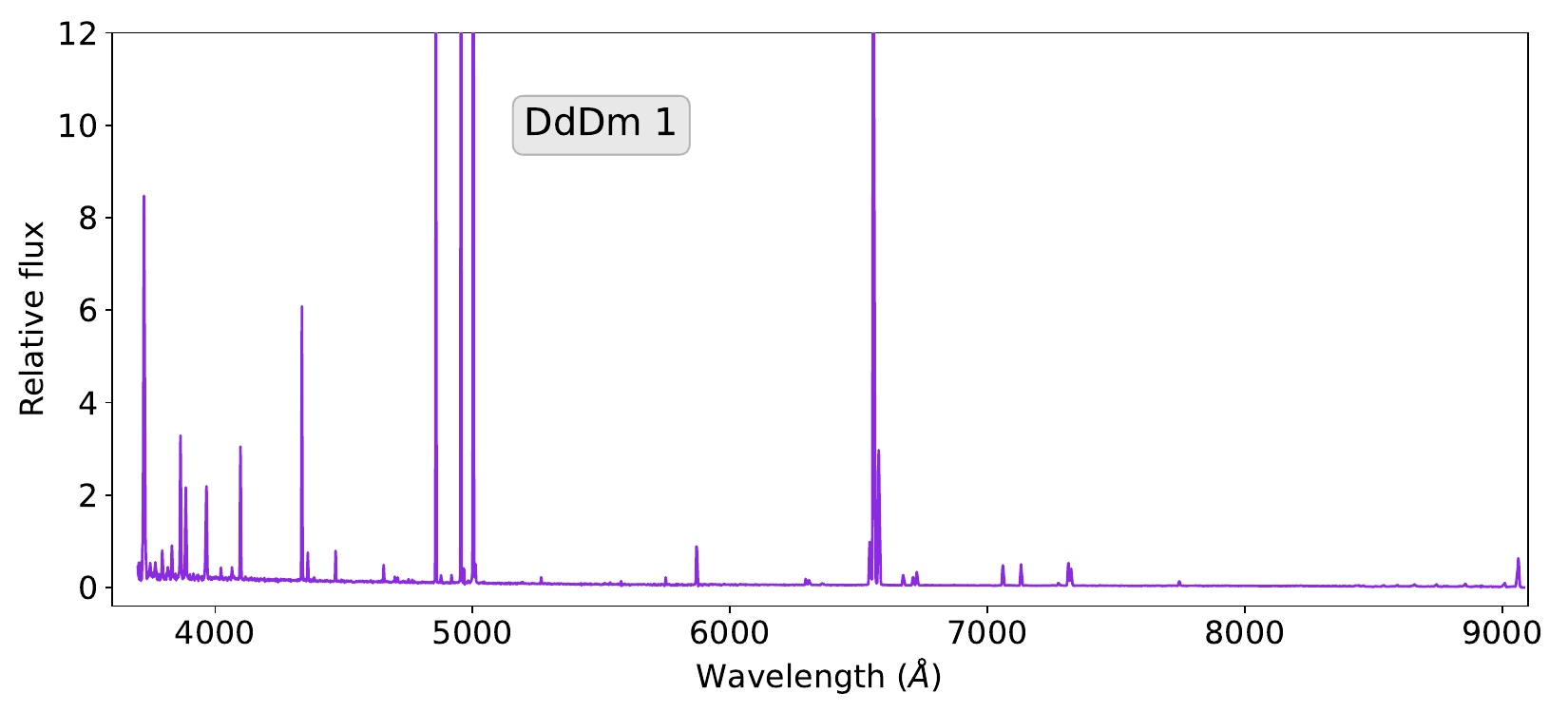}  \llap{\shortstack{%
      \includegraphics[width=0.15\linewidth, clip]{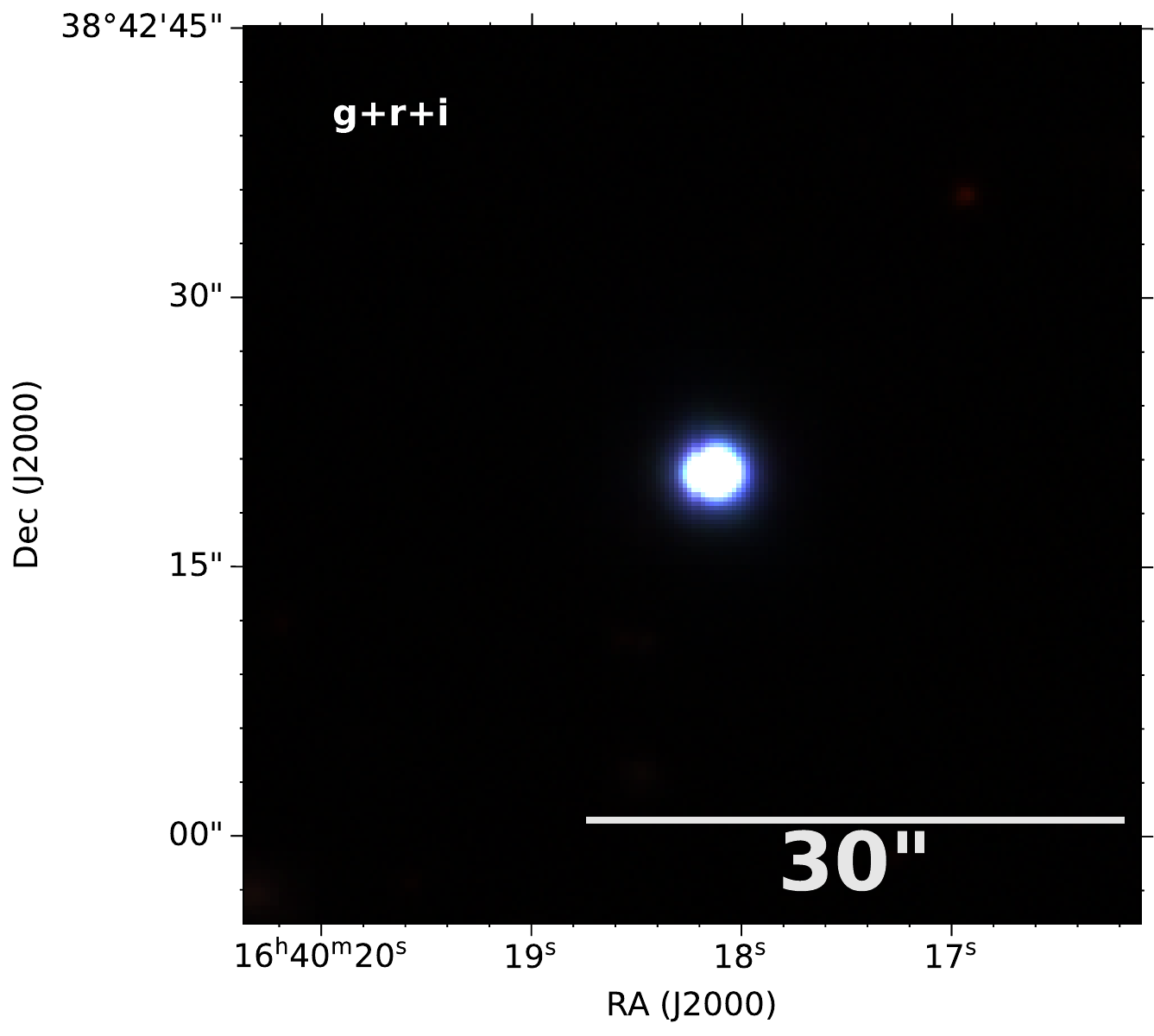} \\
      \rule{0ex}{0.85cm}%
      }
    \rule{0.25cm}{0ex}} \\
  
  \includegraphics[width=0.5\linewidth, trim=5 10 10 10, clip]{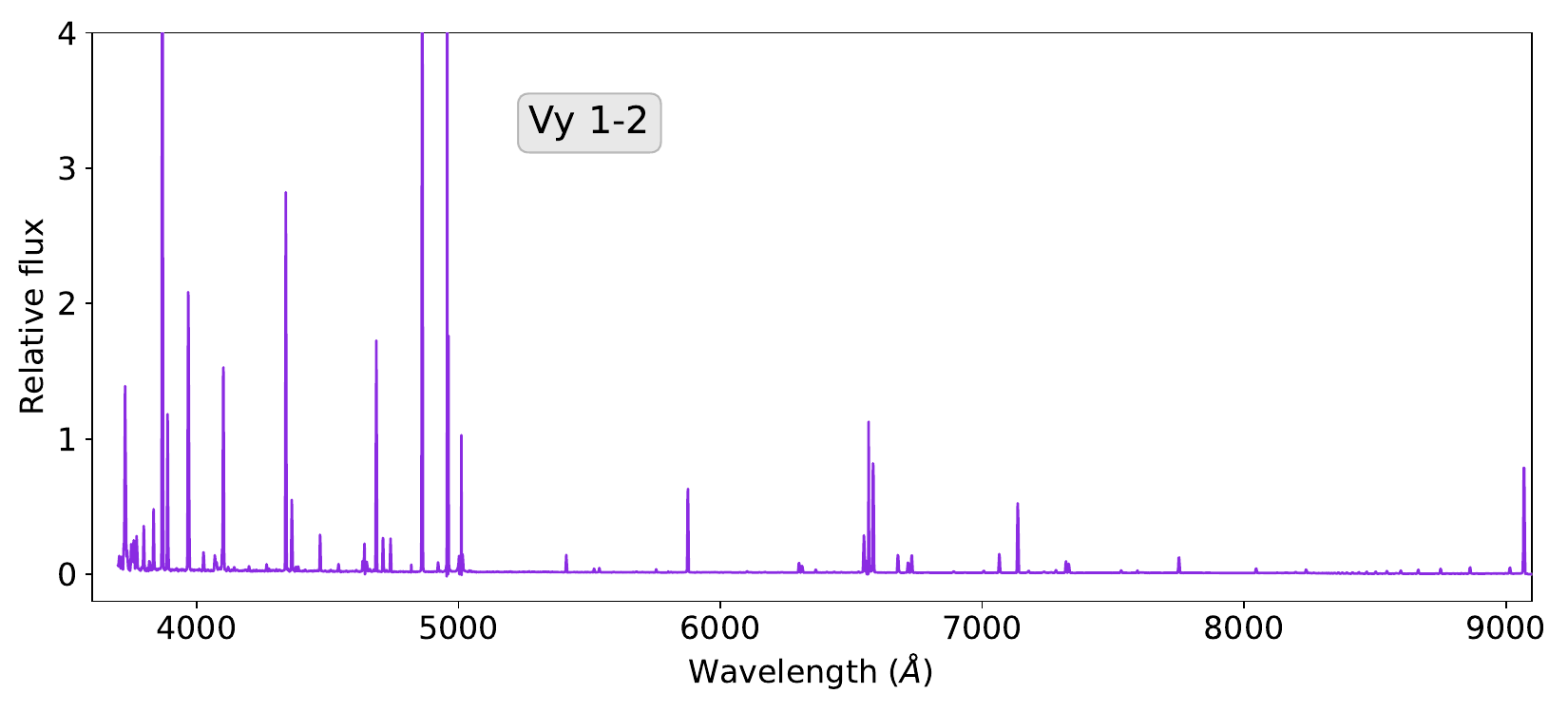} \llap{\shortstack{%
      \includegraphics[width=0.15\linewidth,  clip]{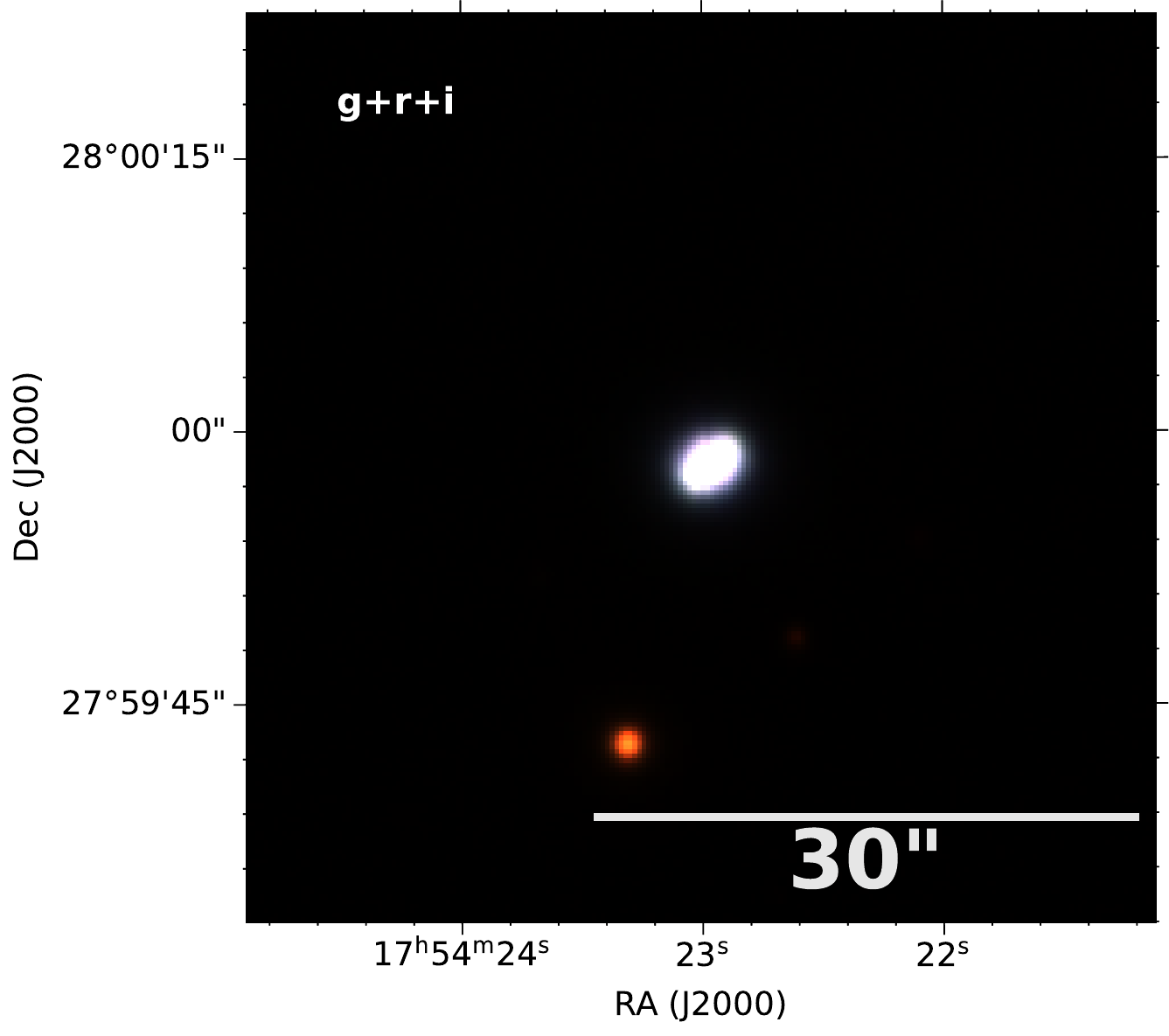}\\
      \rule{0ex}{1.3cm}%
      }
    \rule{0.25cm}{0ex}} & 
  \includegraphics[width=0.5\linewidth, trim=10 10 10 10, clip]{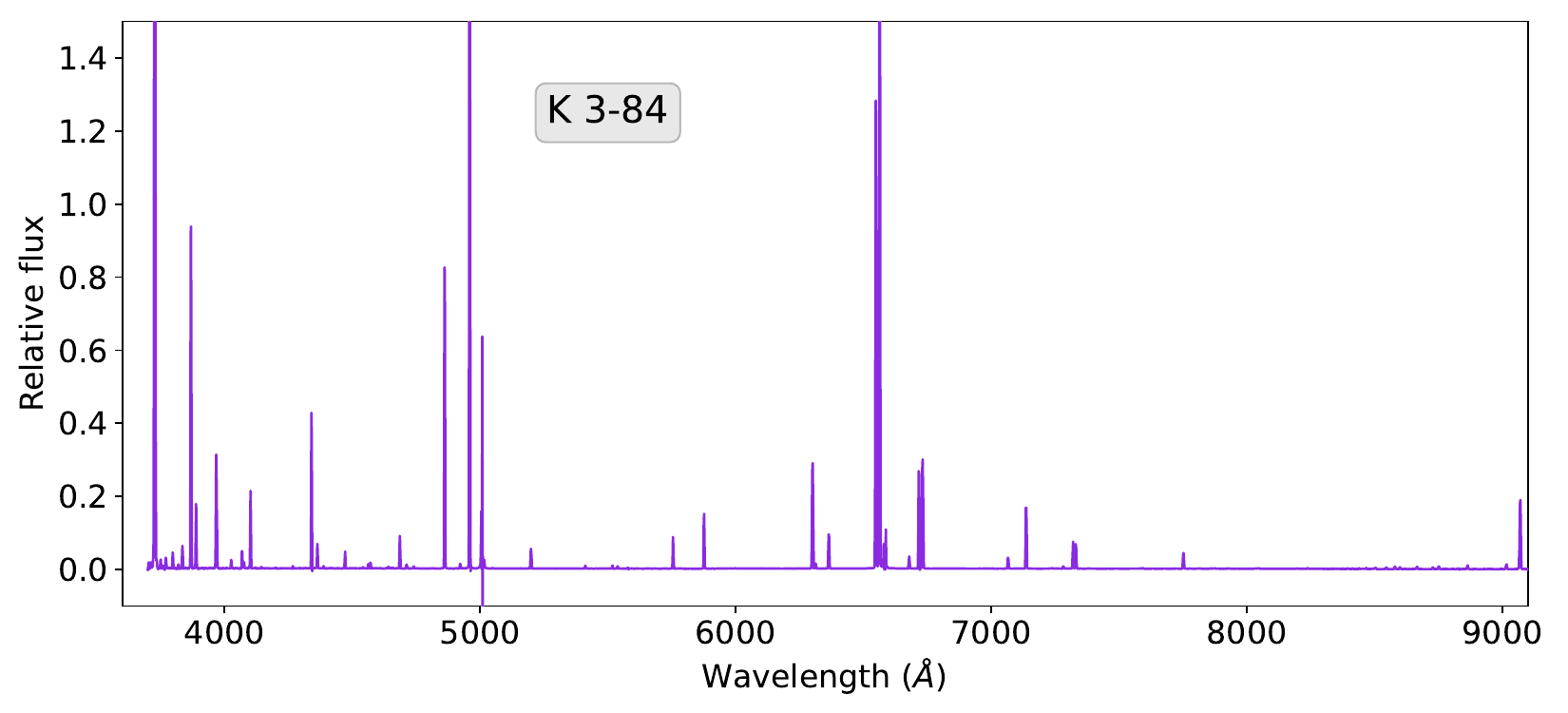}  \llap{\shortstack{%
      \includegraphics[width=0.15\linewidth,  clip]{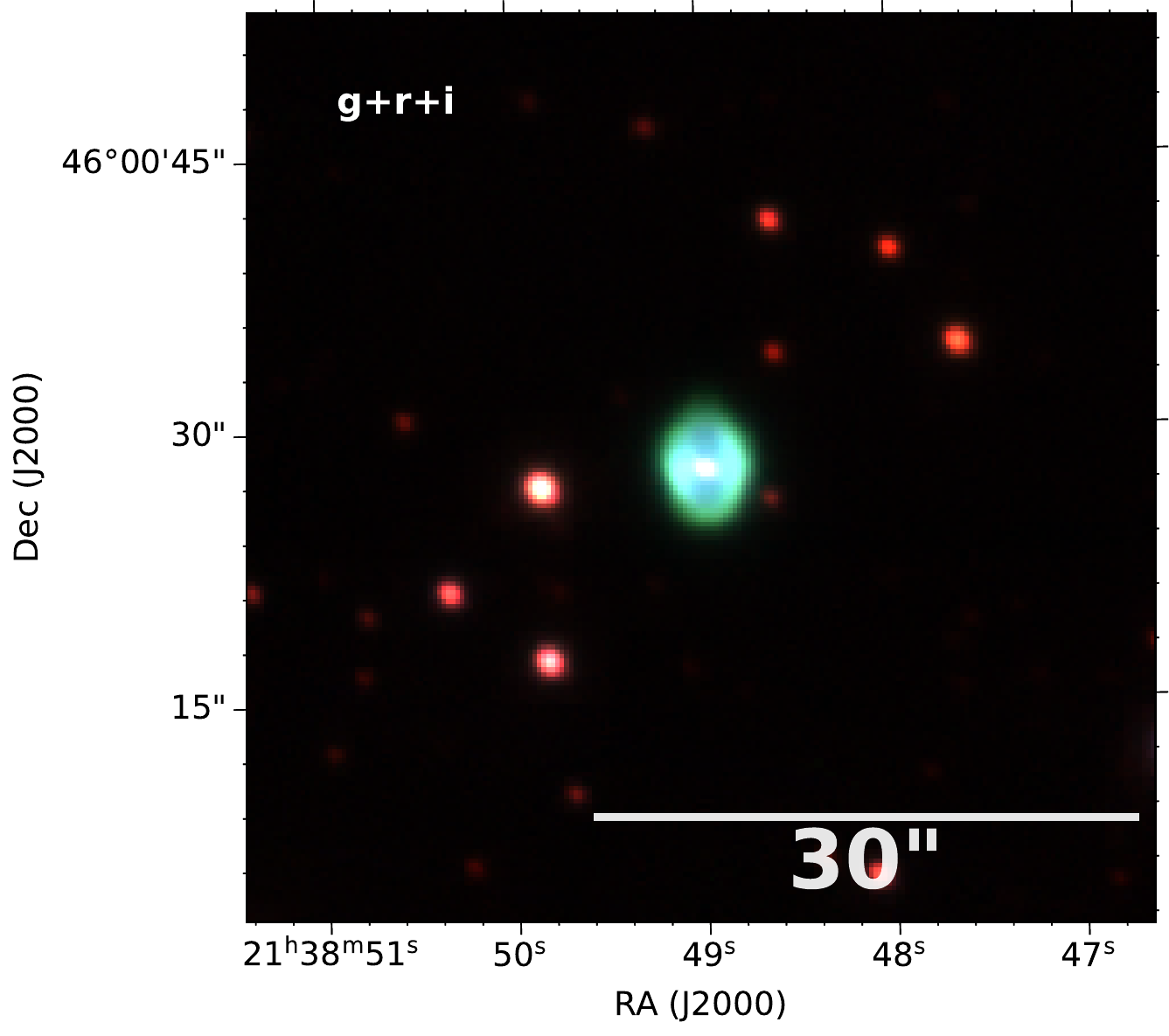}\\
      \rule{0ex}{1.4cm}%
      }
    \rule{0.25cm}{0ex}}
\end{tabular}
\caption{LAMOST spectra and RGB images of the eight true PNe located in the compact PNe zone of the \((G - r)\) versus \((G_{BP} - G_{RP})\) color-color diagram.}
\label{fig:spectra-image-trurPN-better}
\end{figure*}

\begin{figure*}
\centering
\begin{tabular}{ll}
\includegraphics[width=0.5\linewidth, trim=5 10 10 10, clip]{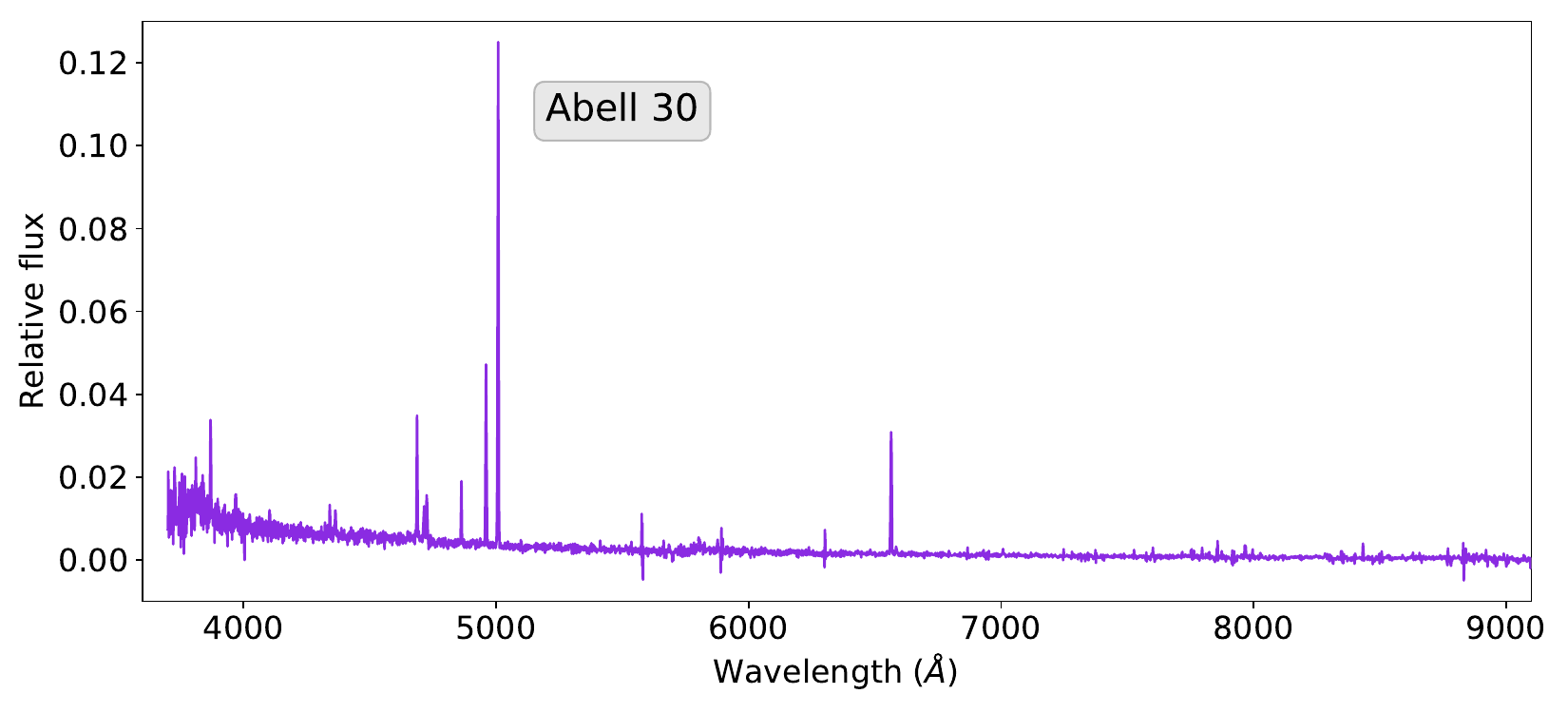} \llap{\shortstack{%
        \includegraphics[width=0.15\linewidth,  clip]{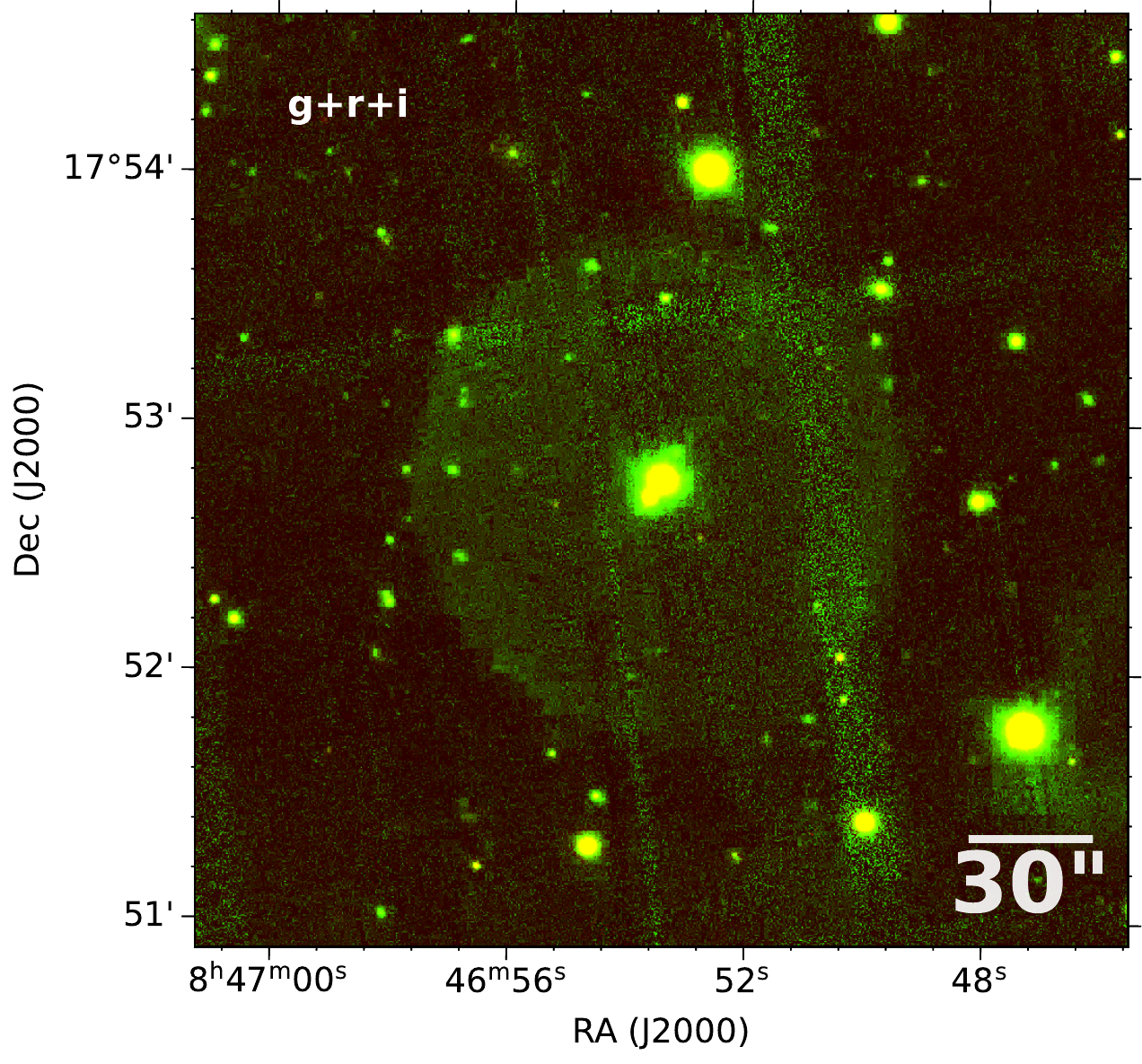}\\
        \rule{0ex}{1.3cm}%
  }
  \rule{0.25cm}{0ex}} &

     \includegraphics[width=0.5\linewidth, trim=10 10 10 10, clip]{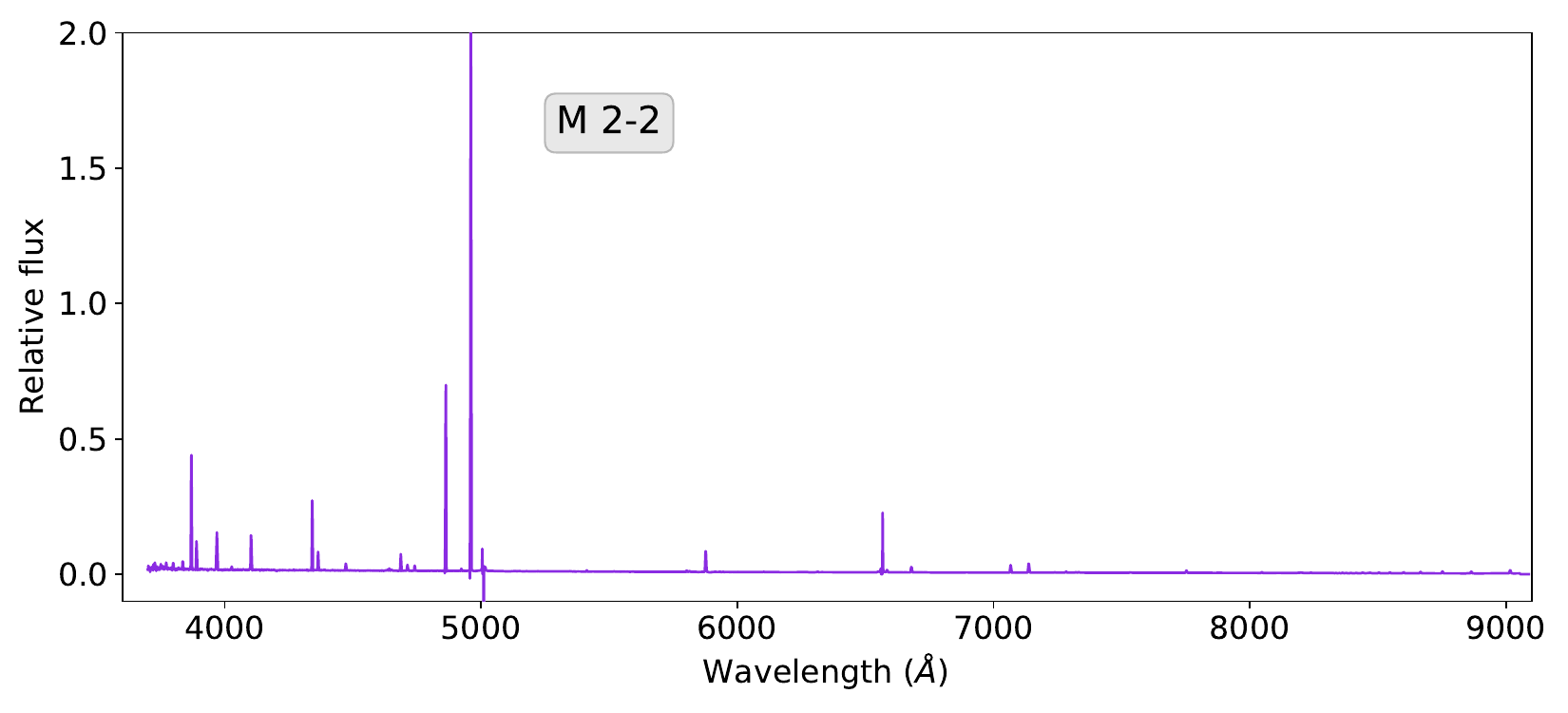} \llap{\shortstack{%
      \includegraphics[width=0.15\linewidth,  clip]{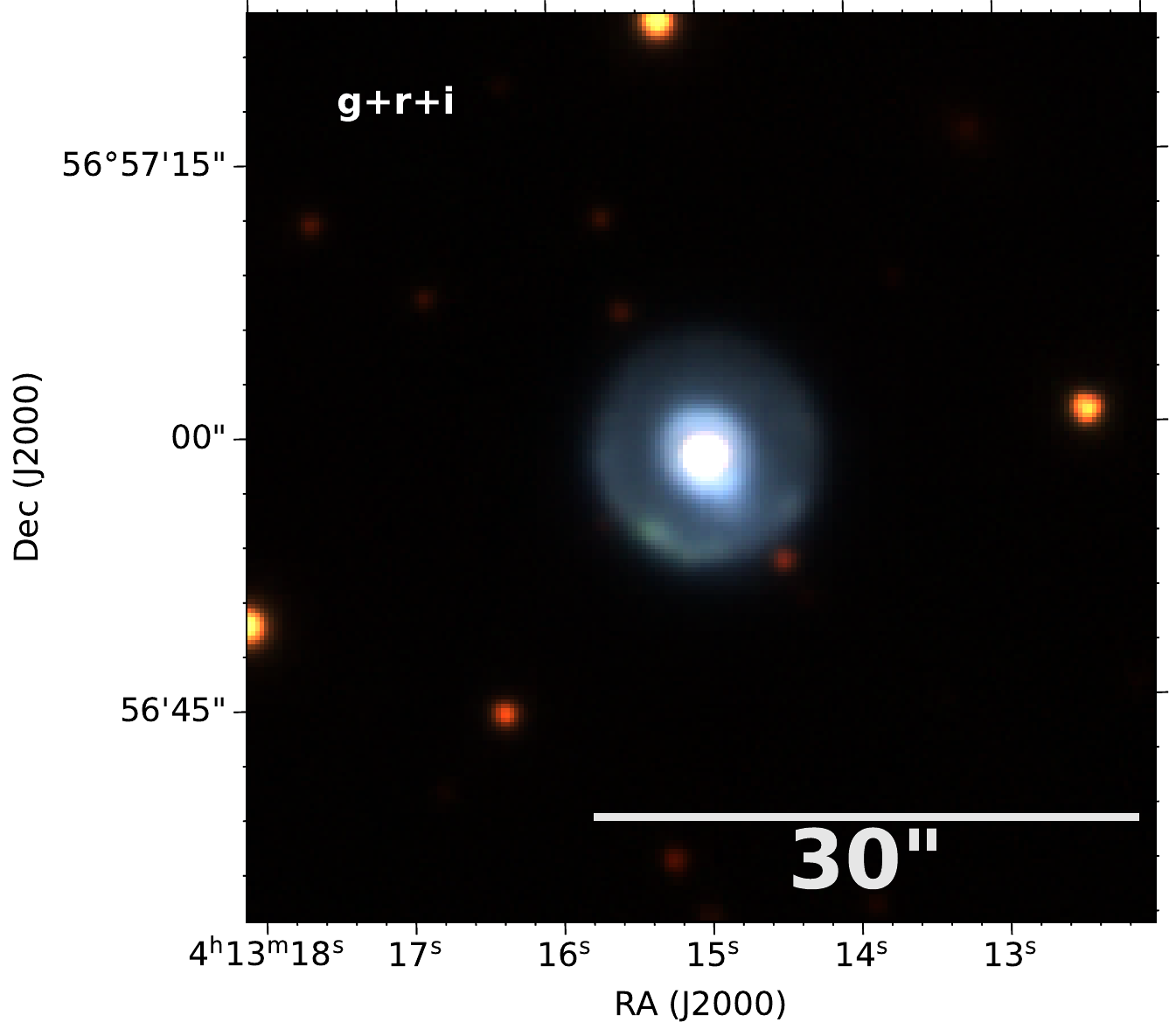}\\
      \rule{0ex}{1.4cm}%
      }
    \rule{0.25cm}{0ex}}
  \end{tabular}
  \caption{Same as Figure~\ref{fig:spectra-image-trurPN-better}, showcasing PN Abel 30, located in the compact PN zone in the $(G - g)$ versus $(G_{BP} - G_{RP})$ color-color diagram and in the very extended zone in the $(G - r)$ versus $(G_{BP} - G_{RP})$ color-color diagram. Also featuring M 2-2, situated in the extended PN zone in the $(G - g)$ versus $(G_{BP} - G_{RP})$ color-color diagram and in the compact zone in the $(G - r)$ versus $(G_{BP} - G_{RP})$ color-color diagram.}
  \label{fig:spectra-image-trurPN-justonecompact}
\end{figure*}

\begin{figure*}
\centering
\begin{tabular}{ll}
  \includegraphics[width=0.5\linewidth, trim=1 50 10 10, clip]{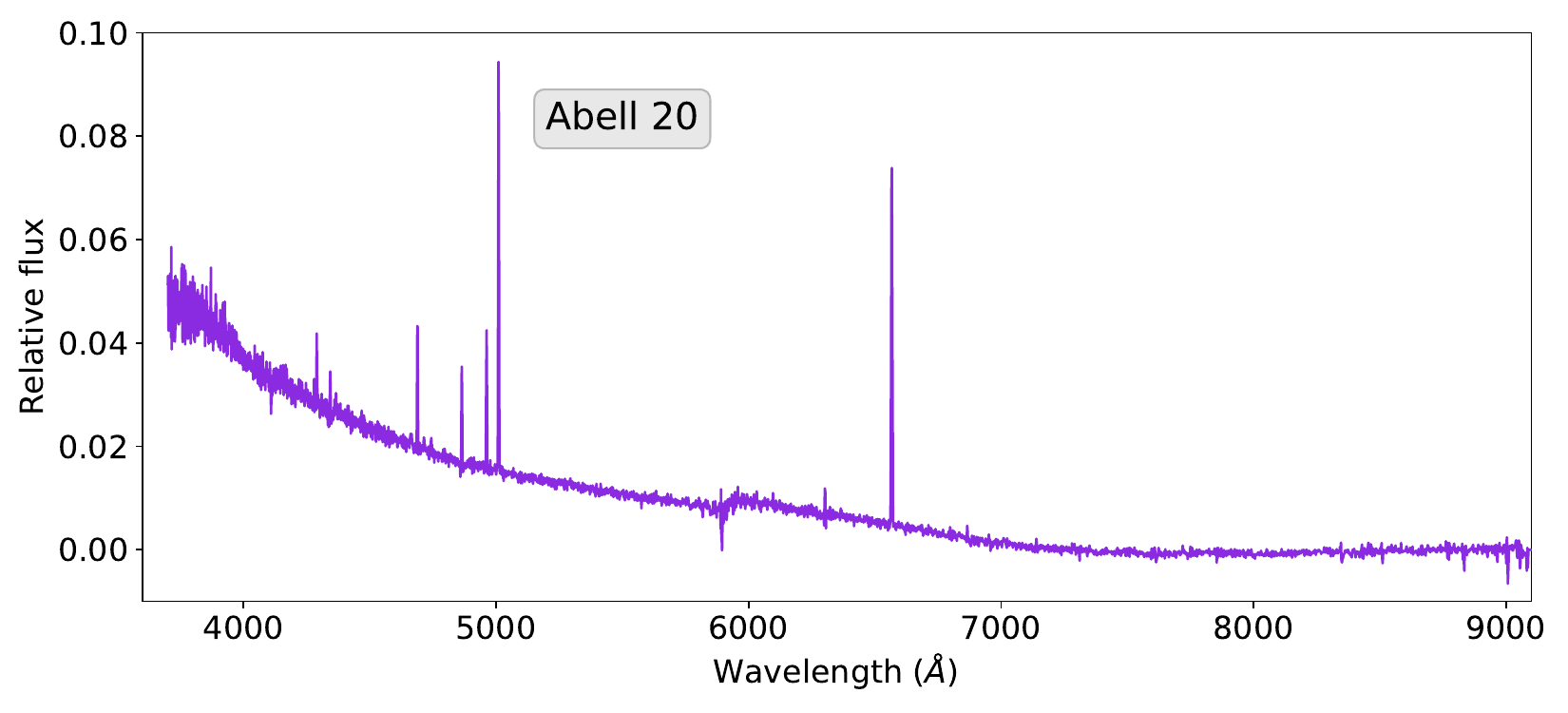} \llap{\shortstack{%
      \includegraphics[width=0.15\linewidth, clip]{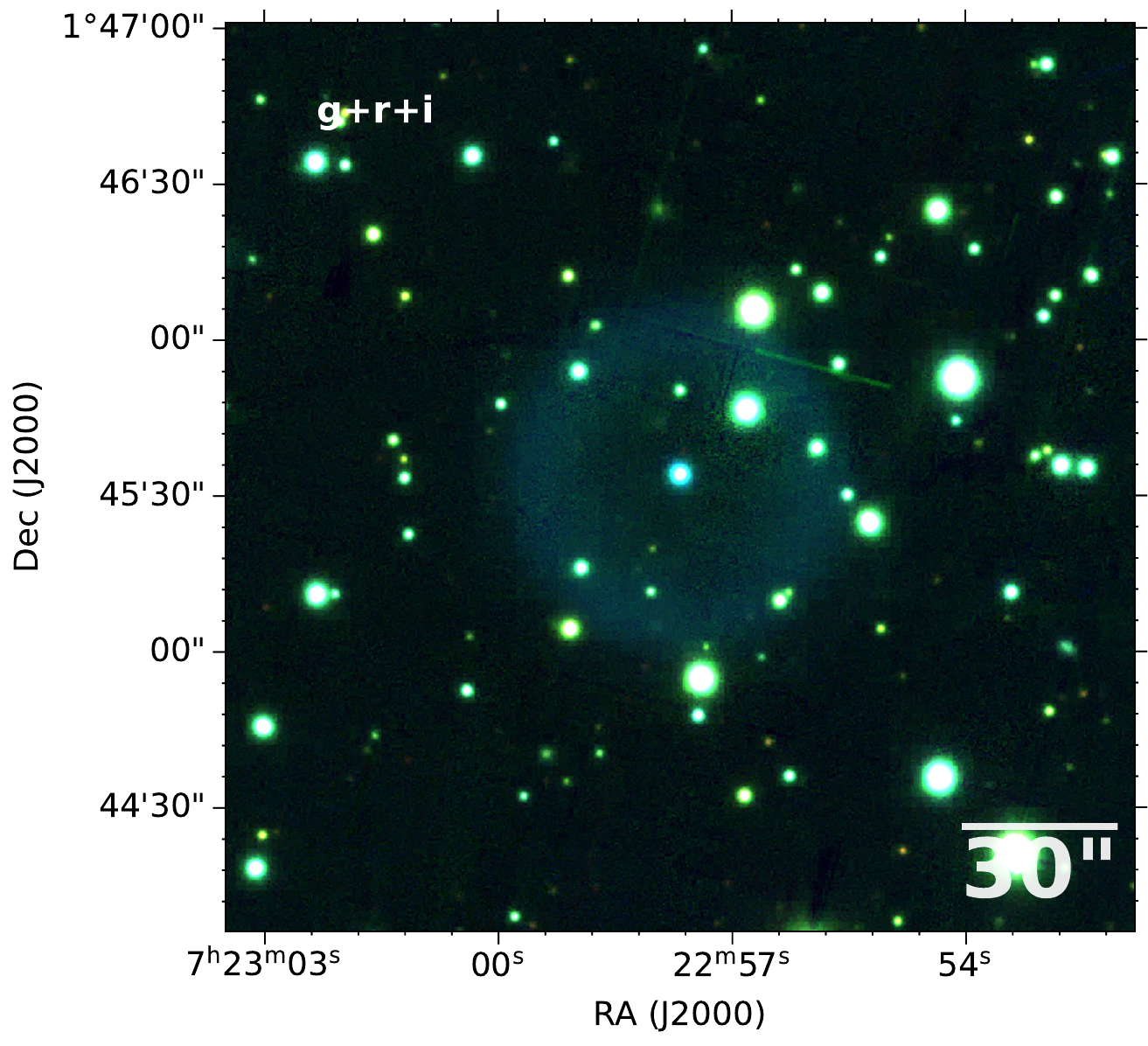}\\
      \rule{0ex}{0.8cm}%
      }
    \rule{0.25cm}{0ex}}  &
  \includegraphics[width=0.5\linewidth, trim=10 50 10 10, clip]{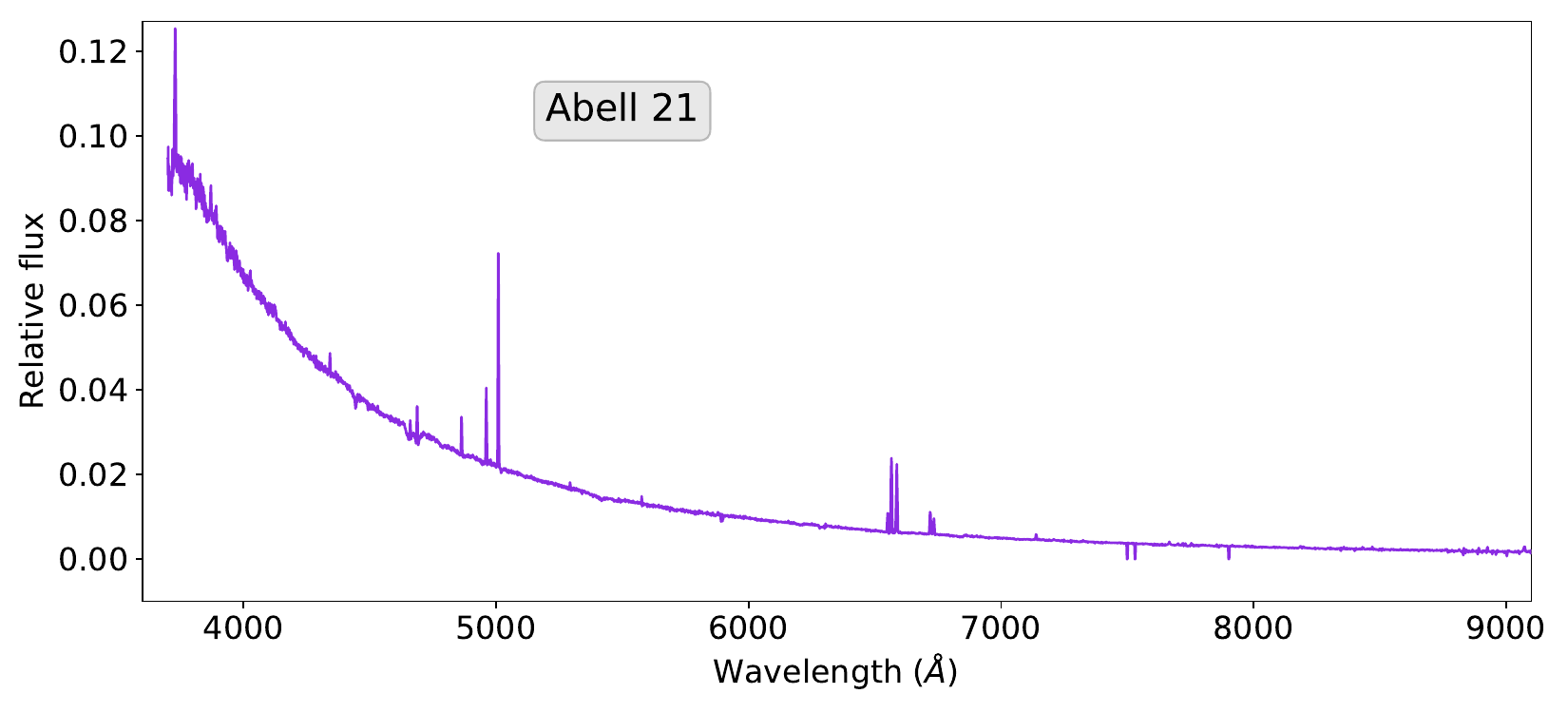} \llap{\shortstack{%
      \includegraphics[width=0.15\linewidth, clip]{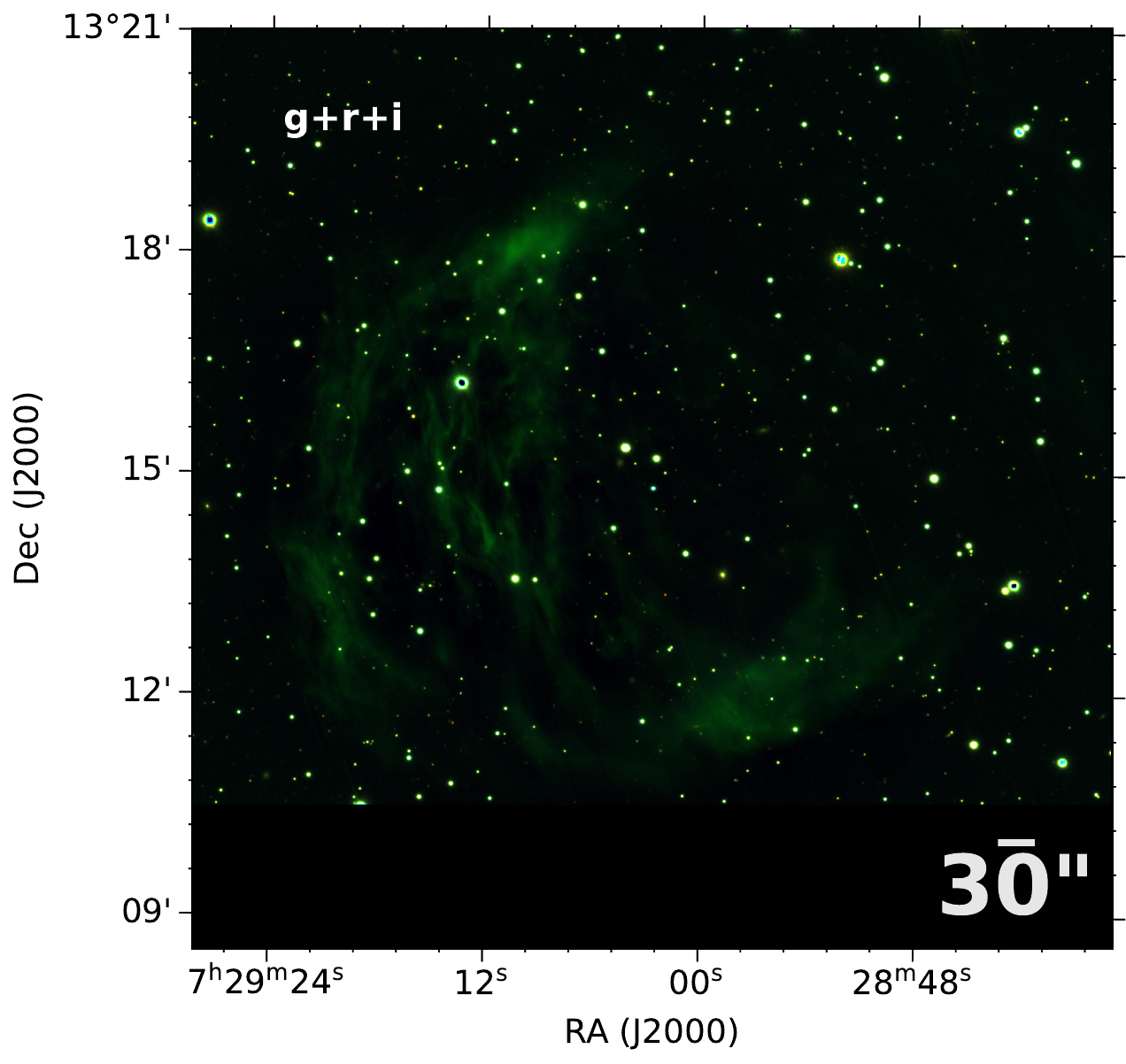} \\
      \rule{0ex}{0.85cm}%
      }
    \rule{0.2cm}{0ex}} \\
  
  \includegraphics[width=0.5\linewidth, trim=1 50 10 10, clip]{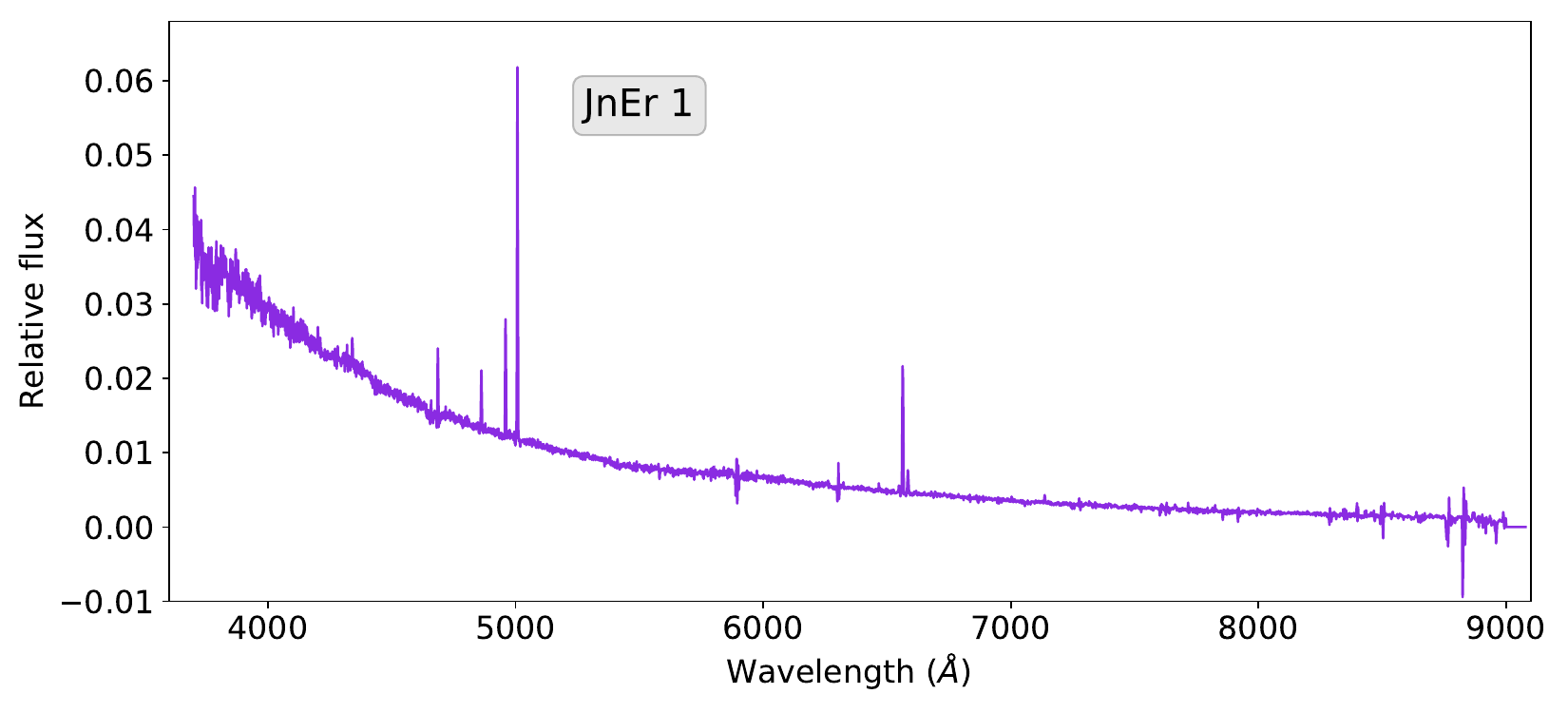} \llap{\shortstack{%
      \includegraphics[width=0.15\linewidth,  clip]{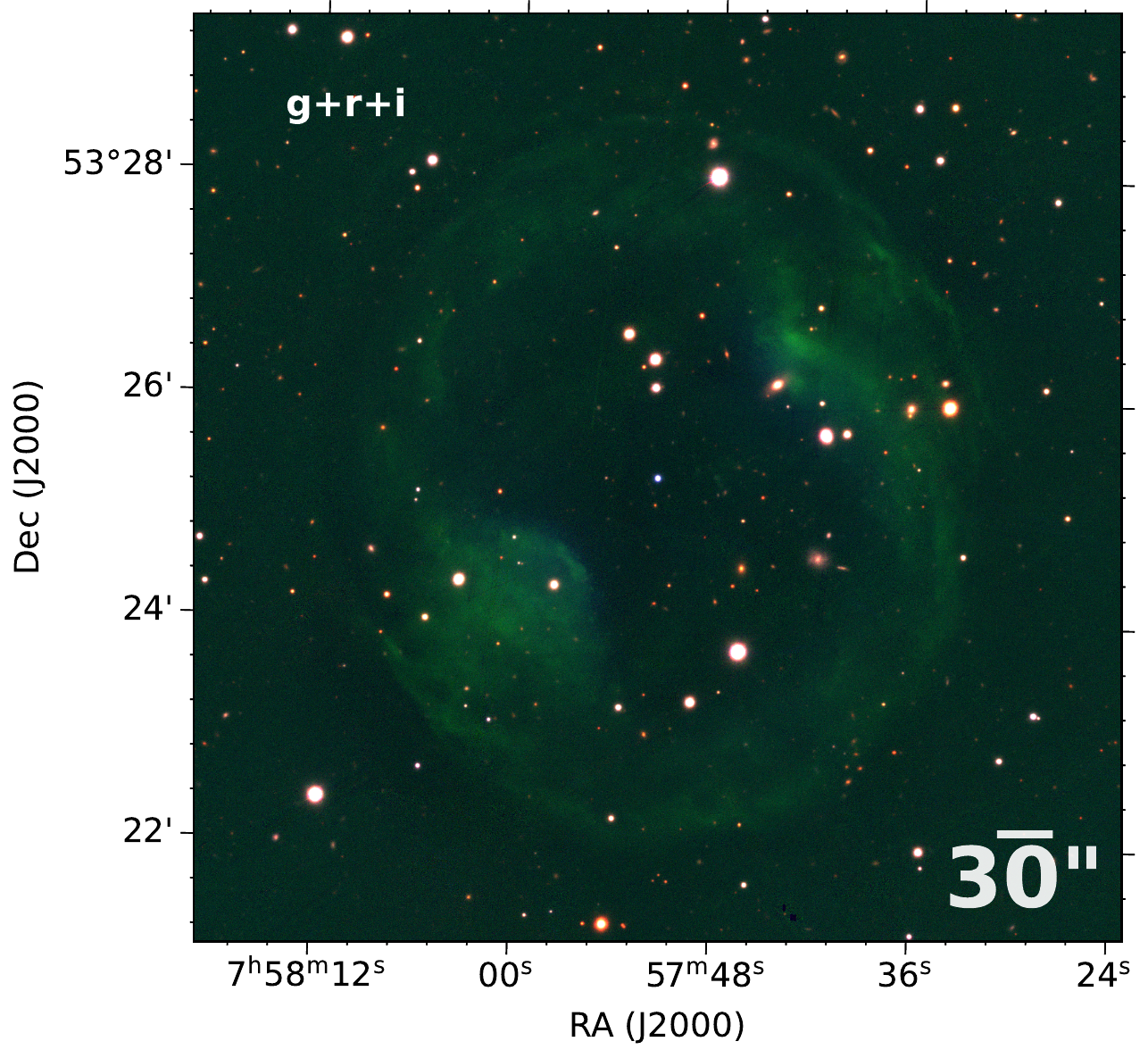}\\
      \rule{0ex}{0.85cm}%
      }
    \rule{0.25cm}{0ex}} &
  \includegraphics[width=0.5\linewidth, trim=1 50 10 10, clip]{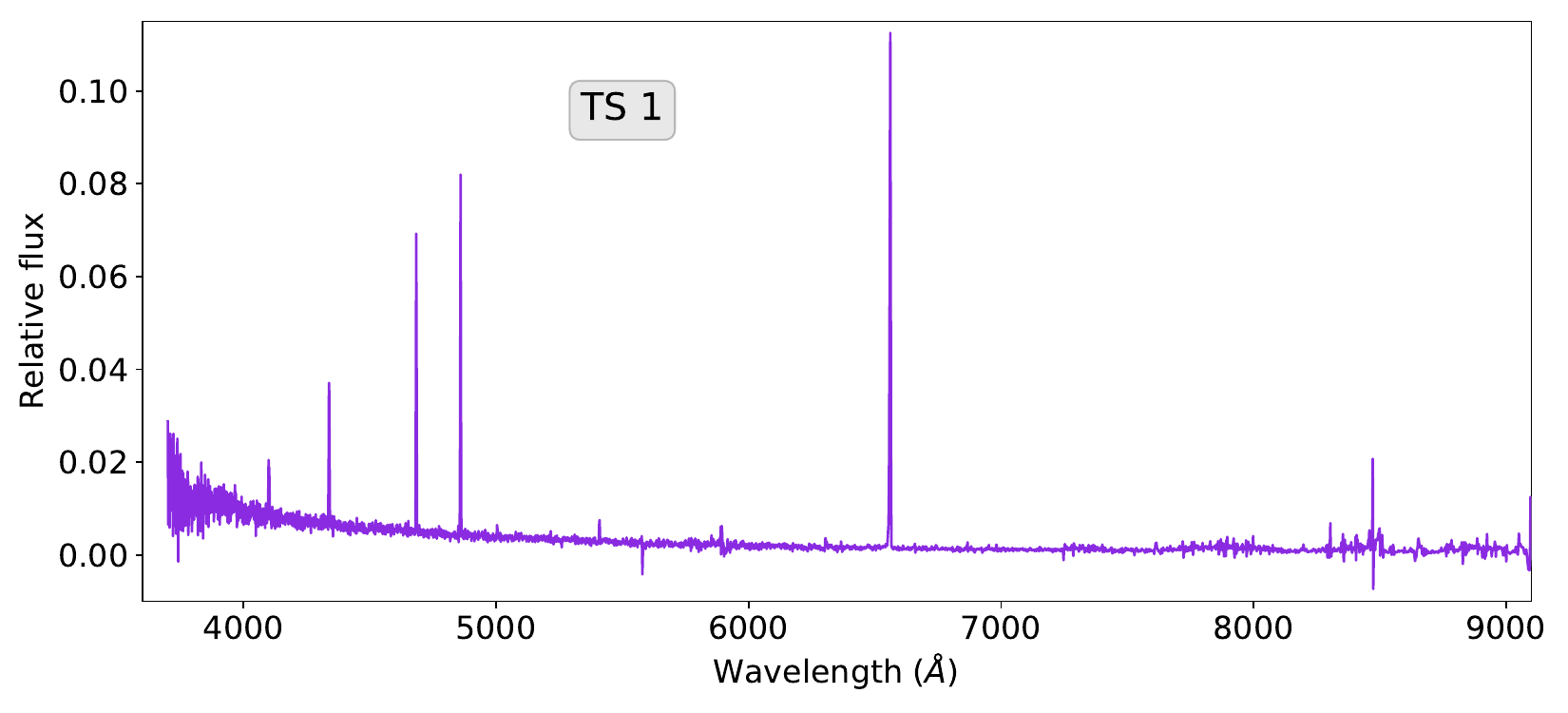} \llap{\shortstack{%
      \includegraphics[width=0.15\linewidth, clip]{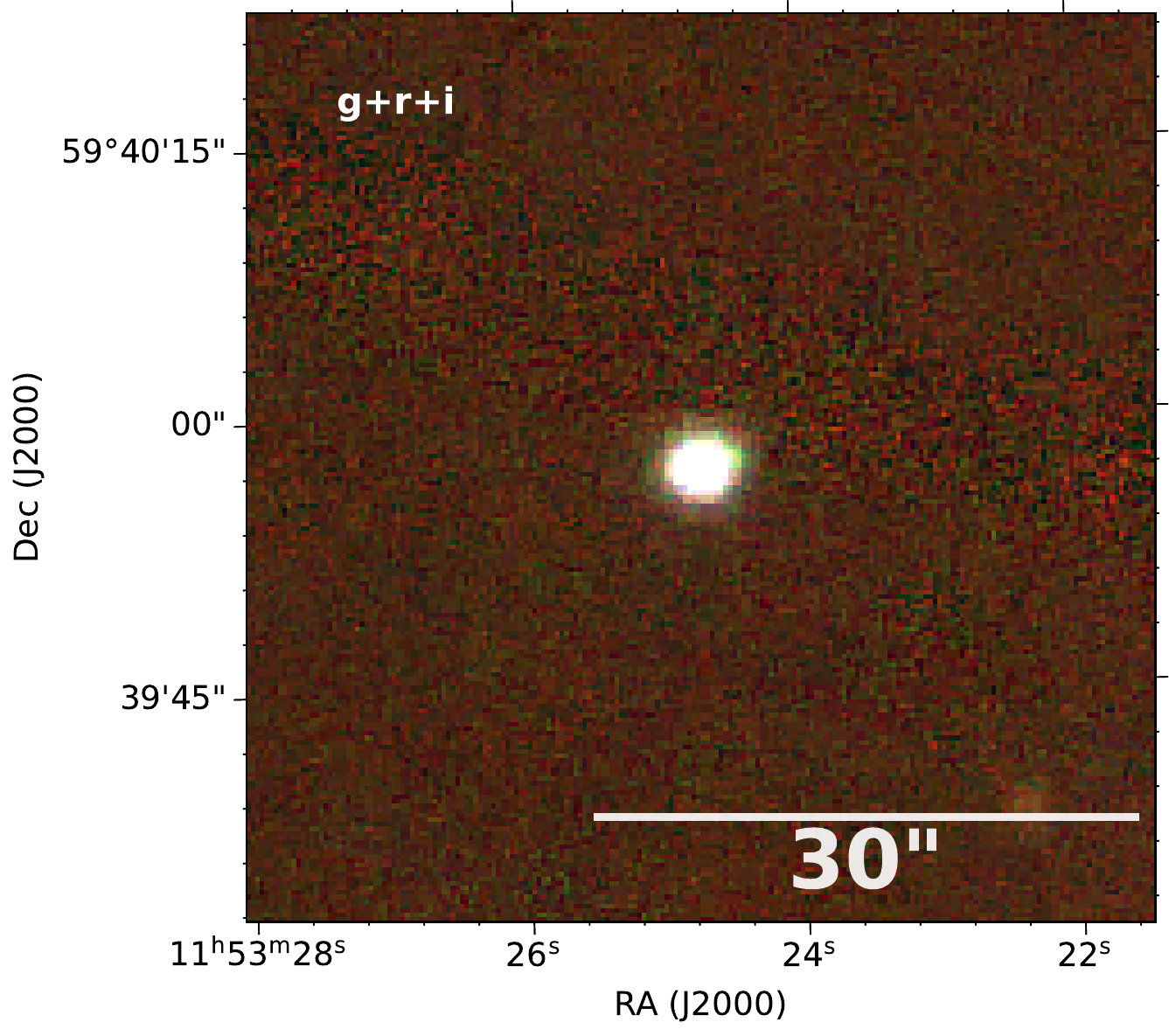}\\
      \rule{0ex}{0.9cm}%
      }
    \rule{0.25cm}{0ex}} \\
  
  \includegraphics[width=0.5\linewidth, trim=10 50 10 10, clip]{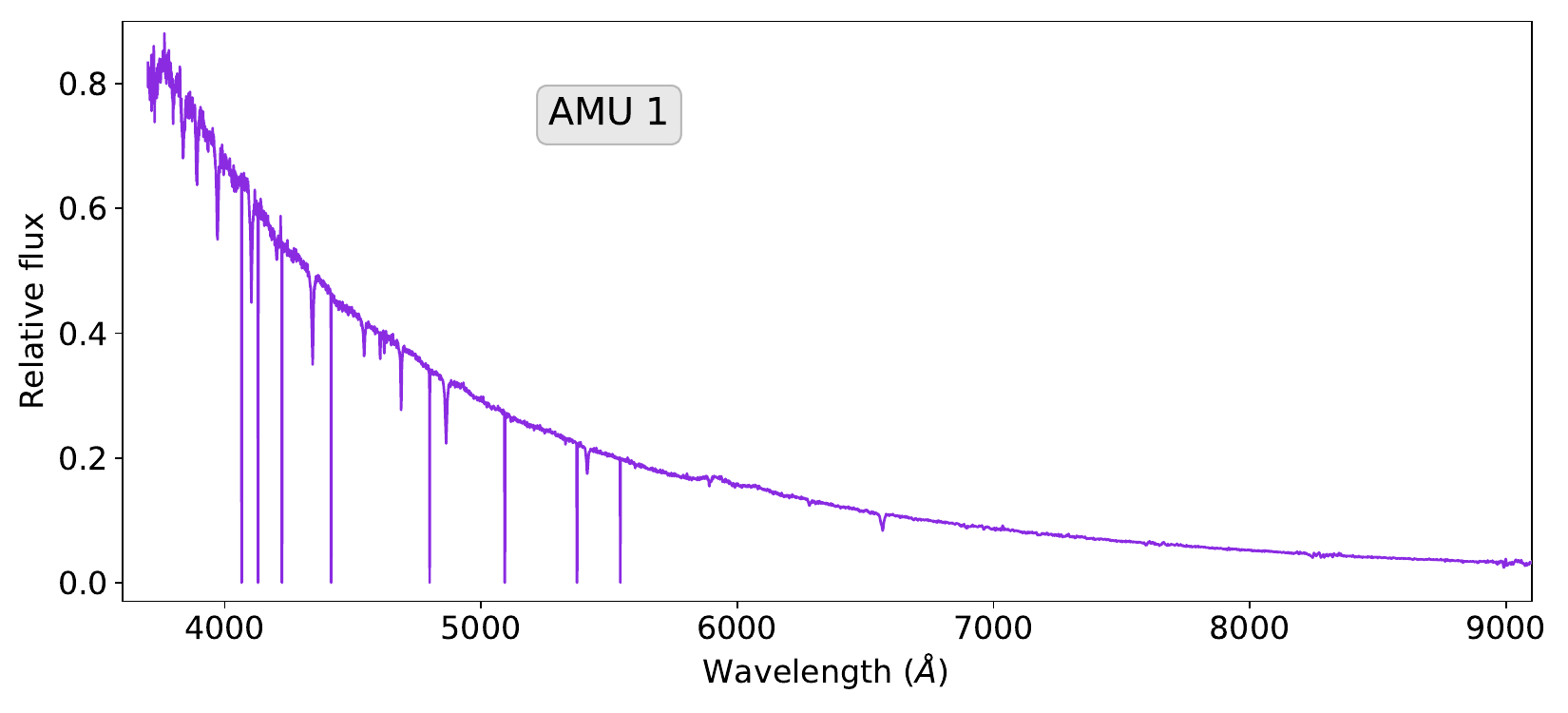} \llap{\shortstack{%
      \includegraphics[width=0.15\linewidth, clip]{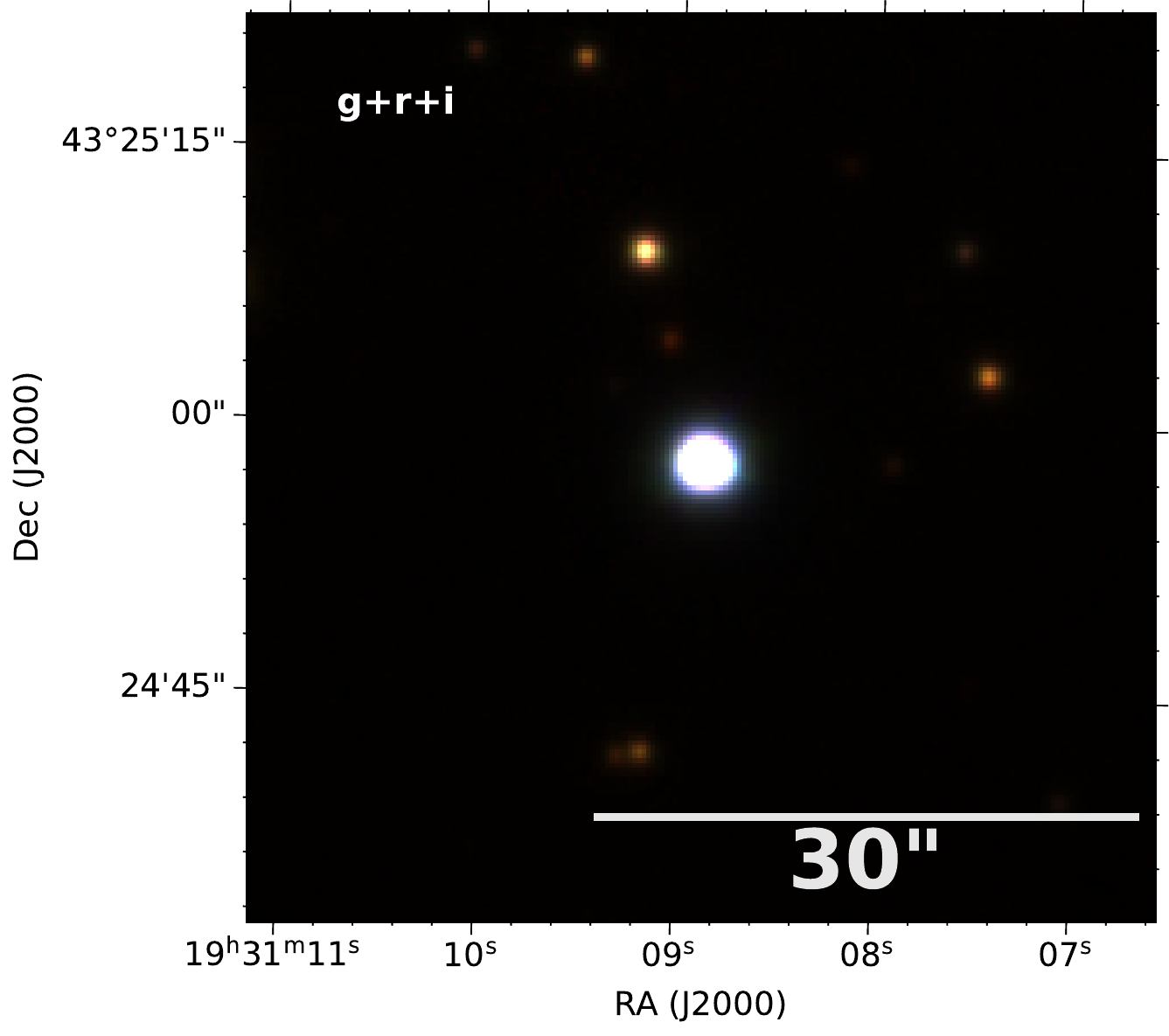} \\
      \rule{0ex}{0.9cm}%
      }
    \rule{0.25cm}{0ex}} &
  \includegraphics[width=0.5\linewidth, trim=5 10 10 10, clip]{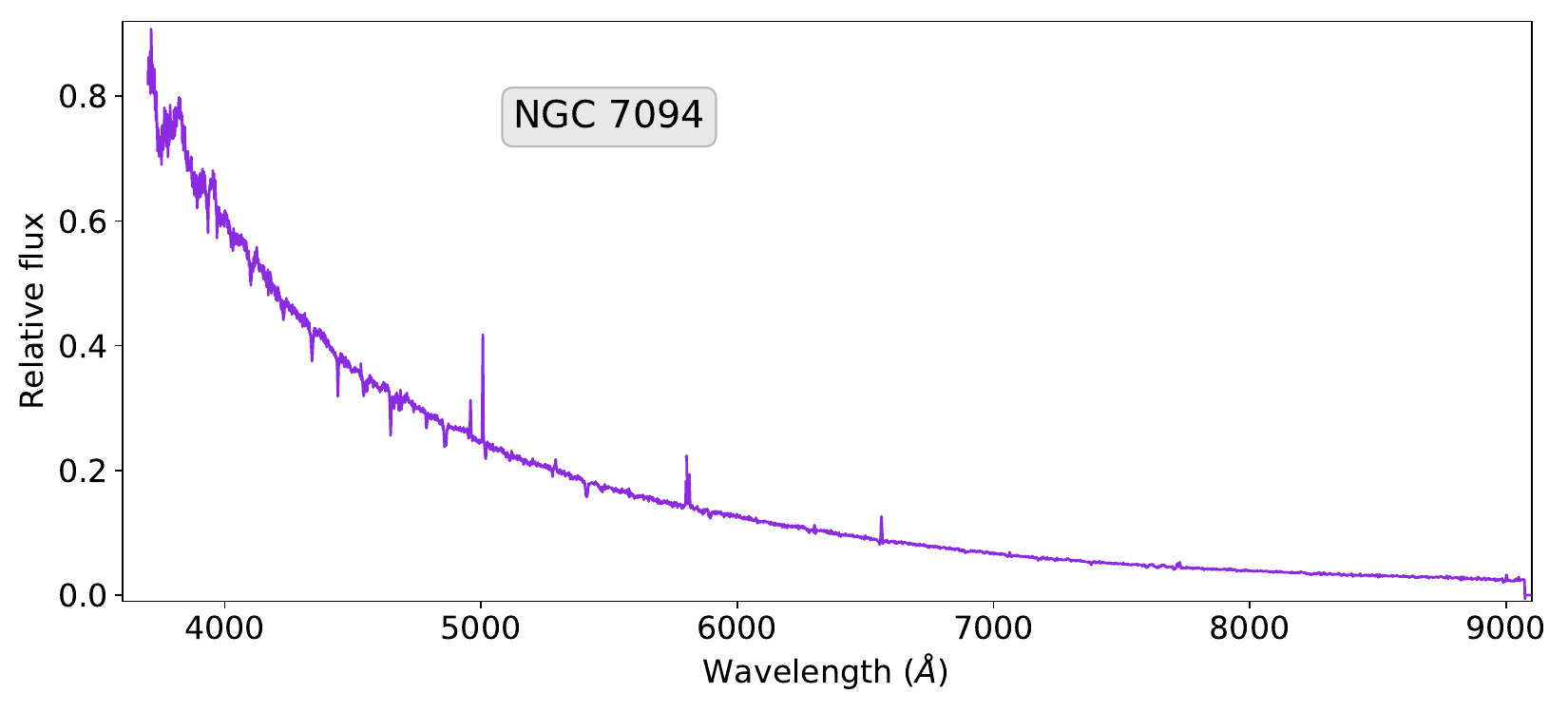} \llap{\shortstack{%
      \includegraphics[width=0.15\linewidth,  clip]{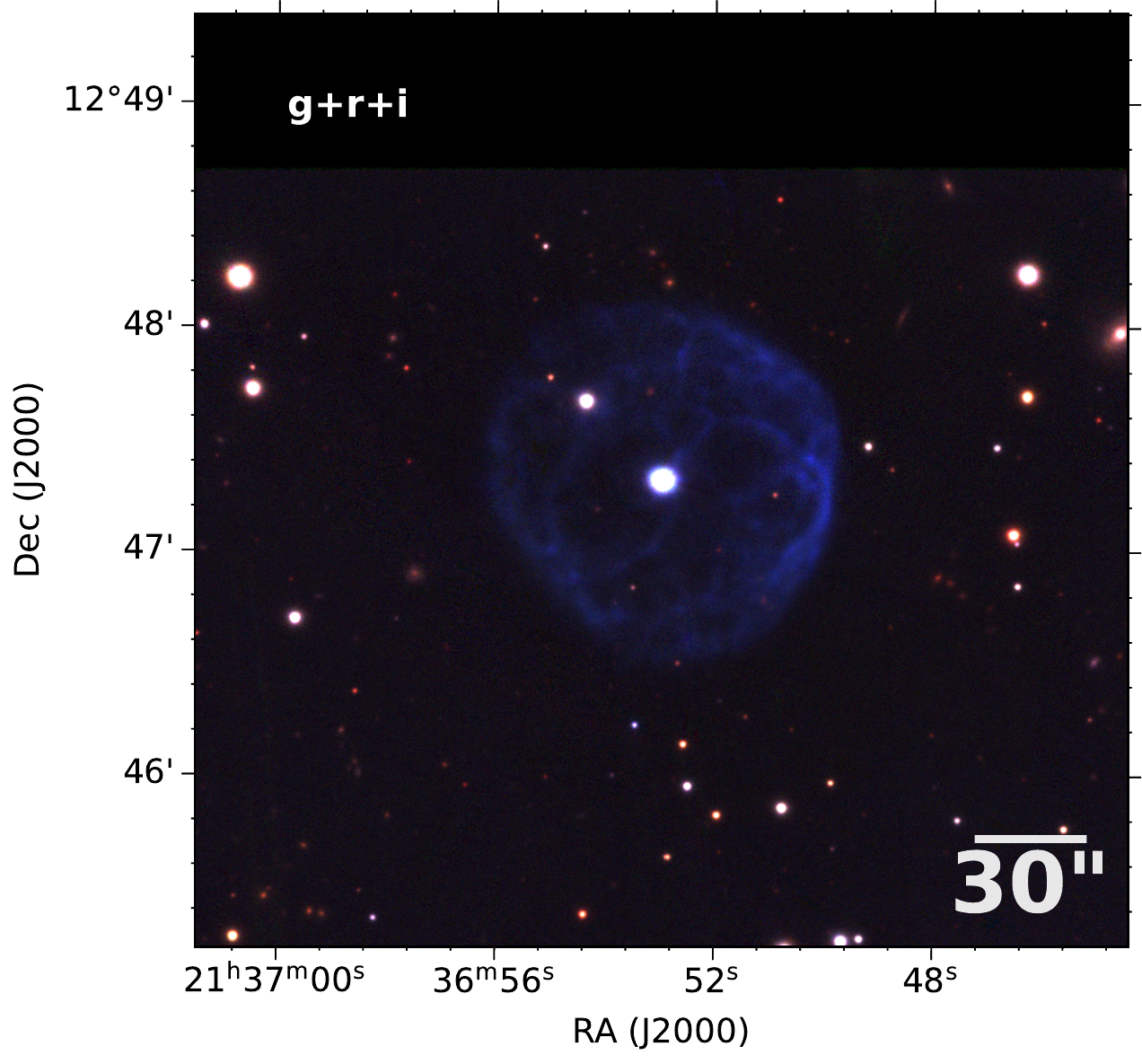}\\
      \rule{0ex}{1.3cm}%
      }
    \rule{0.25cm}{0ex}} \\
  
  \includegraphics[width=0.5\linewidth, trim=10 10 10 10, clip]{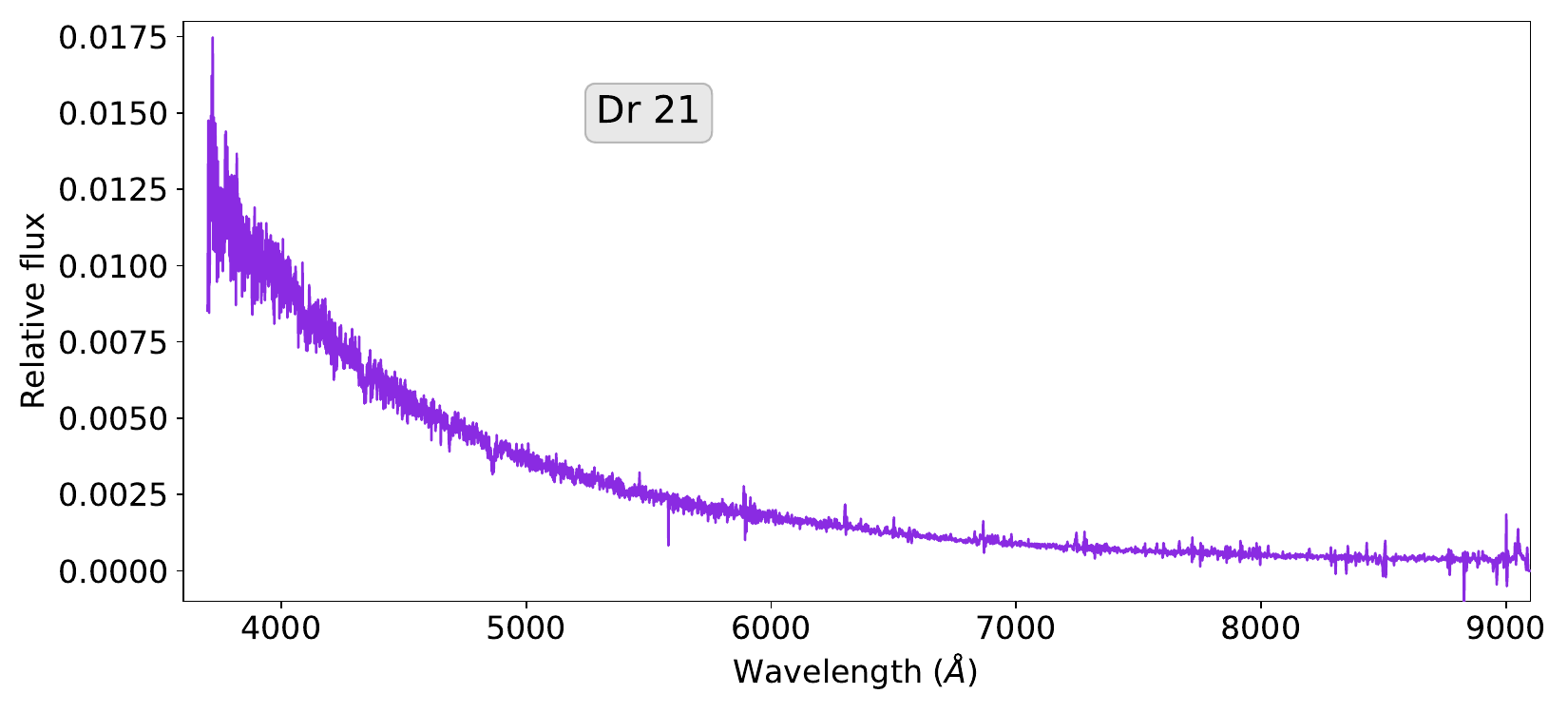} \llap{\shortstack{%
      \includegraphics[width=0.15\linewidth,  clip]{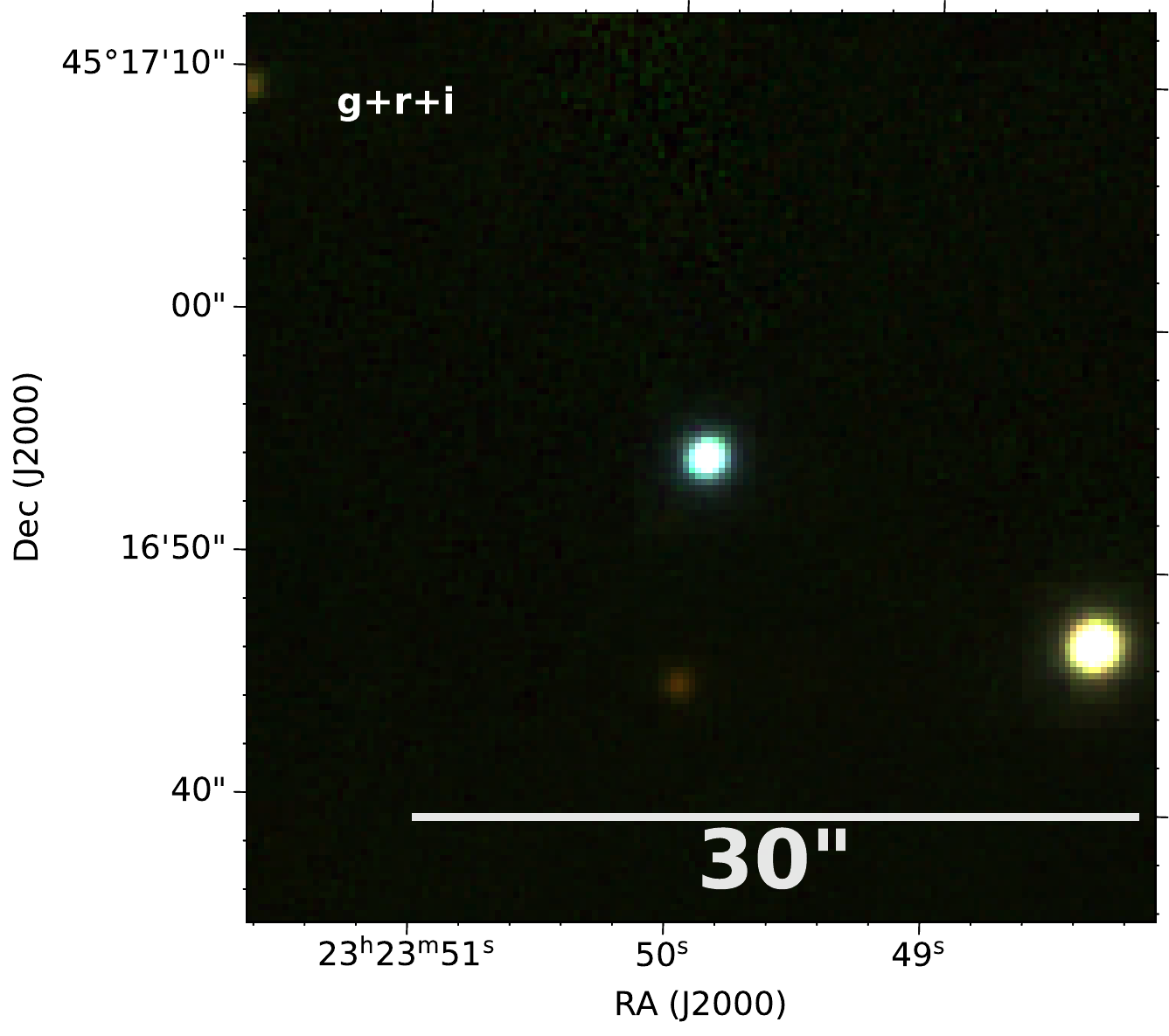}\\
      \rule{0ex}{1.4cm}%
      }
    \rule{0.25cm}{0ex}} &
  \rule{0.5\linewidth}{0pt} \\
\end{tabular}
\caption{Same as Figure~\ref{fig:spectra-image-trurPN-better} but for the eight true PNe located in the very extended zone.}
\label{fig:spectra-image-trurPN-medium}
\end{figure*}

\begin{figure*}
\centering
\begin{tabular}{ll}
  \includegraphics[width=0.5\linewidth, trim=10pt 50pt 10pt 10pt, clip]{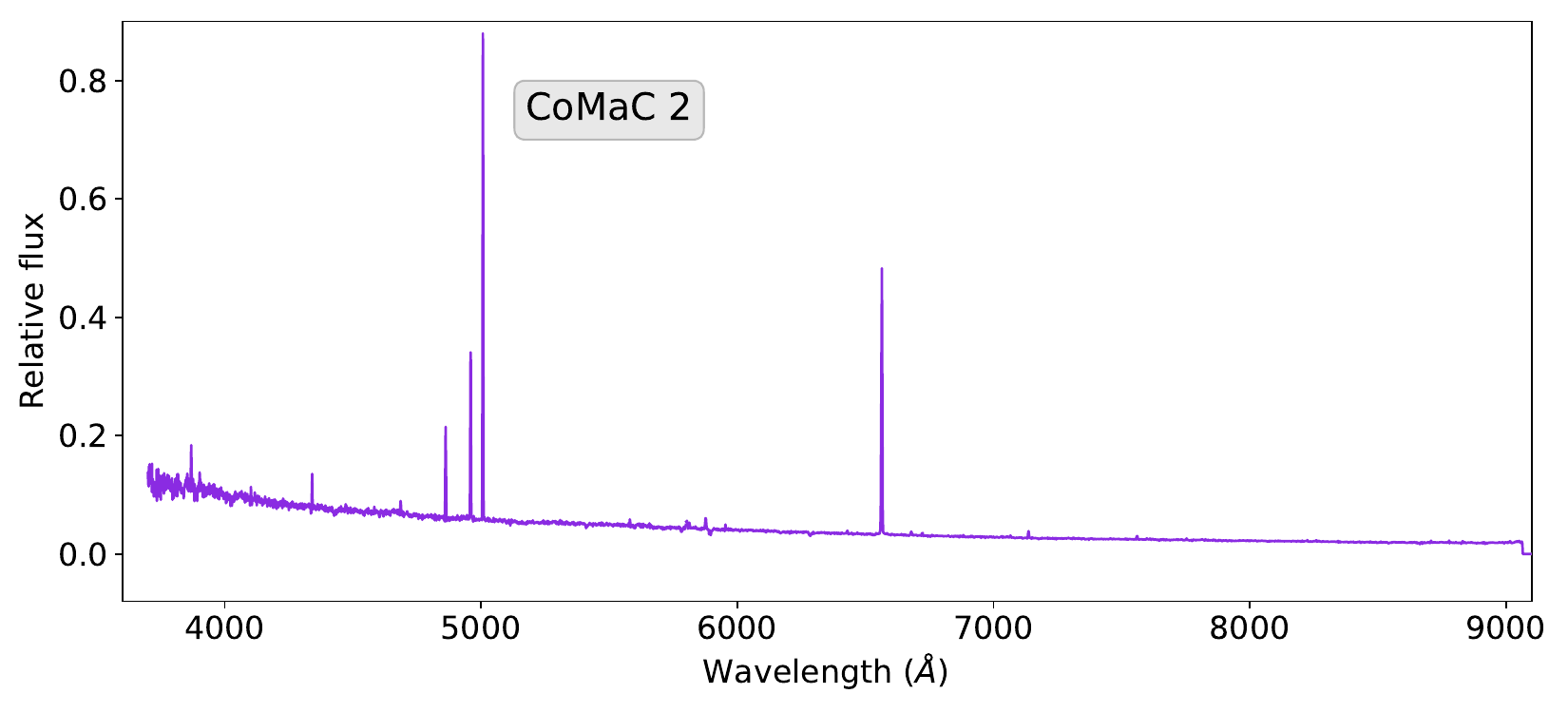} \llap{\shortstack{%
      \includegraphics[width=0.15\linewidth, clip]{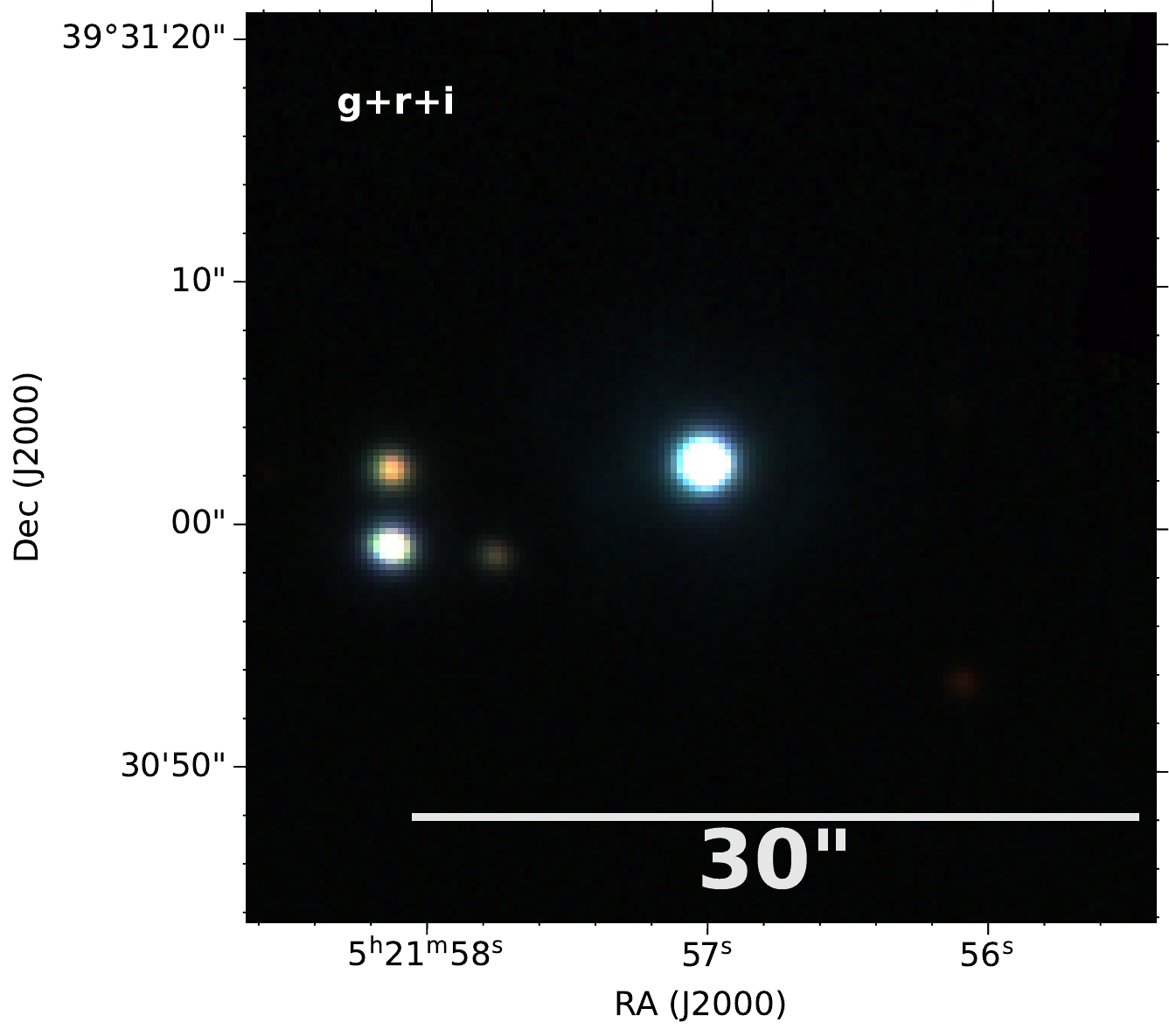} \\
      \rule{0ex}{0.85cm}%
      }
    \rule{0.2cm}{0ex}} &
  \includegraphics[width=0.5\linewidth, trim=1pt 50pt 10pt 10pt, clip]{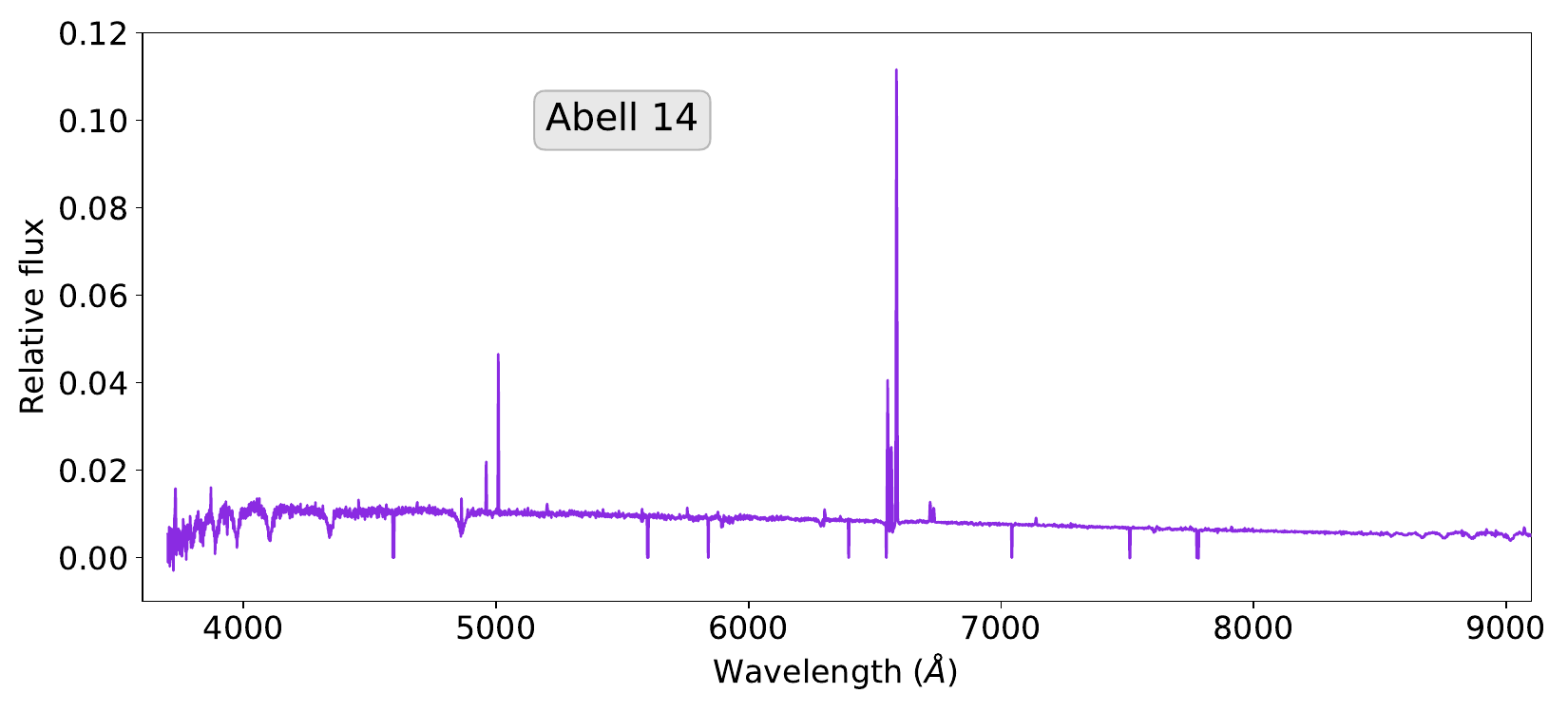} \llap{\shortstack{%
      \includegraphics[width=0.15\linewidth, clip]{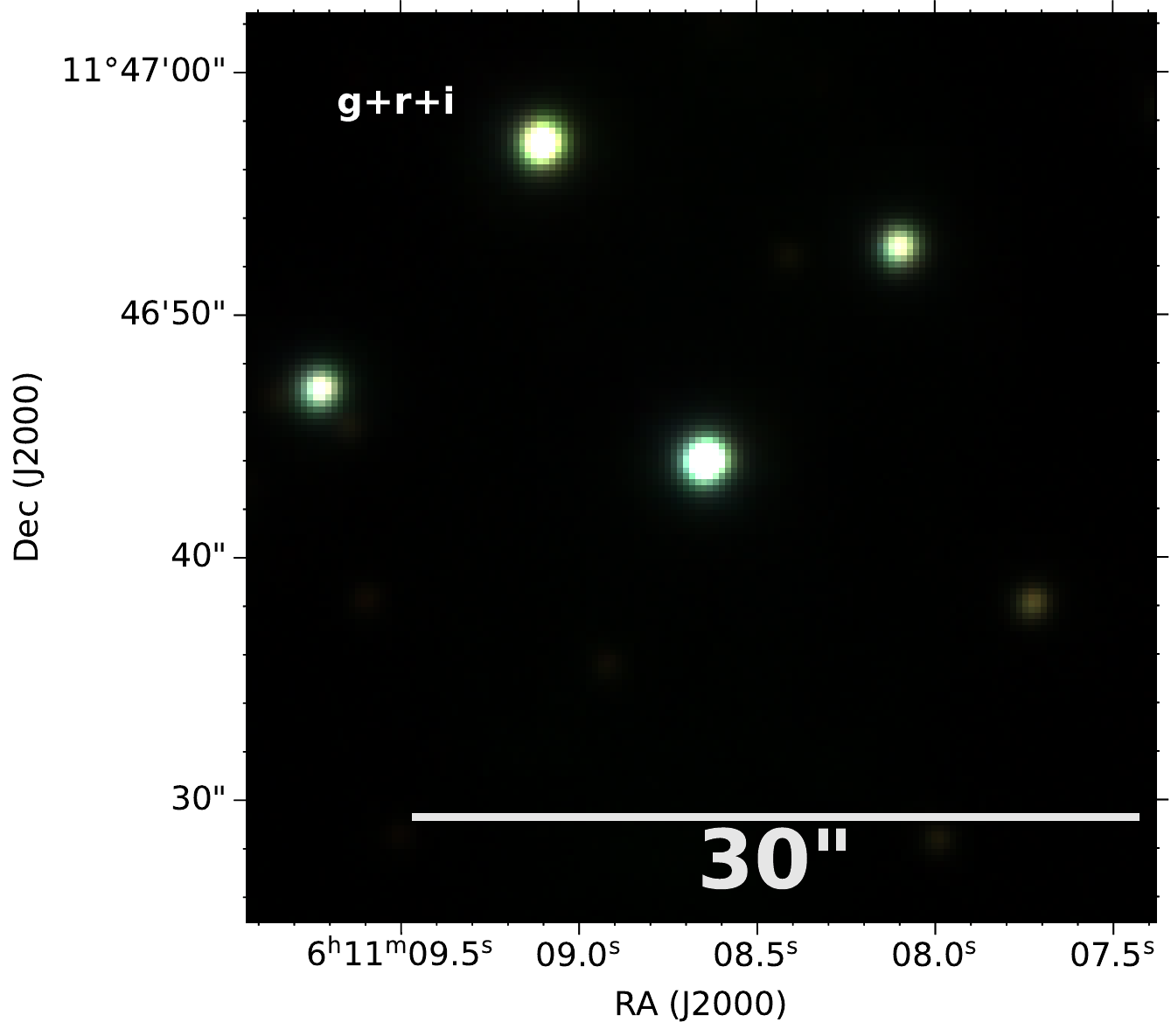}\\
      \rule{0ex}{0.85cm}%
      }
    \rule{0.25cm}{0ex}} \\
  
  \includegraphics[width=0.5\linewidth, trim=10pt 50pt 10pt 10pt, clip]{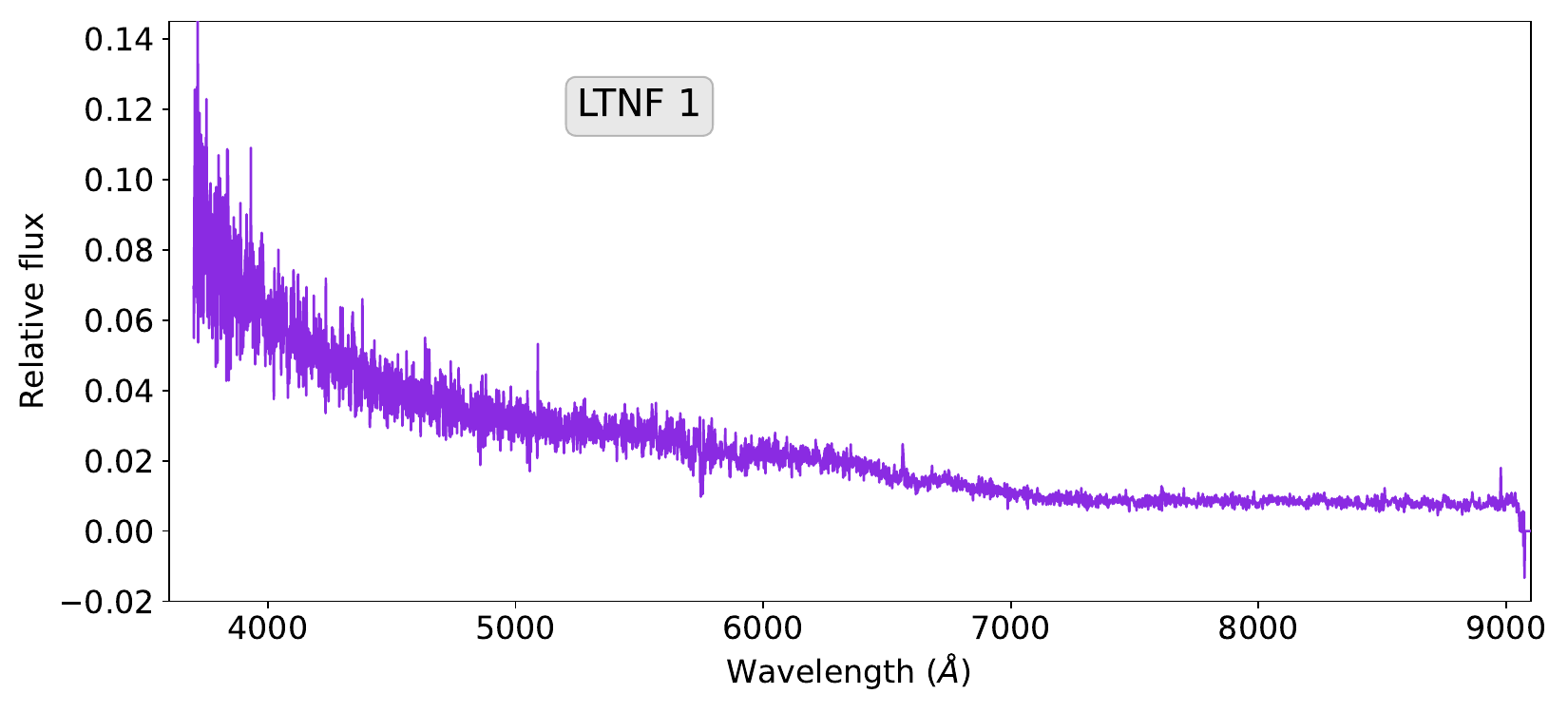} \llap{\shortstack{%
      \includegraphics[width=0.15\linewidth, clip]{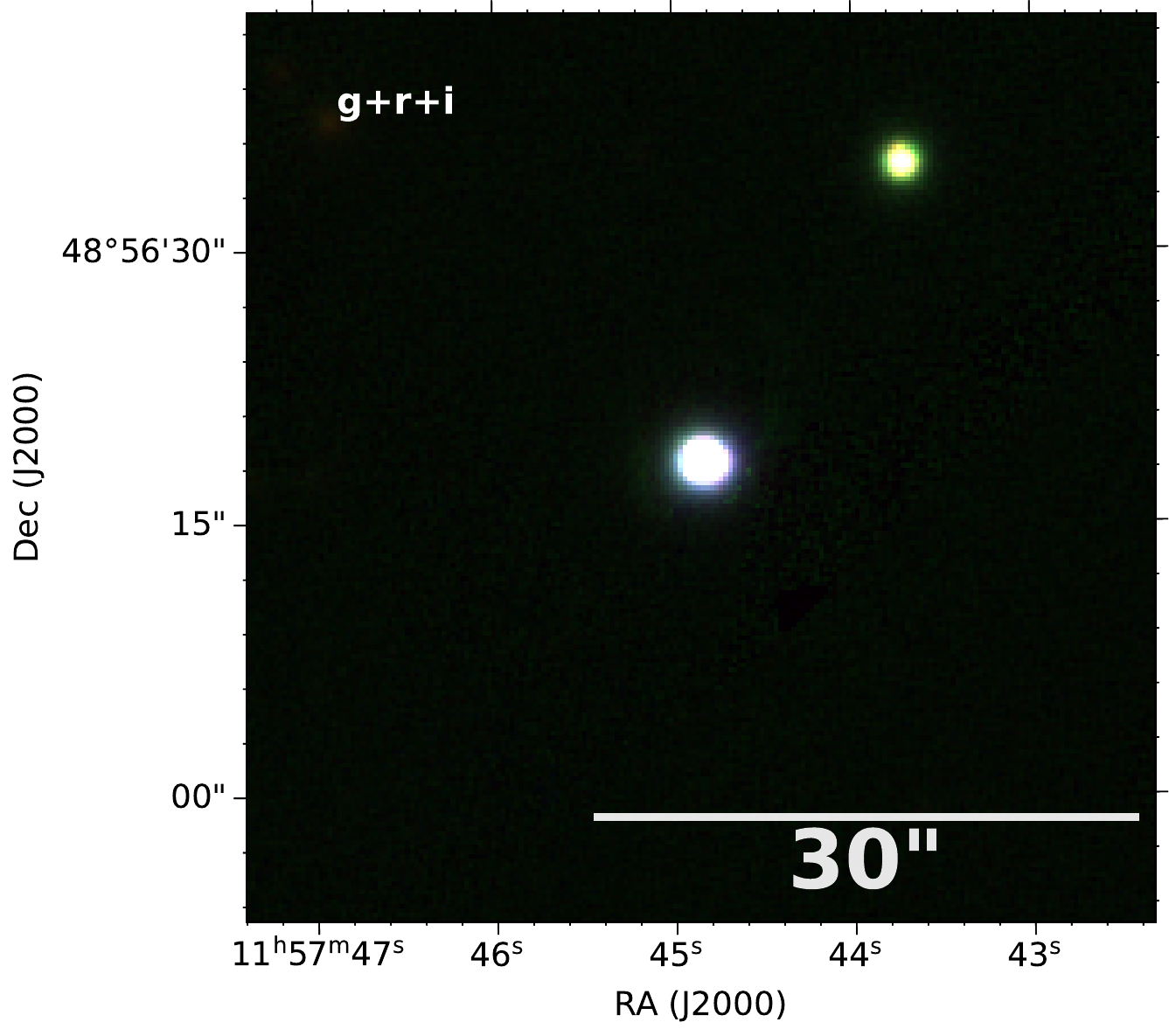}\\
      \rule{0ex}{0.85cm}%
      }
    \rule{0.2cm}{0ex}} &
  \rule{0.5\linewidth}{0pt} \\
\end{tabular}
\caption{Same as Figure~\ref{fig:spectra-image-trurPN-better} but for the three true PNe that are located outside of both compact zones in the color-color diagrams.}
\label{fig:spectra-image-trurPN-outside}
\end{figure*}

We present the LAMOST spectra and false-color images based on the $g$, $r$, and $i$ filters of Pan-STARRS for the true PNe featured in this study. We have organized the spectra and images into four figures: Figure~\ref{fig:spectra-image-trurPN-better} showcases the most compact objects within our sample, specifically the PNe located within the limited compact zone identified in the color-color diagrams. Figure~\ref{fig:spectra-image-trurPN-justonecompact} exhibits the PNe that met at least one of the color criteria for compact PNe. Figure~\ref{fig:spectra-image-trurPN-medium} presents the PNe situated in the zone where we found the very extended PNe. And Figure~\ref{fig:spectra-image-trurPN-outside} highlights PNe that fall outside our established selection criteria. Additional detailed information about the confirmed true PNe can be found in Table~\ref{tab:TruePN-inf}. 

\begin{table*}
	\centering
  \begin{threeparttable}
	\caption{List of true PNe in our study with their respective names, coordinates (Right Ascension and Declination), angular sizes (in arcseconds) obtained from the HASH catalog, \((G - g)\), \((G - r)\), and \((G_{BP} - G_{RP})\) colors. Additionally, the comments column describes the region classification in the diagrams. }
	\label{tab:TruePN-inf}

	\begin{tabular}{lcccccccccc} 
                \hline
		\hline
		Name     &         \(\mathrm{RA (J2000)}\) & \(\mathrm{Dec (J2000)}\) & PN diameter (HASH) & $G - g$ & $G - r$ & $BP-RP$ & Comments$^{a}$\\
          \hline
M 1-4 & 03:41:43.42 & 52:17:00.3 & 4.2 & 2.81 & 2.60 & 0.70 & Cpn(g), Cpn(r)  \\
IC 2003 & 03:56:21.99 & 33:52:30.7 & 10.0 & 2.02 & 1.06 & -0.16 & Cpn(g), Cpn(r)  \\
M 2-2 & 04:13:15.03 & 56:56:58.4 & 13.0 & 0.03 & 0.45 & 0.89 & Cpn(r), Epn(g)\\
K 3-67 & 04:39:47.93 & 36:45:42.5 & -- & 3.78 & 3.50 & 0.41 & Cpn(g), Cpn(r)   \\
CoMaC 2 & 05:21:57.03 & 39:31:02.5 & 12.0 & -0.46 & -0.08 & 0.95 & Other \\
K 3-69 & 05:41:22.13 & 39:15:08.0 & -- & 0.70 & 0.80 & 0.64 & Cpn(g), Cpn(r) \\
K 3-70 & 05:58:45.35 & 25:18:43.9 & 5.5 & 2.77 & 4.07 & 0.98 & Cpn(g), Cpn(r) \\
Abell 14 & 06:11:08.66 & 11:46:43.8 & 40.0 & -0.32 & -0.03 & 0.76 & Other \\
Abell 20 & 07:22:57.67 & 01:45:34.0 & 67.3 & 0.14 & -0.21 & -0.38 & Epn(g),  Epn(r) \\
Abell 21 & 07:29:02.71 & 13:14:48.5 & 750.0 & 0.17 & -0.29 & -0.50 & Epn(g), Epn(r)\\
JnEr 1 & 07:57:51.62 & 53:25:16.9 & 394.0 & 0.17 & -0.30 & -0.55 & Epn(g), Epn(r) \\
Abell 30 & 08:46:53.46 & 17:52:46.3 & 127.0 & 0.89 & -0.16 & -0.18 & Cpn(g), Epn(r) \\
TS 1 & 11:53:24.74 & 59:39:57.0 & 9.2 & 0.30 & -0.04 & -0.29 & Epn(g), Epn(r)\\
LTNF 1 & 11:57:44.85 & 48:56:18.1 & 230.0 & -0.18 & -0.69 & 0.41 & Other \\
DdDm 1 & 16:40:18.15 & 38:42:20.0 & 1.4 & 1.32 & 1.10 & -0.02 & Cpn(g), Cpn(r) \\
Vy 1-2 & 17:54:23.04 & 27:59:57.0 & 6.0 & 3.88 & 2.39 & -0.44 & Cpn(g), Cpn(r) \\
AMU 1 & 19:31:08.88 & 43:24:57.7 & 294.0 & 0.12 & -0.26 & -0.40 & Epn(g), Epn(r) \\
NGC 7094 & 21:36:52.97 & 12:47:18.9 & 102.5 & 0.10 & -0.24 & -0.39 & Epn(g), Epn(r) \\
K 3-84 & 21:38:49.00 & 46:00:27.7 & 8.0 & 2.43 & 3.45 & 0.47 & Cpn(g), Cpn(r) \\
Dr 21 & 23:23:49.89 & 45:16:54.1 & 450.0 & 0.10 & -0.24 & -0.34 & Epn(g), Epn(r)\\
\hline
	\end{tabular}
     \begin{tablenotes}
            \item \textbf{Note.} $^{a}$ In this column, Cpn(g) - Compact PN zone in \((G - g)\) vs \((G_{BP} - G_{RP})\) color-color diagram, Cpn(r) - Compact PN zone in \((G - r)\) vs \((G_{BP} - G_{RP})\) color-color diagram, Epn(g) - Extended PN zone in \((G - g)\) vs \((G_{BP} - G_{RP})\) color-color diagram,
            Epn(r) - Extended PN zone in \((G - r)\) vs \((G_{BP} - G_{RP})\) colo-color diagram, Other - PNe that do not meet the specified criteria.
        \end{tablenotes}
 \end{threeparttable}
 
\end{table*}

\clearpage
\section{Likely and Probable PNe}
\label{sec:spectra-lpPNe}
\begin{figure*}
\centering
\begin{tabular}{ll}
  \includegraphics[width=0.5\linewidth, trim=1 50 10 10, clip]{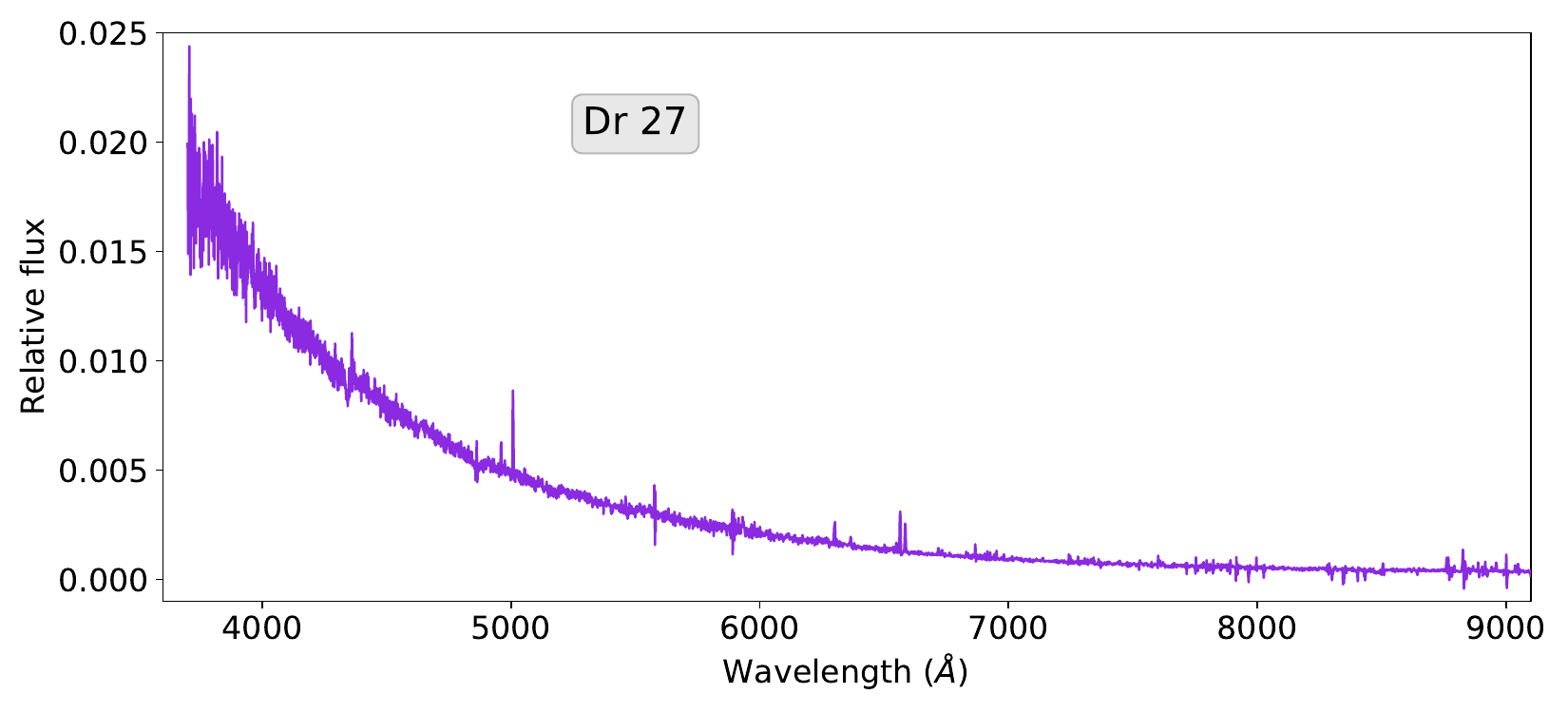} \llap{\shortstack{%
      \includegraphics[width=0.15\linewidth, clip]{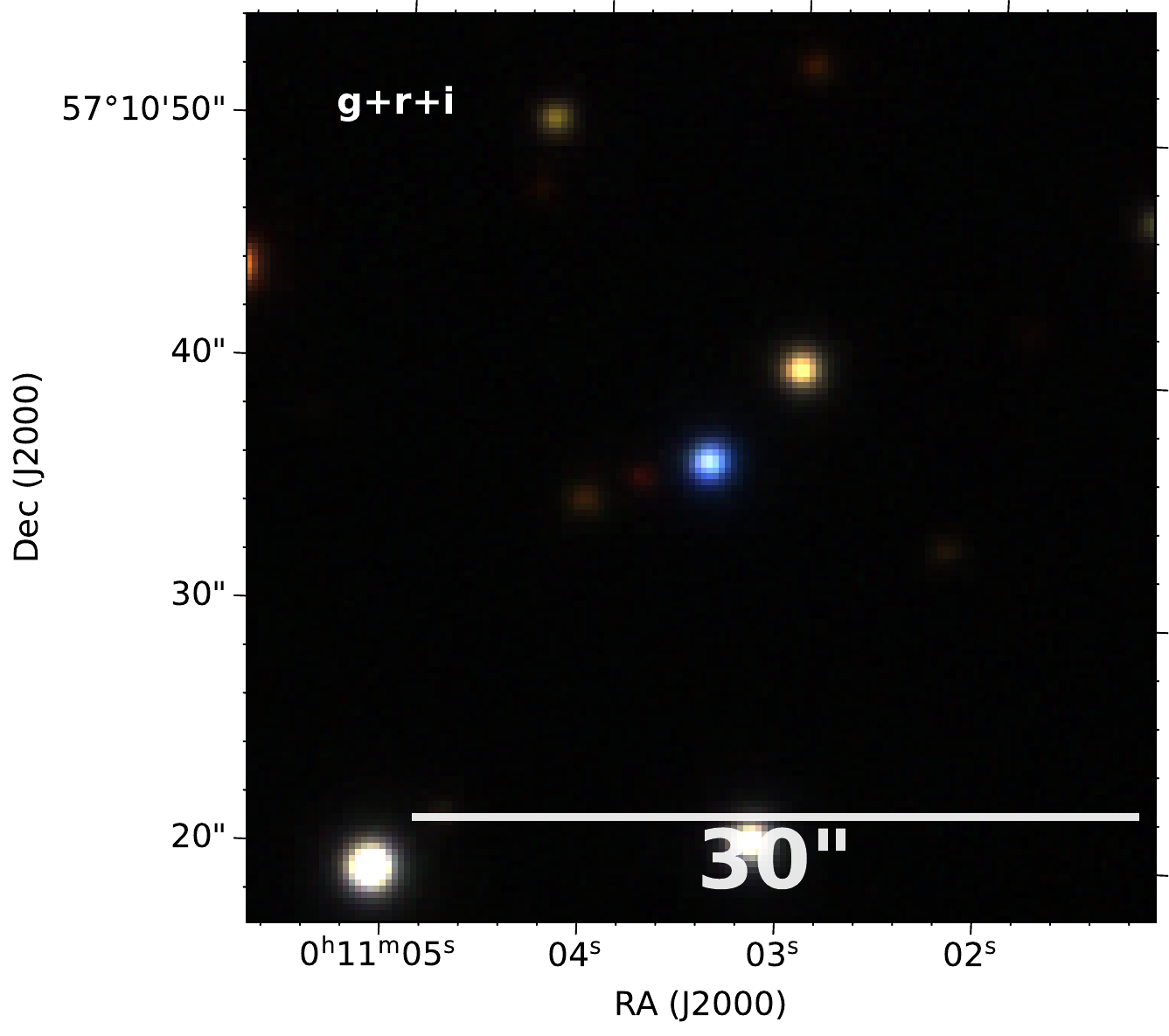}\\
      \rule{0ex}{0.85cm}%
      }
    \rule{0.25cm}{0ex}}  &
      \includegraphics[width=0.5\linewidth, trim=10 50 10 10, clip]{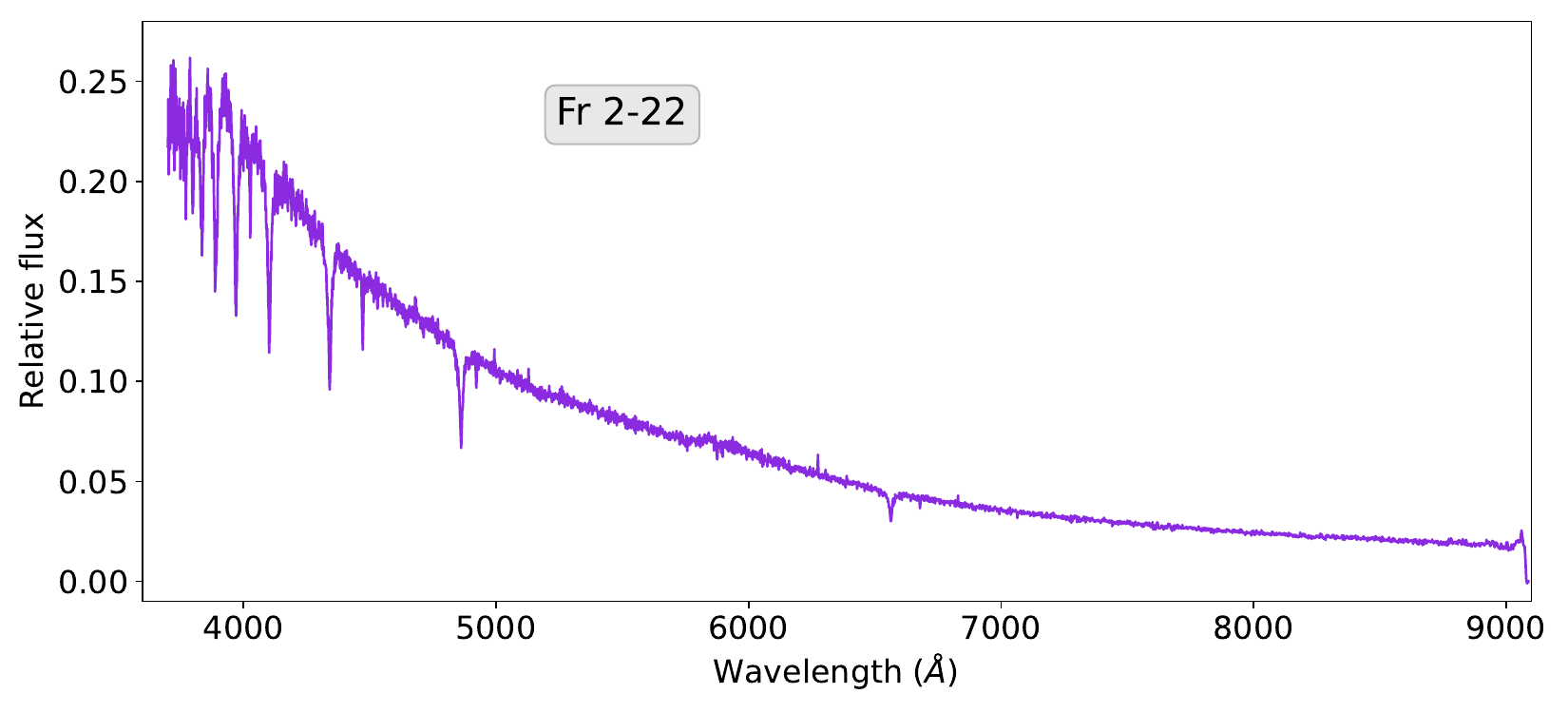} \llap{\shortstack{%
          \includegraphics[width=0.15\linewidth, clip]{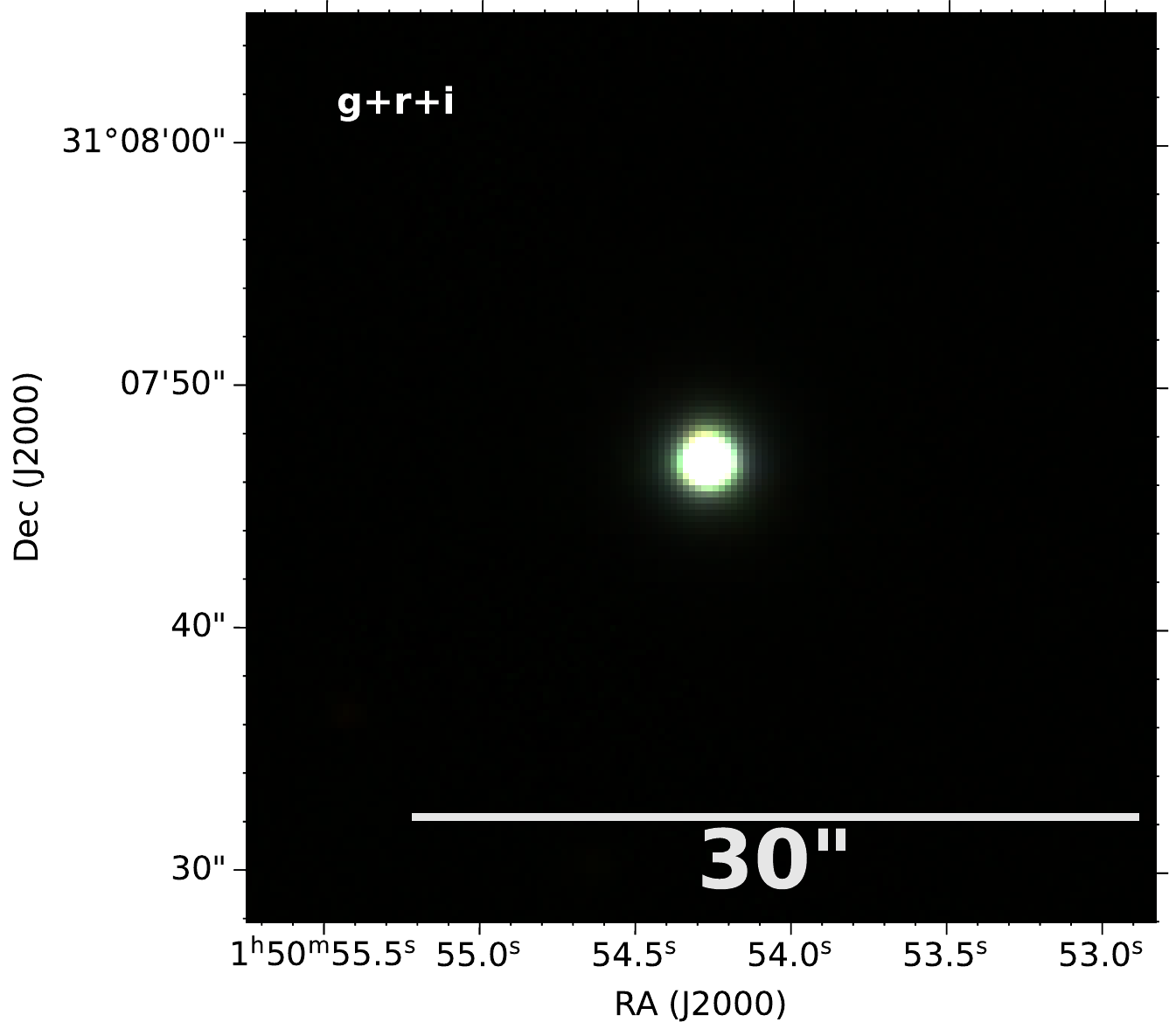} \\
          \rule{0ex}{0.85cm}%
      }
  \rule{0.2cm}{0ex}} \\
\includegraphics[width=0.5\linewidth, trim=1 50 10 10, clip]{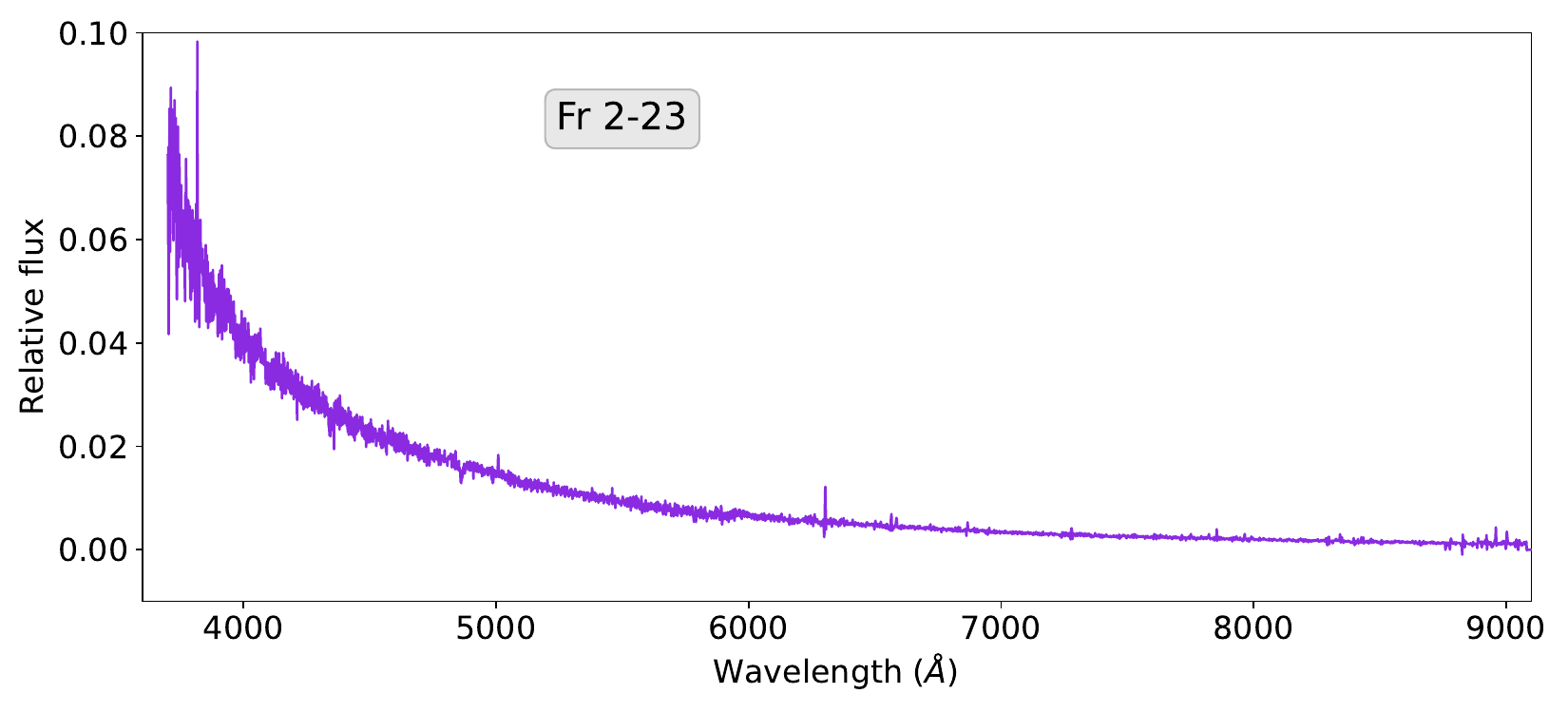} \llap{\shortstack{%
        \includegraphics[width=0.15\linewidth,  clip]{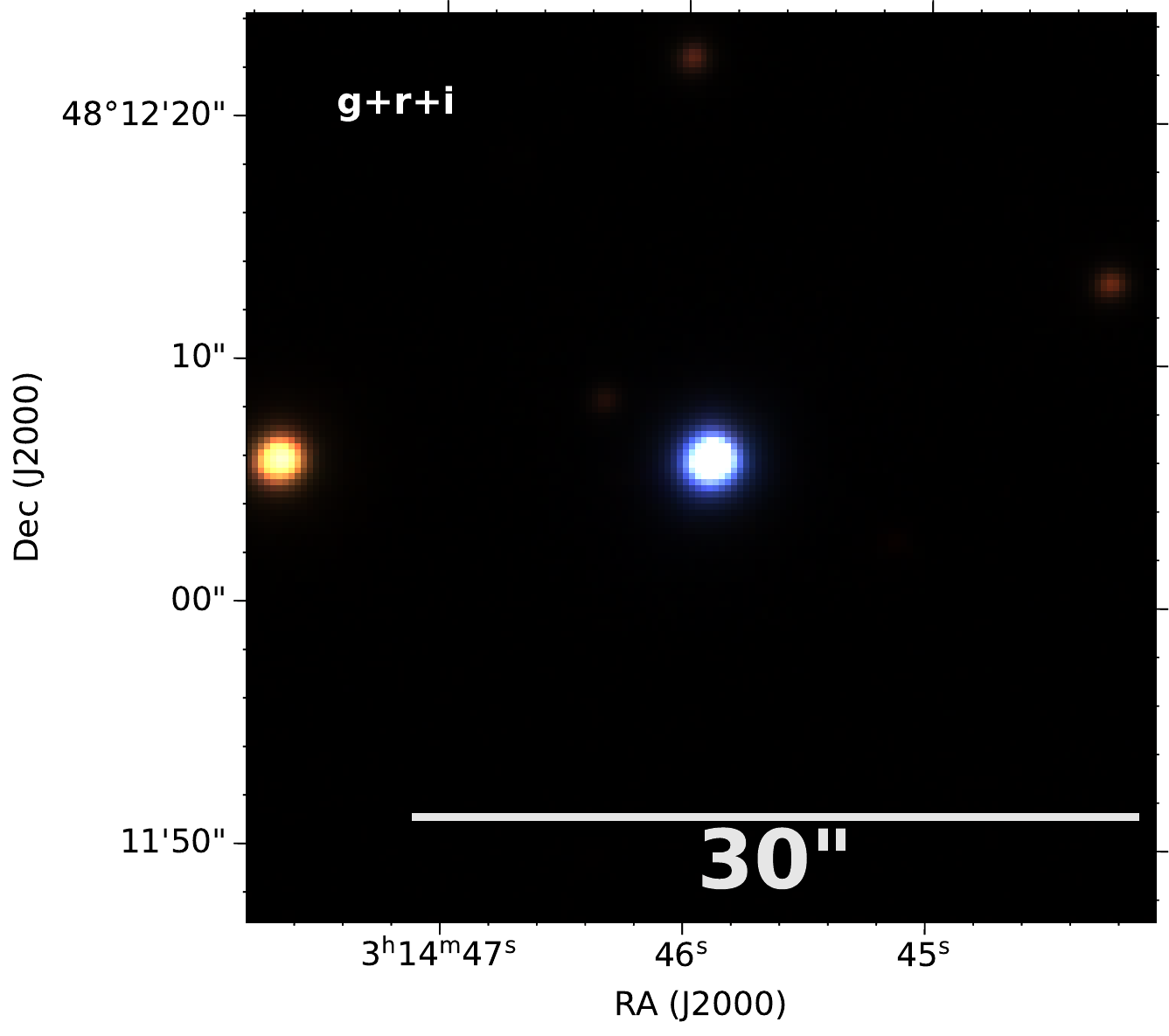}\\
        \rule{0ex}{0.85cm}%
      }
  \rule{0.25cm}{0ex}} &
  \includegraphics[width=0.5\linewidth, trim=10 50 10 10, clip]{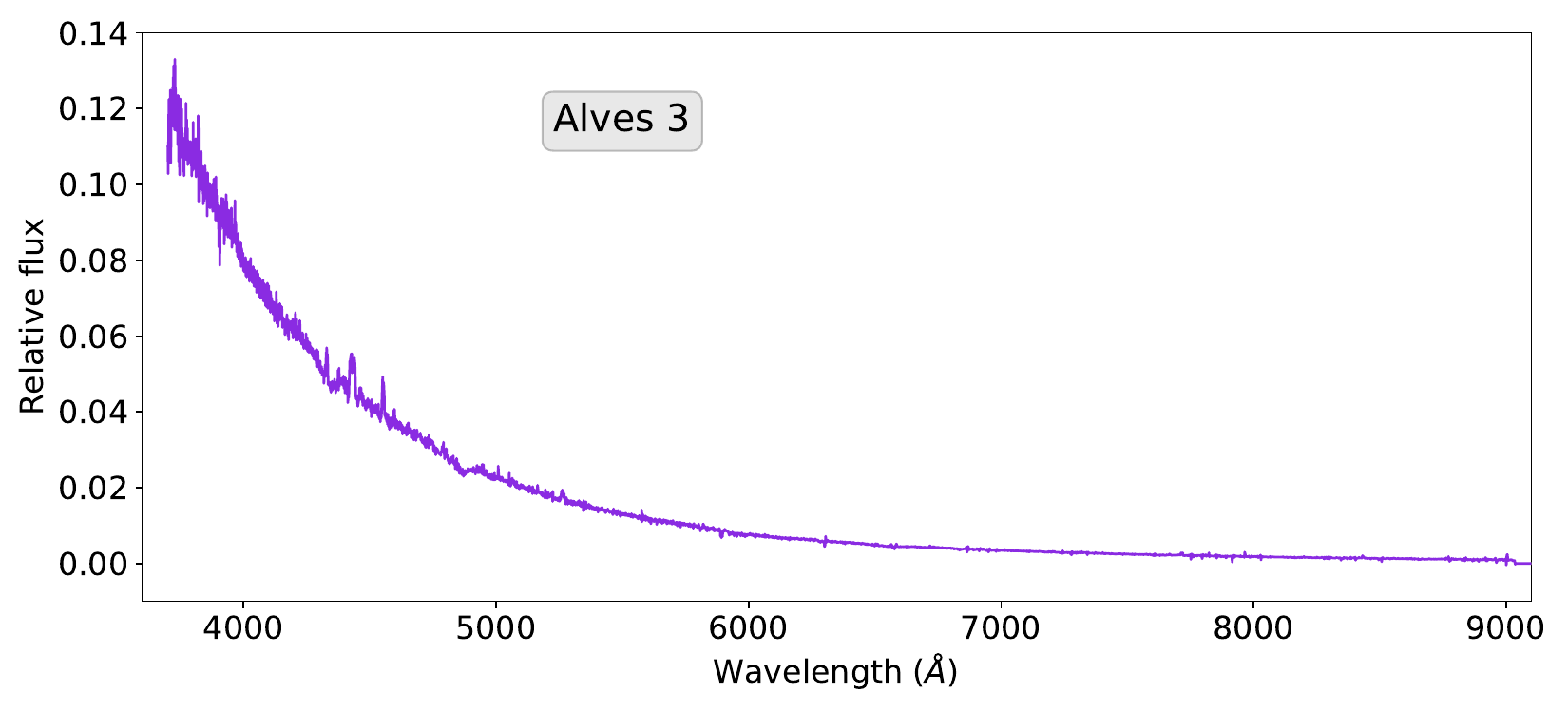} \llap{\shortstack{%
      \includegraphics[width=0.15\linewidth, clip]{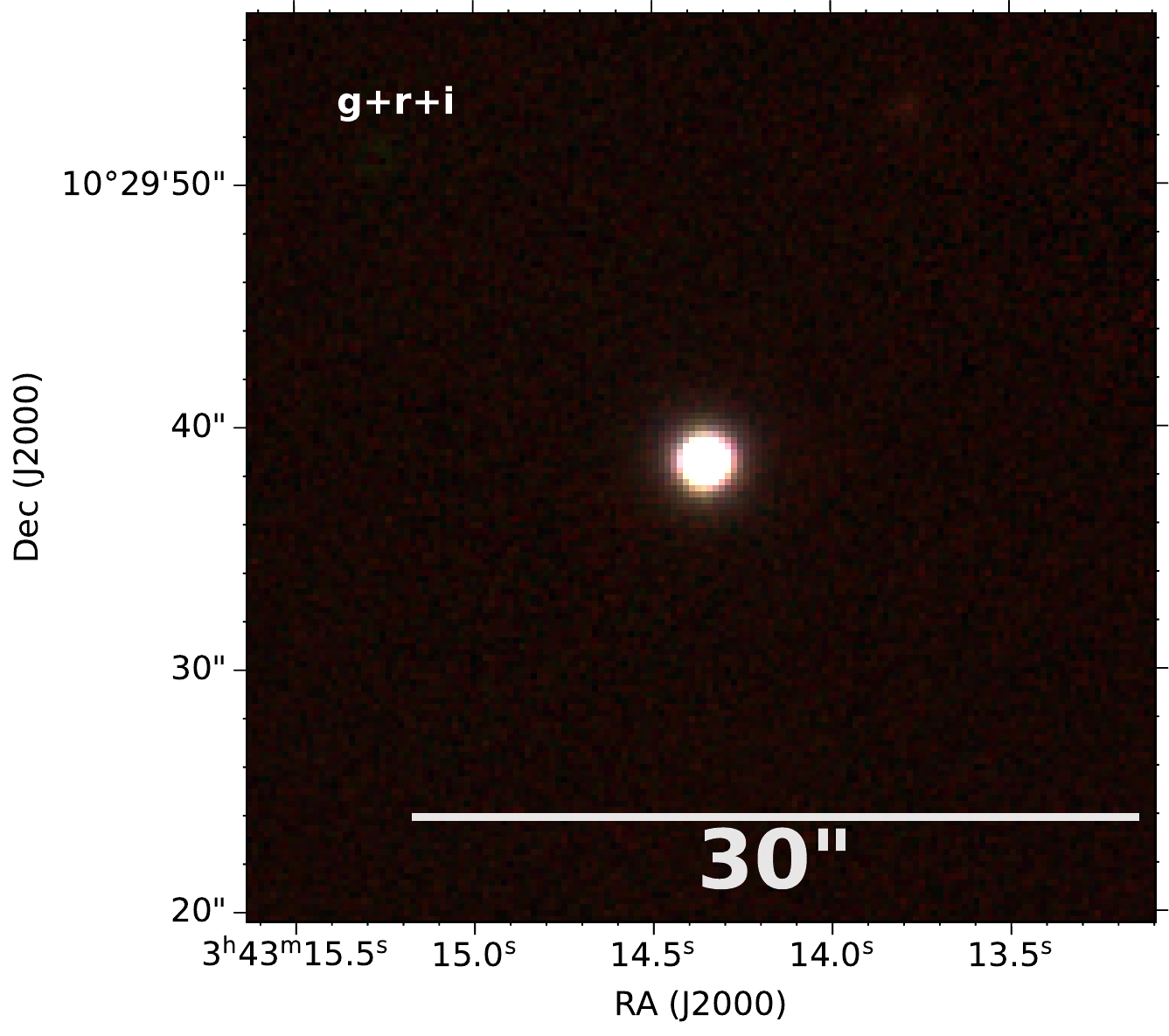}\\
      \rule{0ex}{0.85cm}%
      }
    \rule{0.2cm}{0ex}} \\
  \includegraphics[width=0.5\linewidth, trim=1 50 10 10, clip]{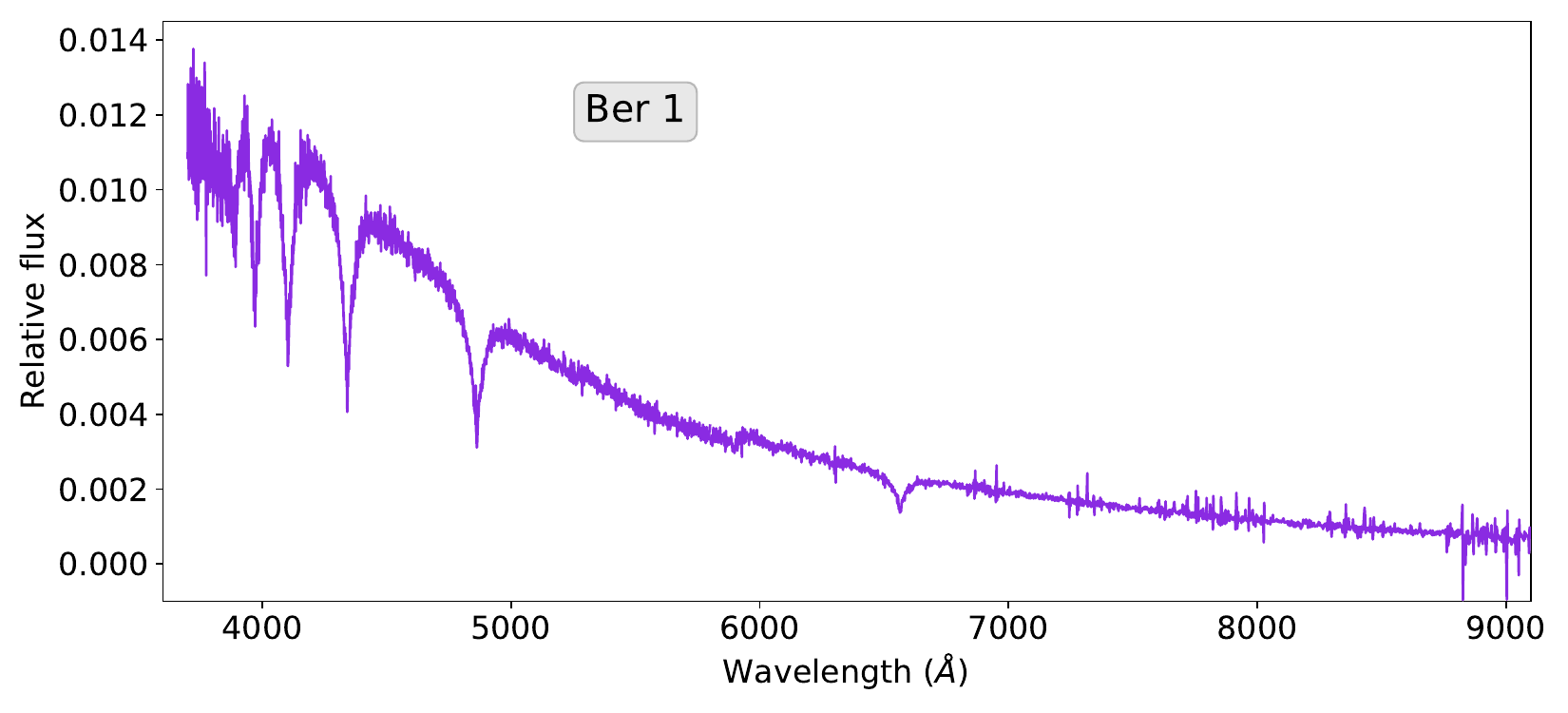} \llap{\shortstack{%
        \includegraphics[width=0.15\linewidth, clip]{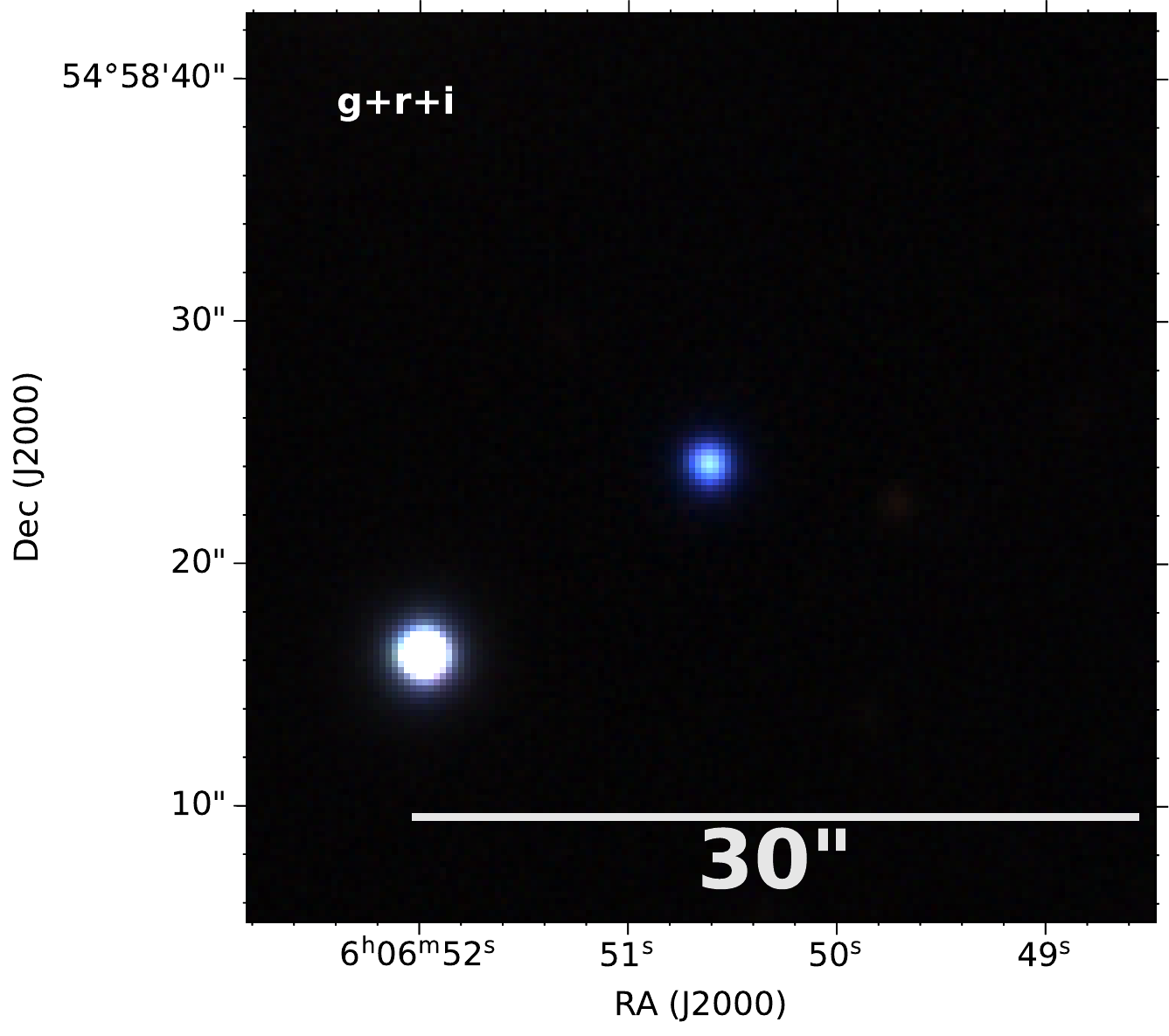}\\
        \rule{0ex}{0.75cm}%
      }
    \rule{0.25cm}{0ex}} & \includegraphics[width=0.5\linewidth, trim=10 50 10 10, clip]{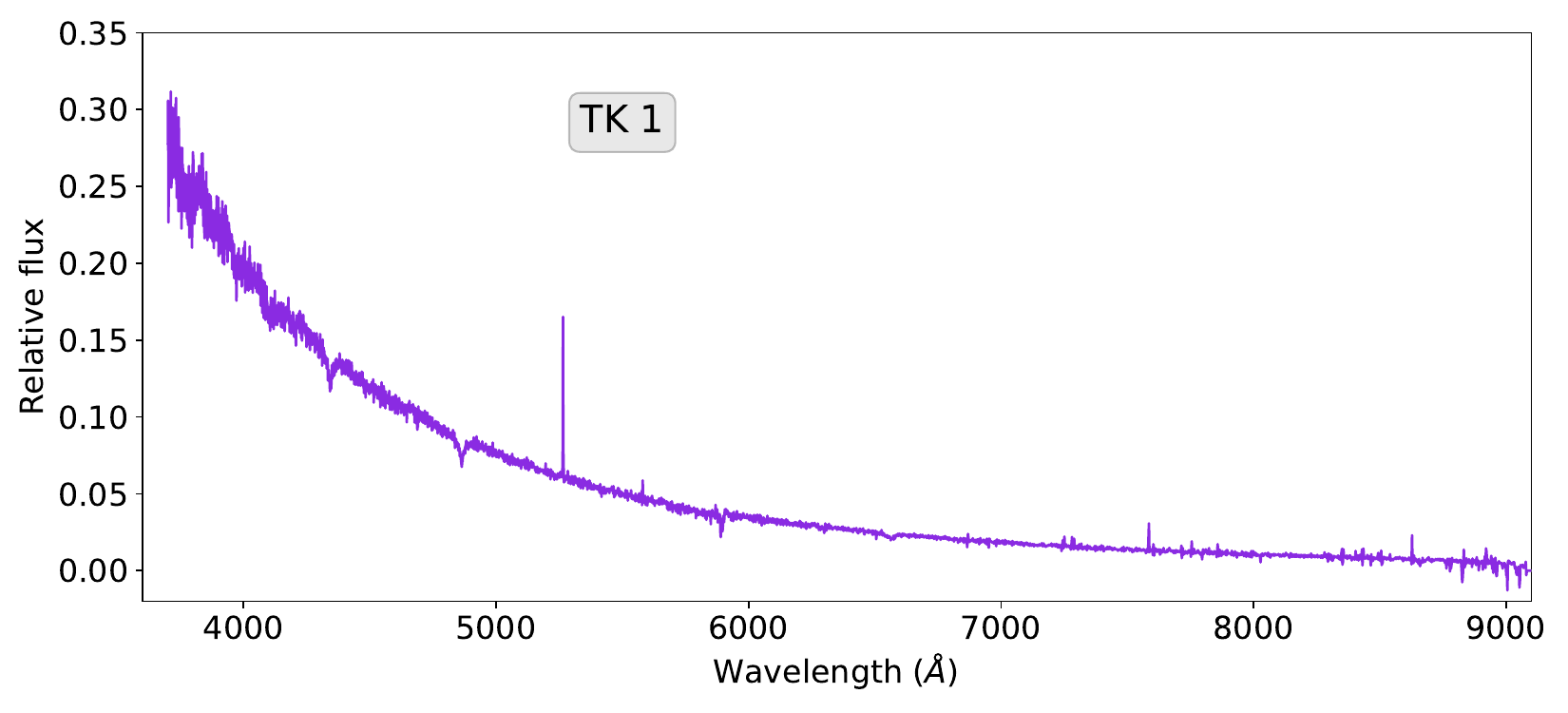}  \llap{\shortstack{%
      \includegraphics[width=0.15\linewidth, clip]{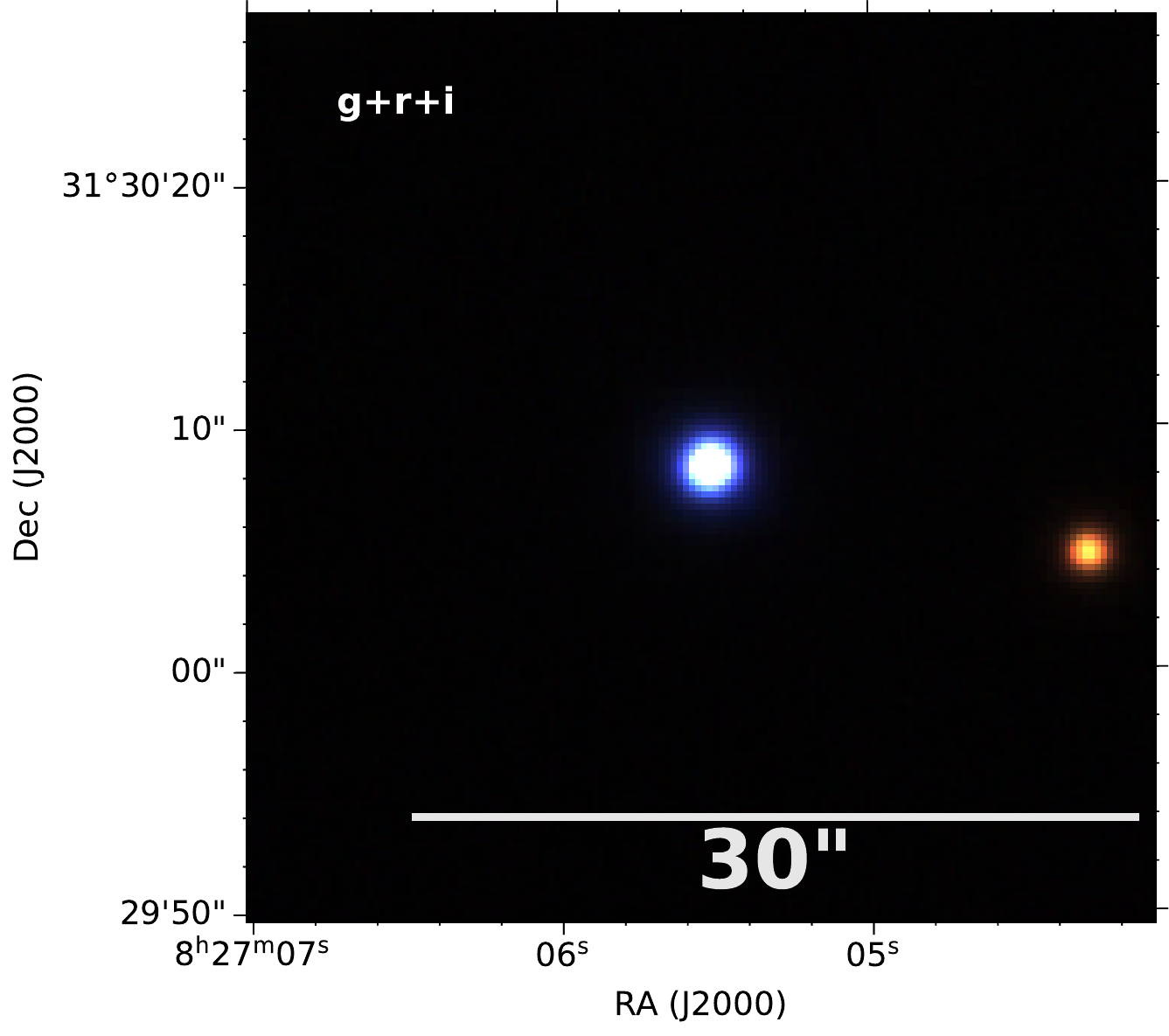} \\
      \rule{0ex}{0.85cm}%
      }
    \rule{0.25cm}{0ex}}\\
  \includegraphics[width=0.5\linewidth, trim=1 50 10 10, clip]{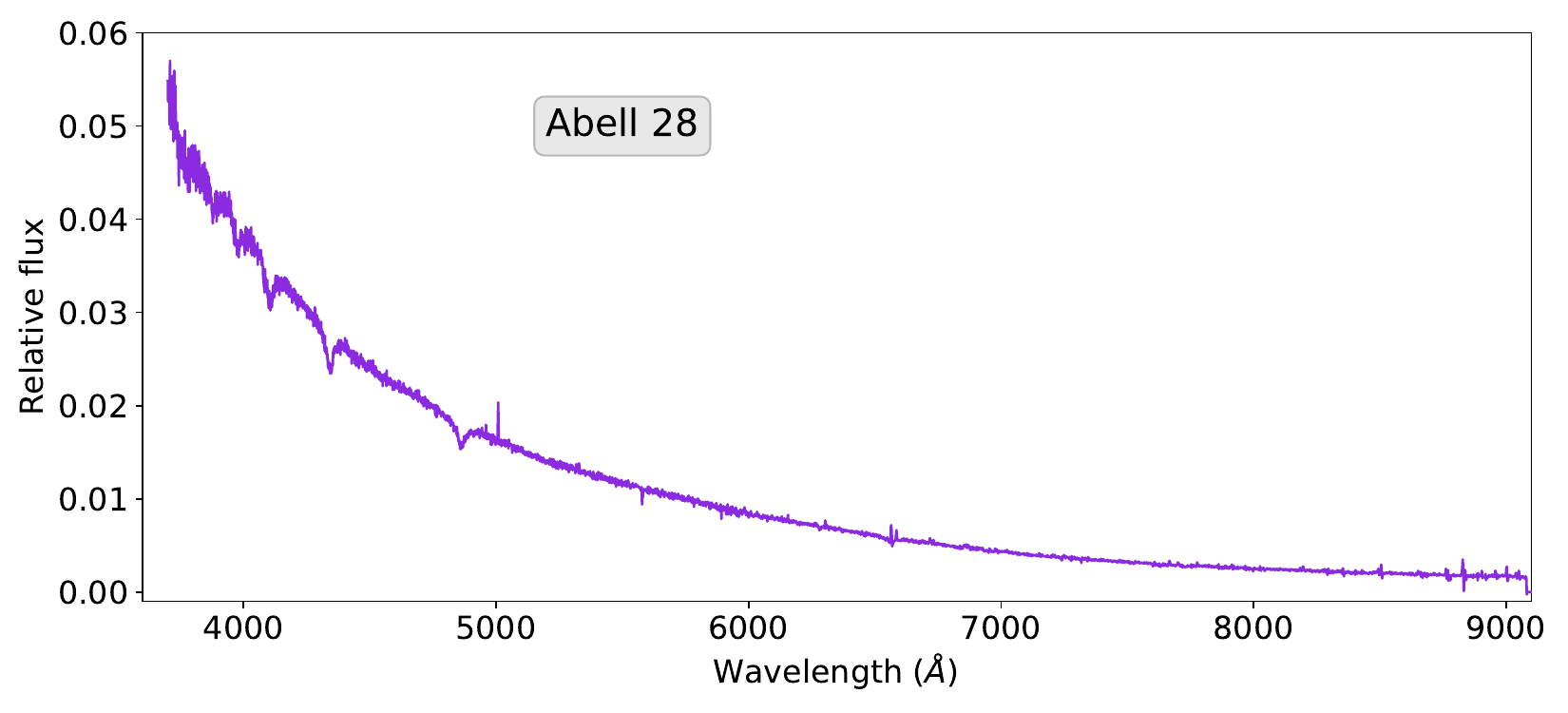} \llap{\shortstack{%
        \includegraphics[width=0.15\linewidth,  clip]{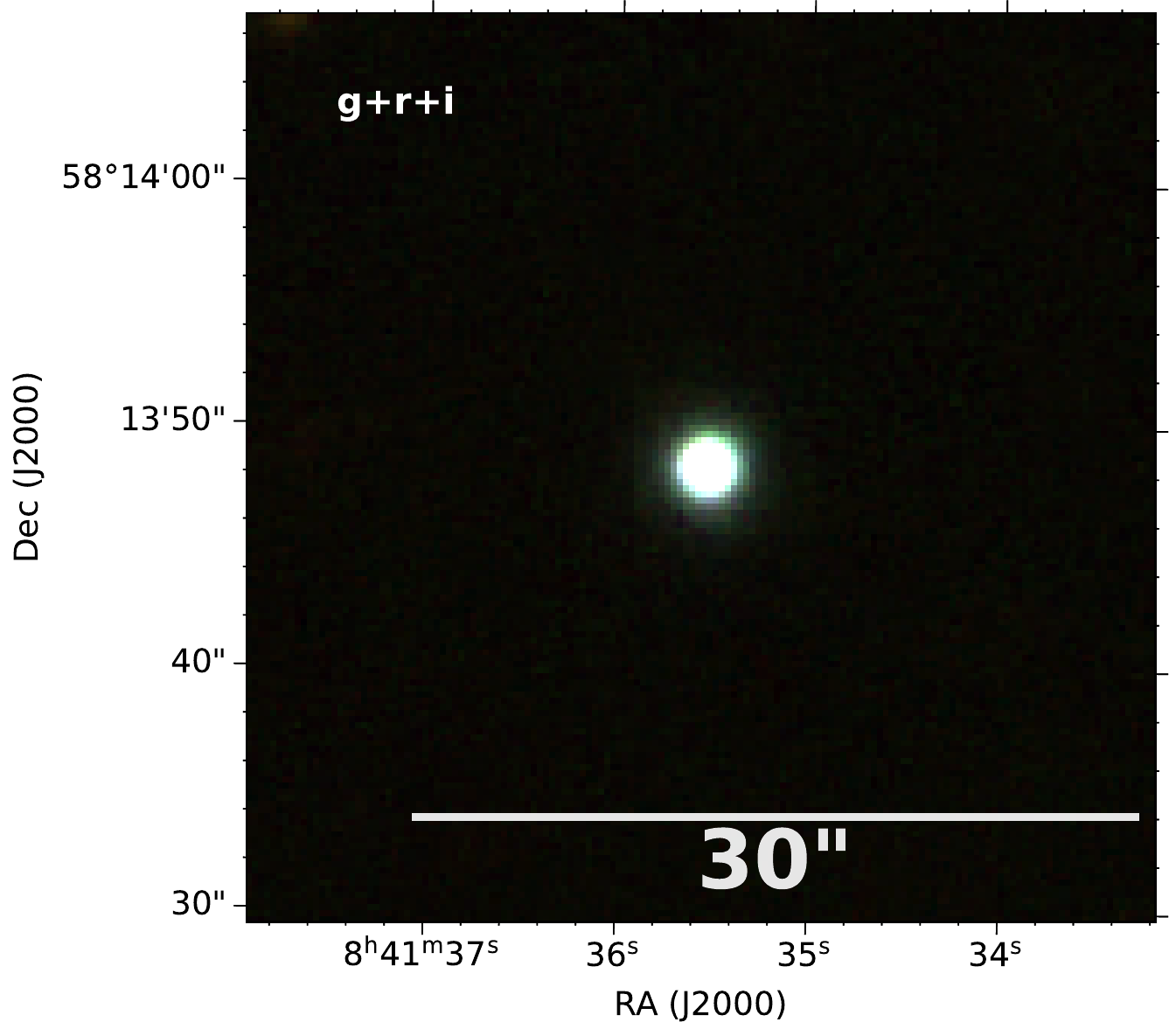}\\
        \rule{0ex}{0.85cm}%
      }
    \rule{0.25cm}{0ex}} & \includegraphics[width=0.5\linewidth, trim=10 50 10 10, clip]{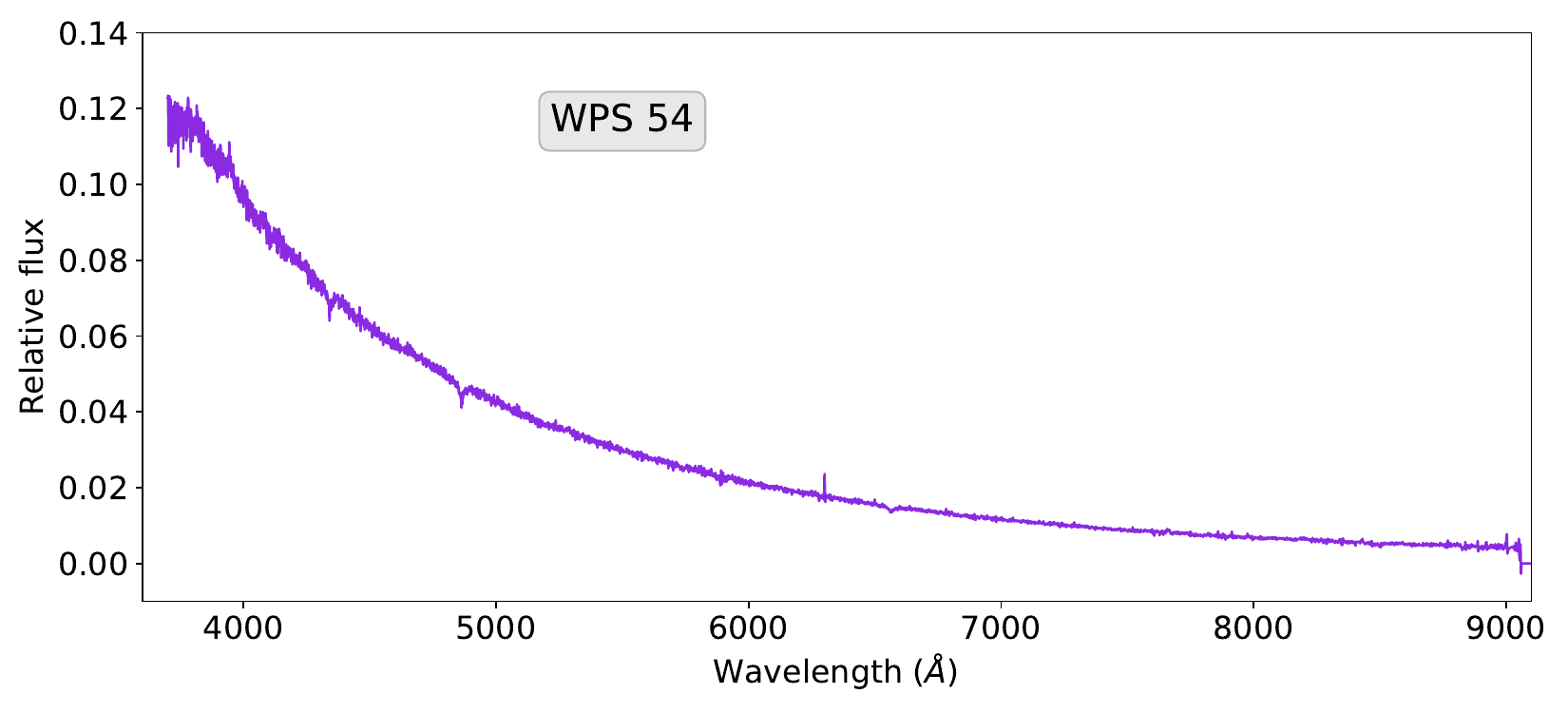}  \llap{\shortstack{%
      \includegraphics[width=0.15\linewidth,  clip]{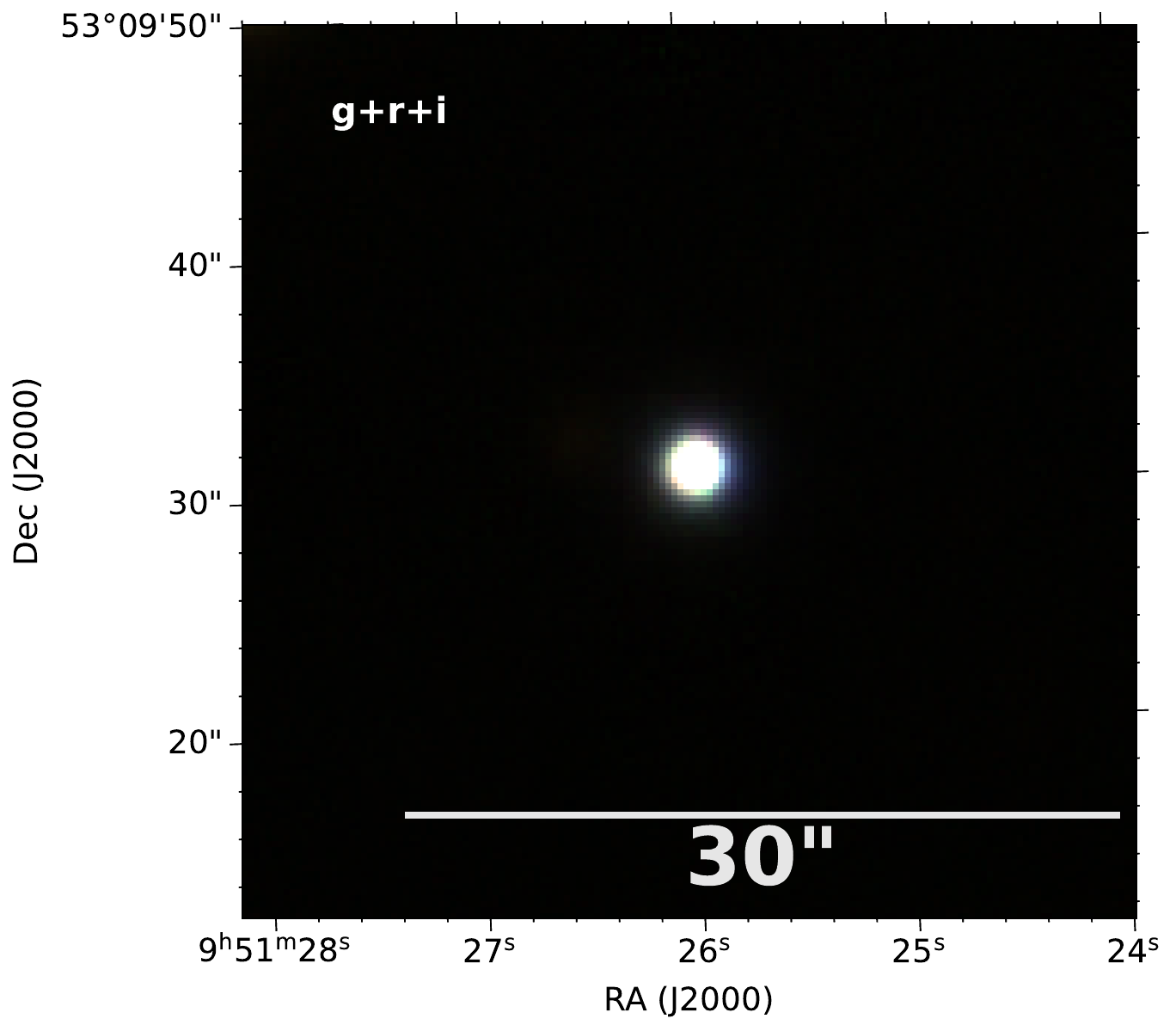}\\
      \rule{0ex}{0.85cm}%
      }
    \rule{0.25cm}{0ex}}\\
  \includegraphics[width=0.5\linewidth, trim=1 50 10 10, clip]{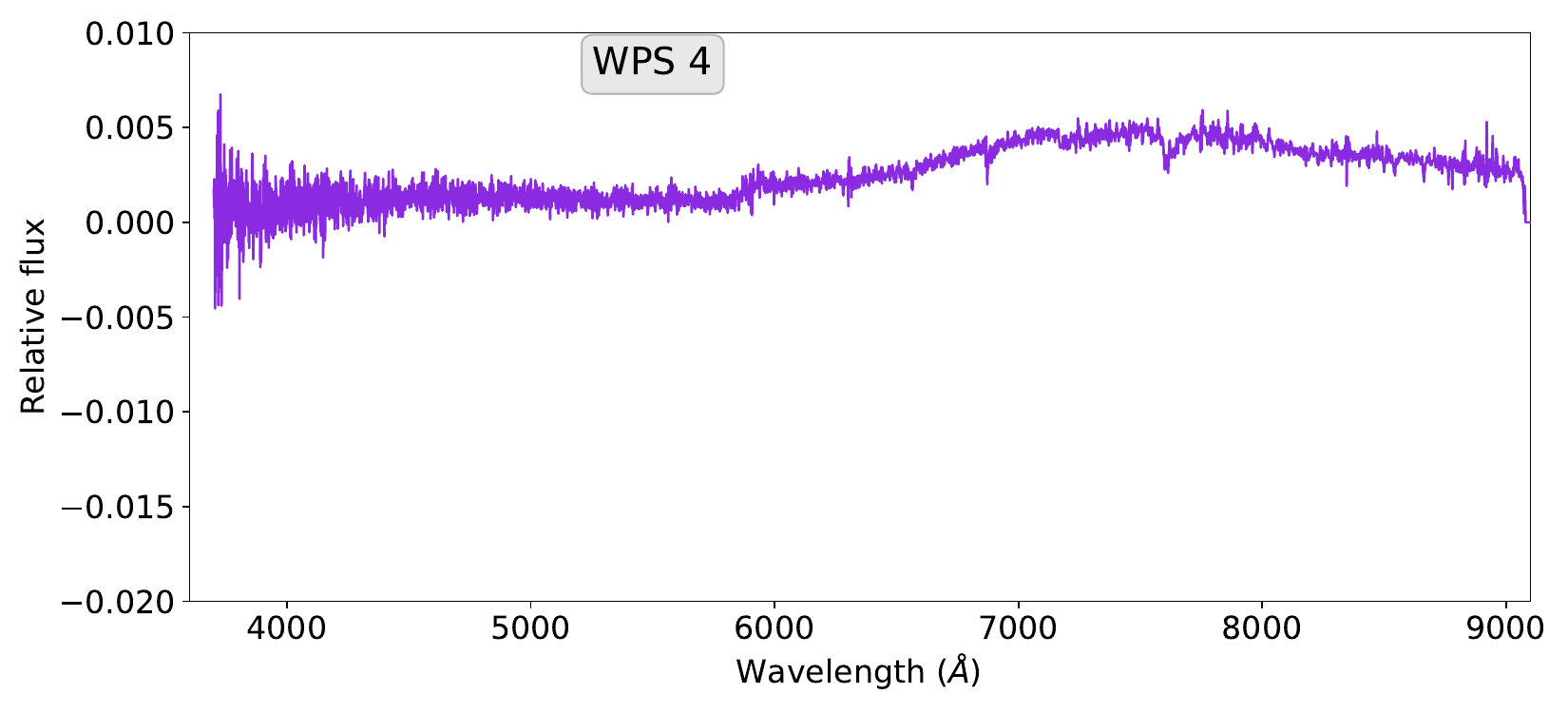} \llap{\shortstack{%
        \includegraphics[width=0.15\linewidth, clip]{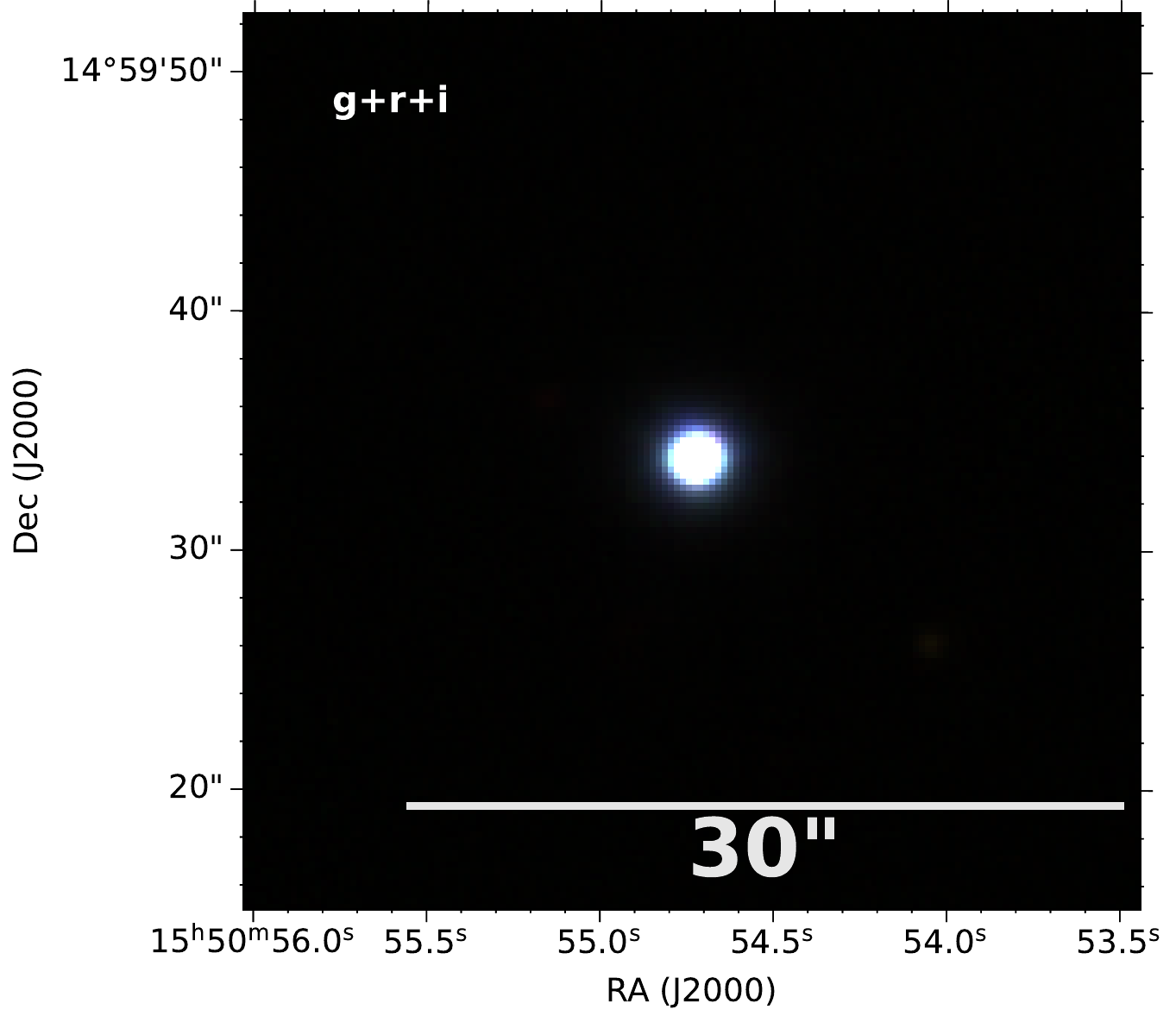}\\
        \rule{0ex}{0.0001cm}%
      }
    \rule{0.25cm}{0ex}} & \includegraphics[width=0.5\linewidth, trim=10 50 10 10, clip]{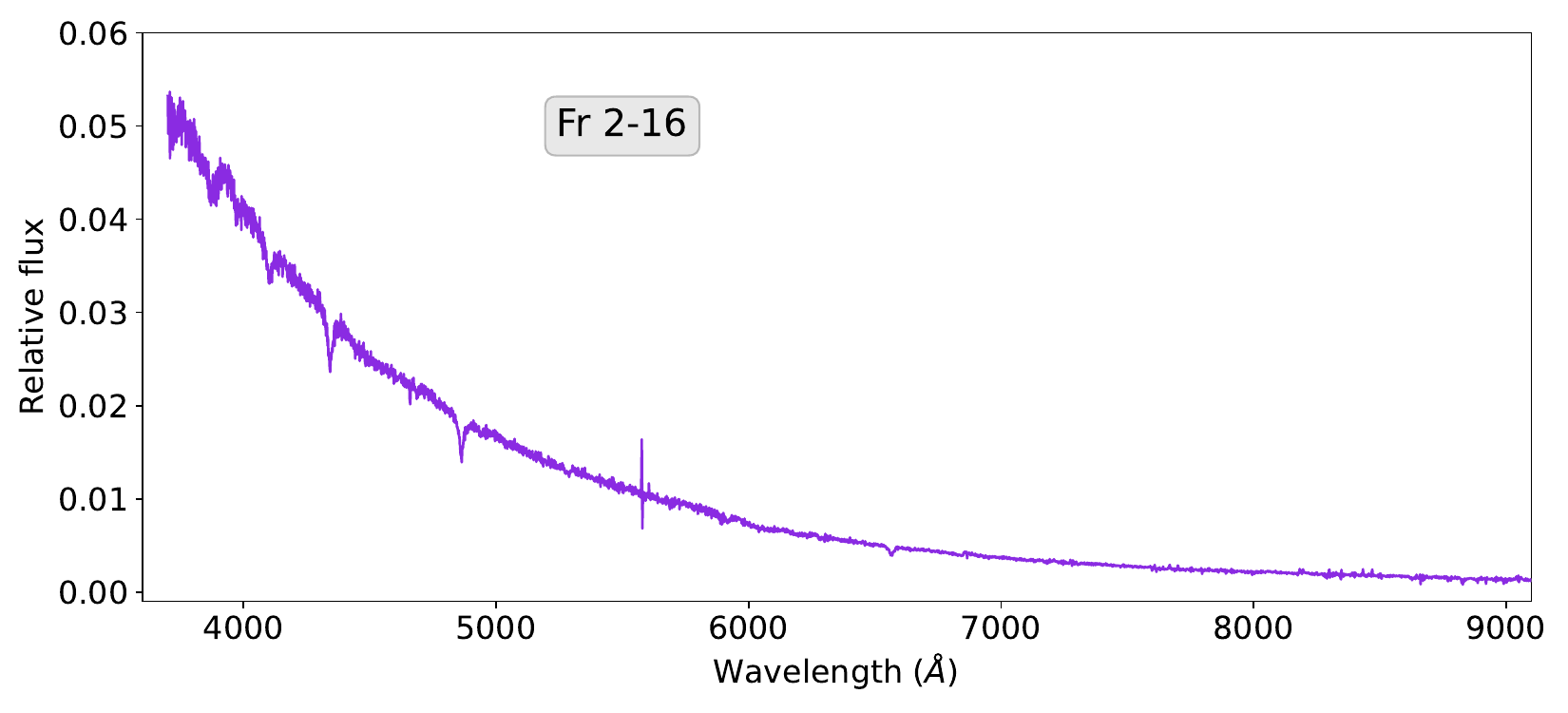}  \llap{\shortstack{%
      \includegraphics[width=0.15\linewidth, clip]{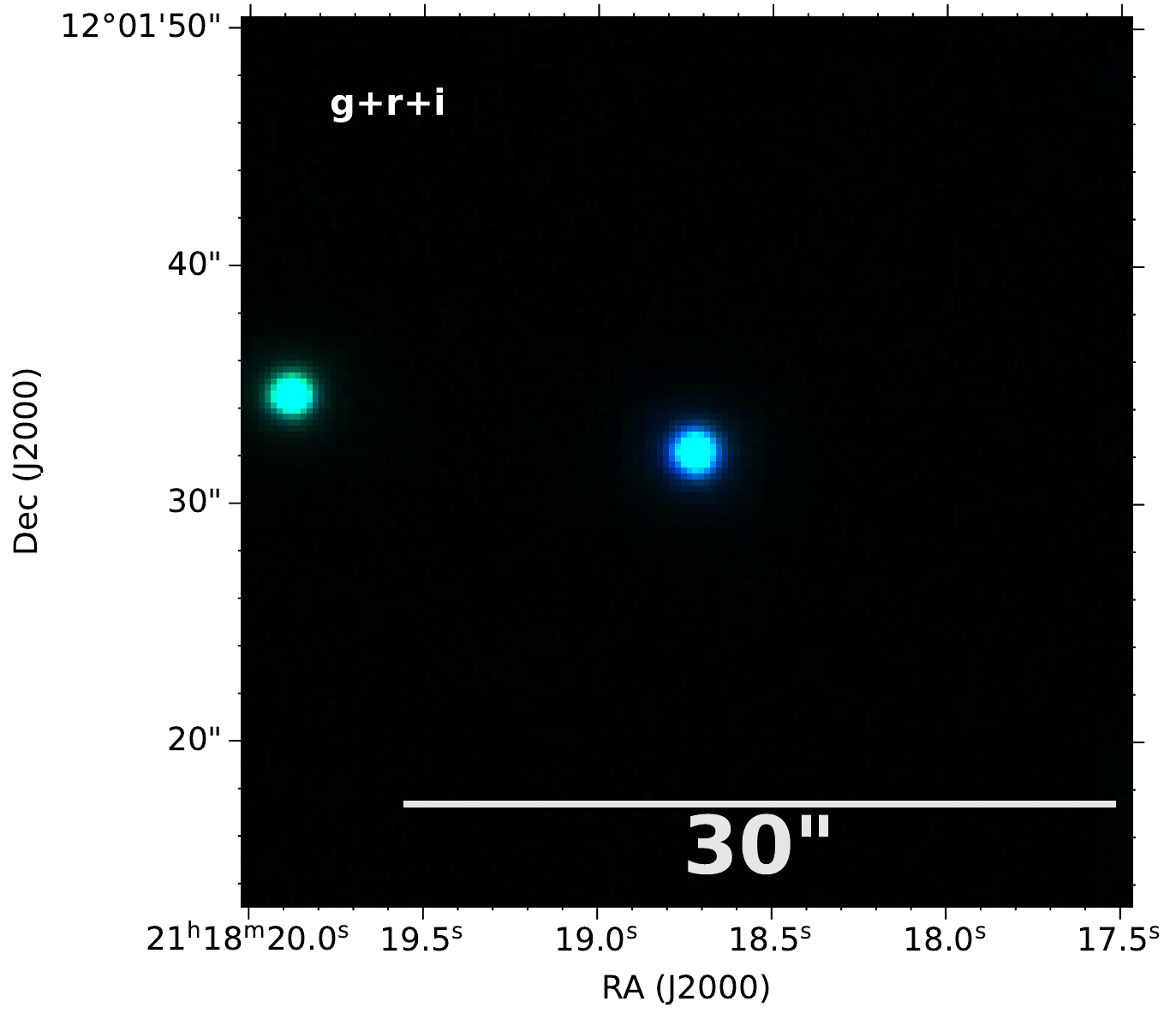} \\
      \rule{0ex}{0.85cm}%
      }
    \rule{0.25cm}{0ex}}\\
  \includegraphics[width=0.5\linewidth, trim=5 10 10 10, clip]{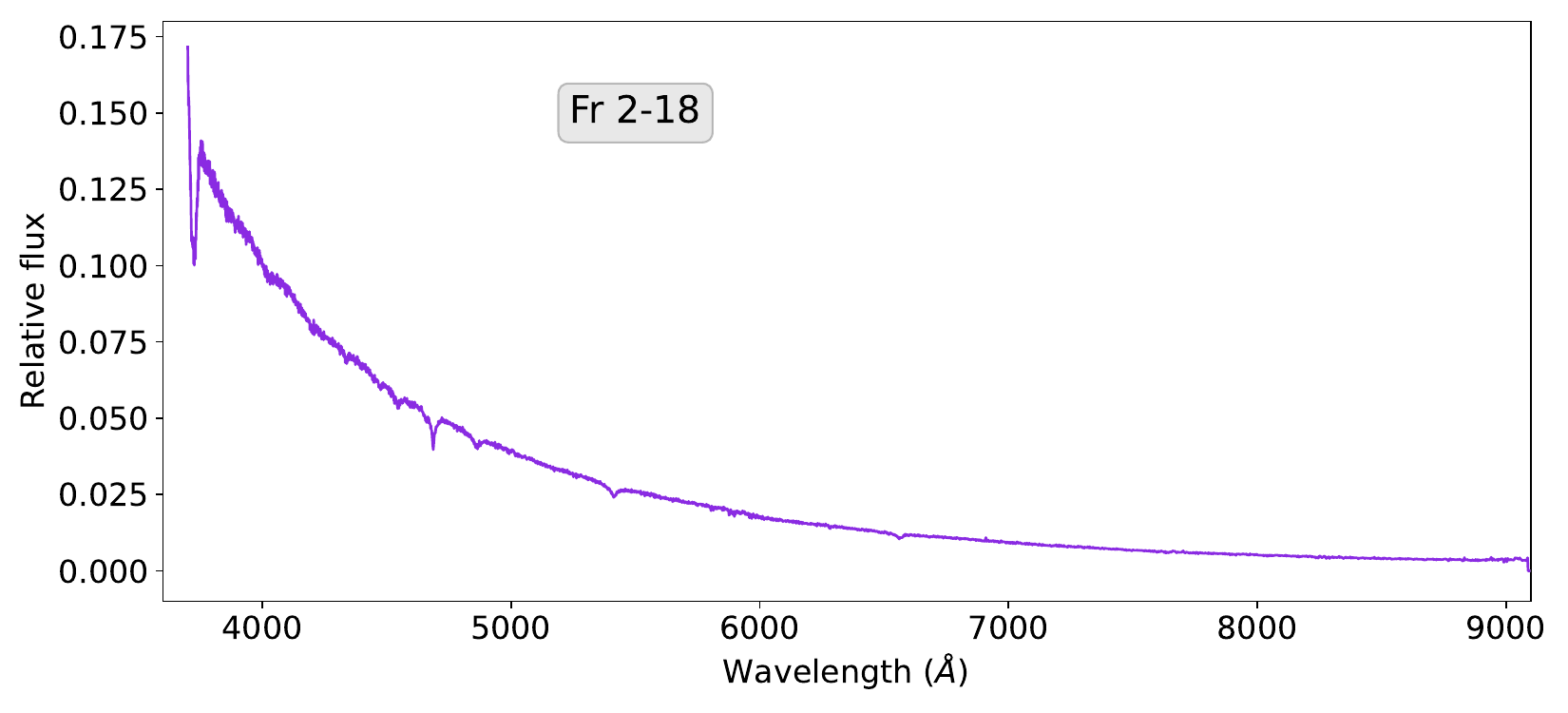} \llap{\shortstack{%
        \includegraphics[width=0.15\linewidth,  clip]{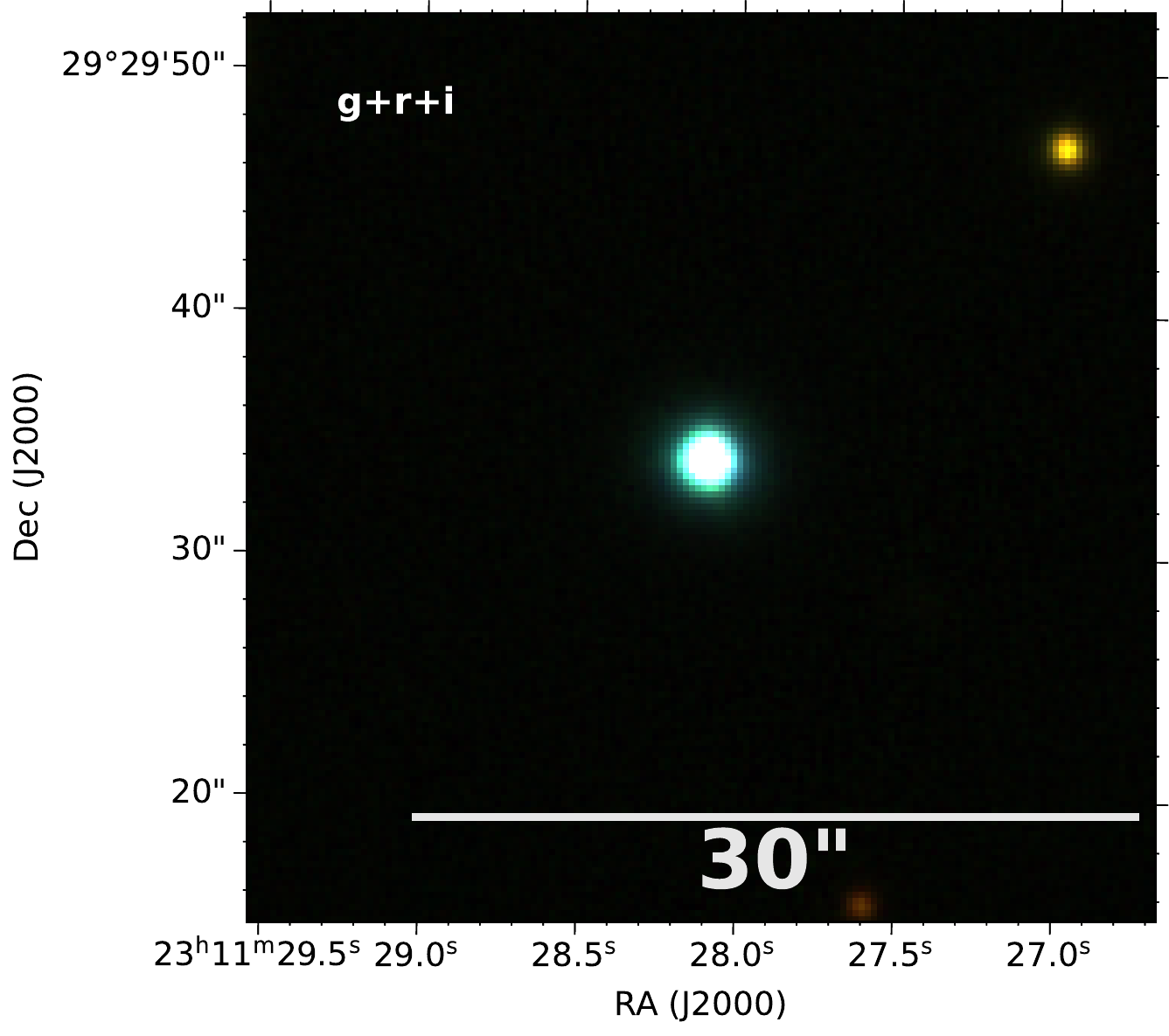}\\
        \rule{0ex}{1.3cm}%
      }
    \rule{0.25cm}{0ex}} & \includegraphics[width=0.5\linewidth, trim=10 10 10 10, clip]{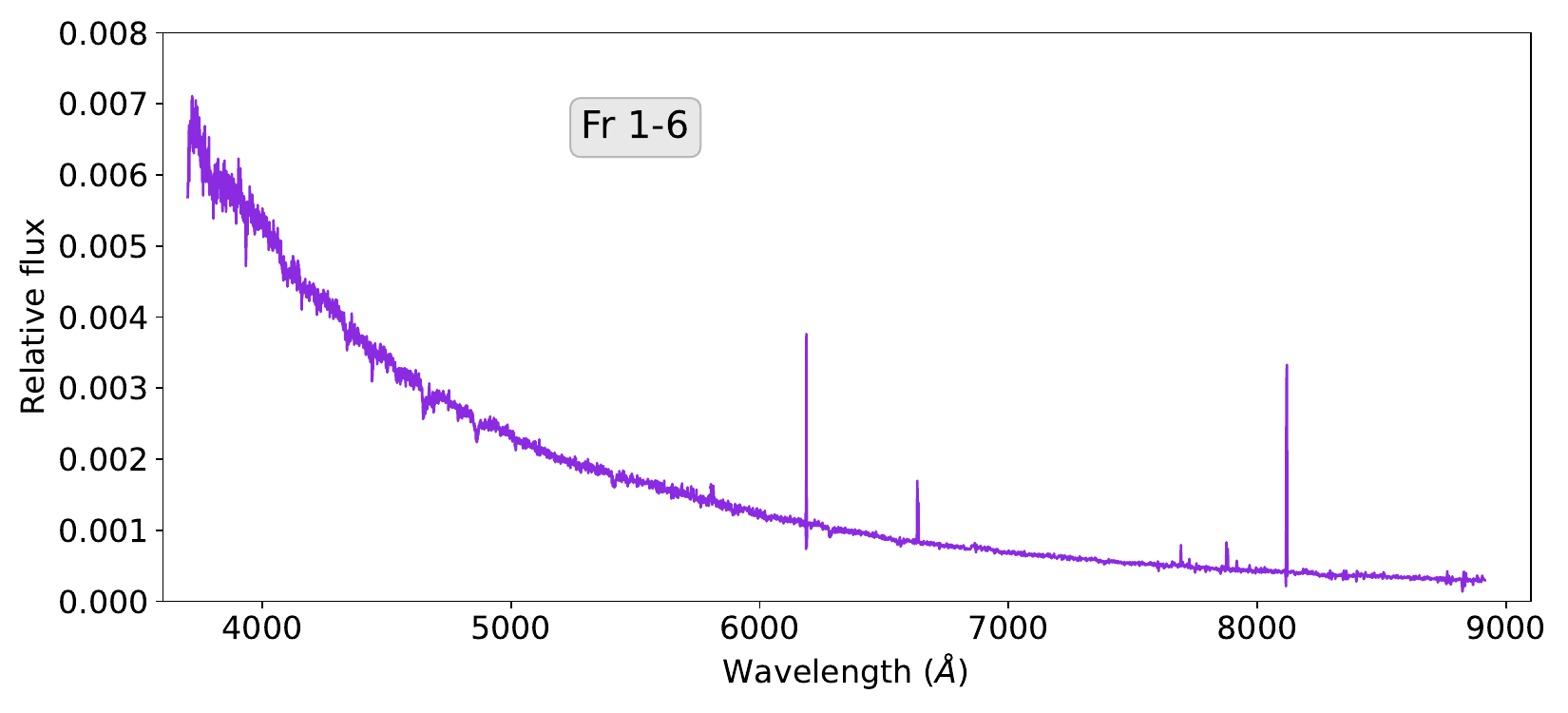}  \llap{\shortstack{%
      \includegraphics[width=0.15\linewidth,  clip]{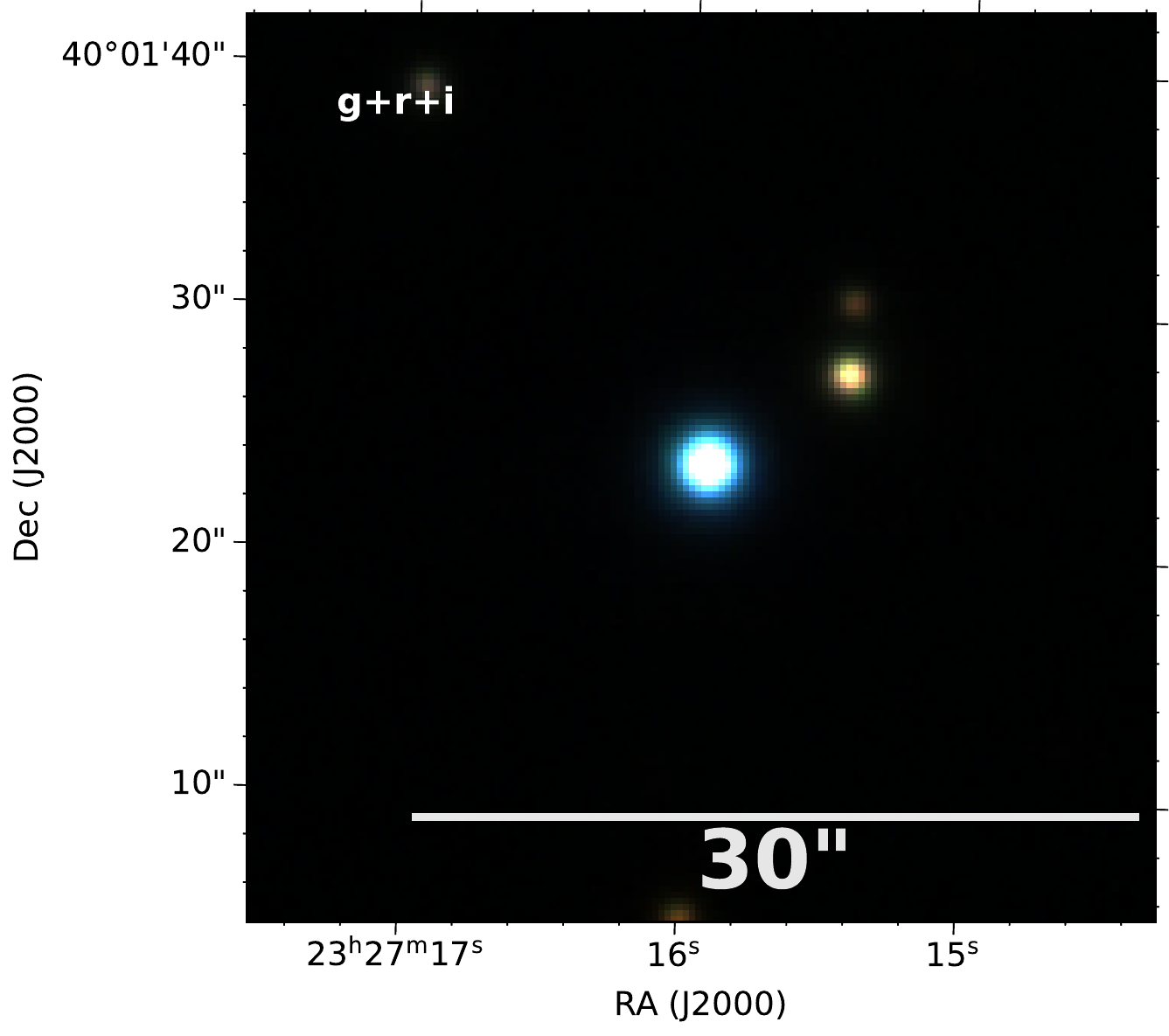}\\
      \rule{0ex}{1.4cm}%
      }
    \rule{0.25cm}{0ex}}
  \end{tabular}
  \caption{Same as Figure~\ref{fig:spectra-image-trurPN-better} but for the eight likely and probable PNe located in the very extended zone in the color-color diagram.}
  \label{fig:spectra-image-LikelyPN-medium}
\end{figure*}

\begin{figure*}
\centering
\begin{tabular}{l}
\includegraphics[width=0.5\linewidth, trim=5 10 10 10, clip]{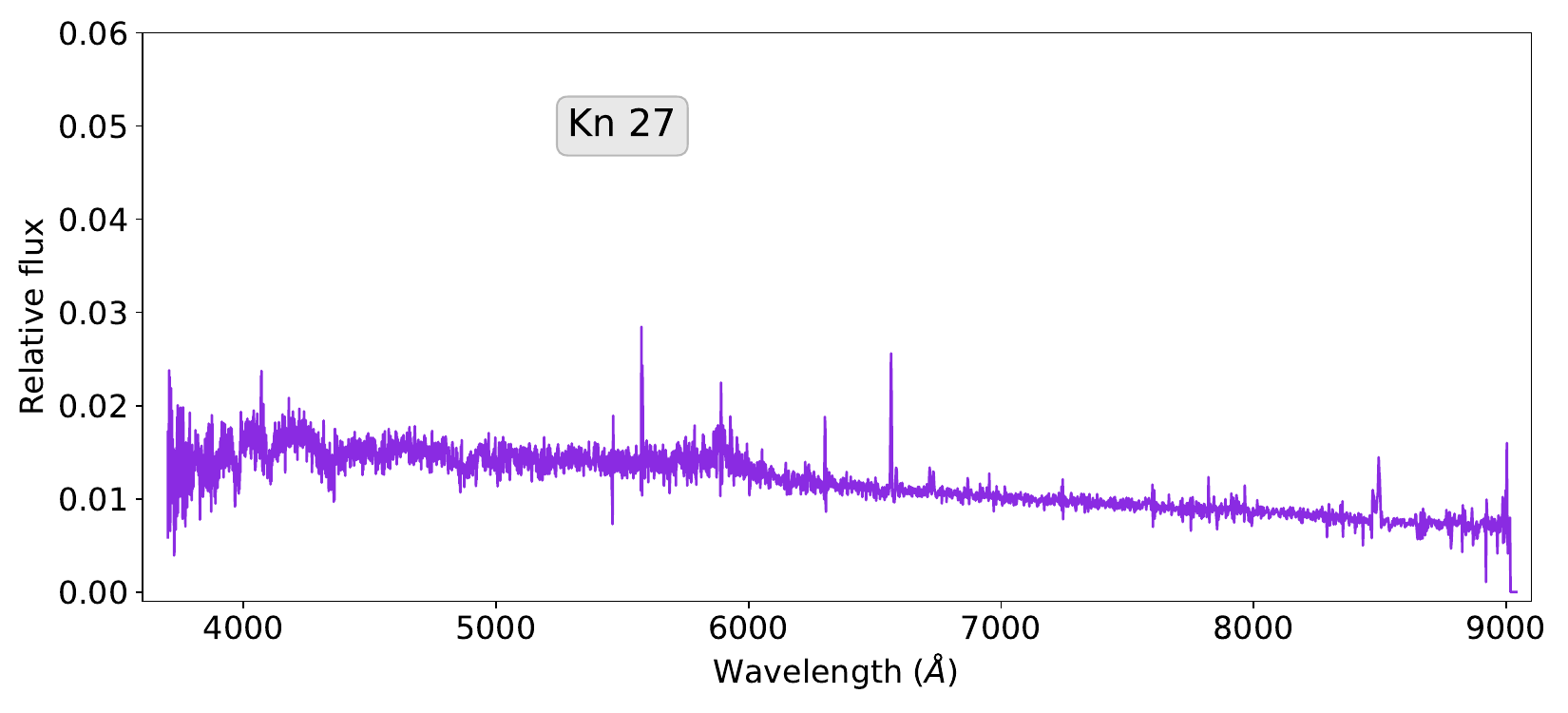} \llap{\shortstack{%
        \includegraphics[width=0.15\linewidth, clip]{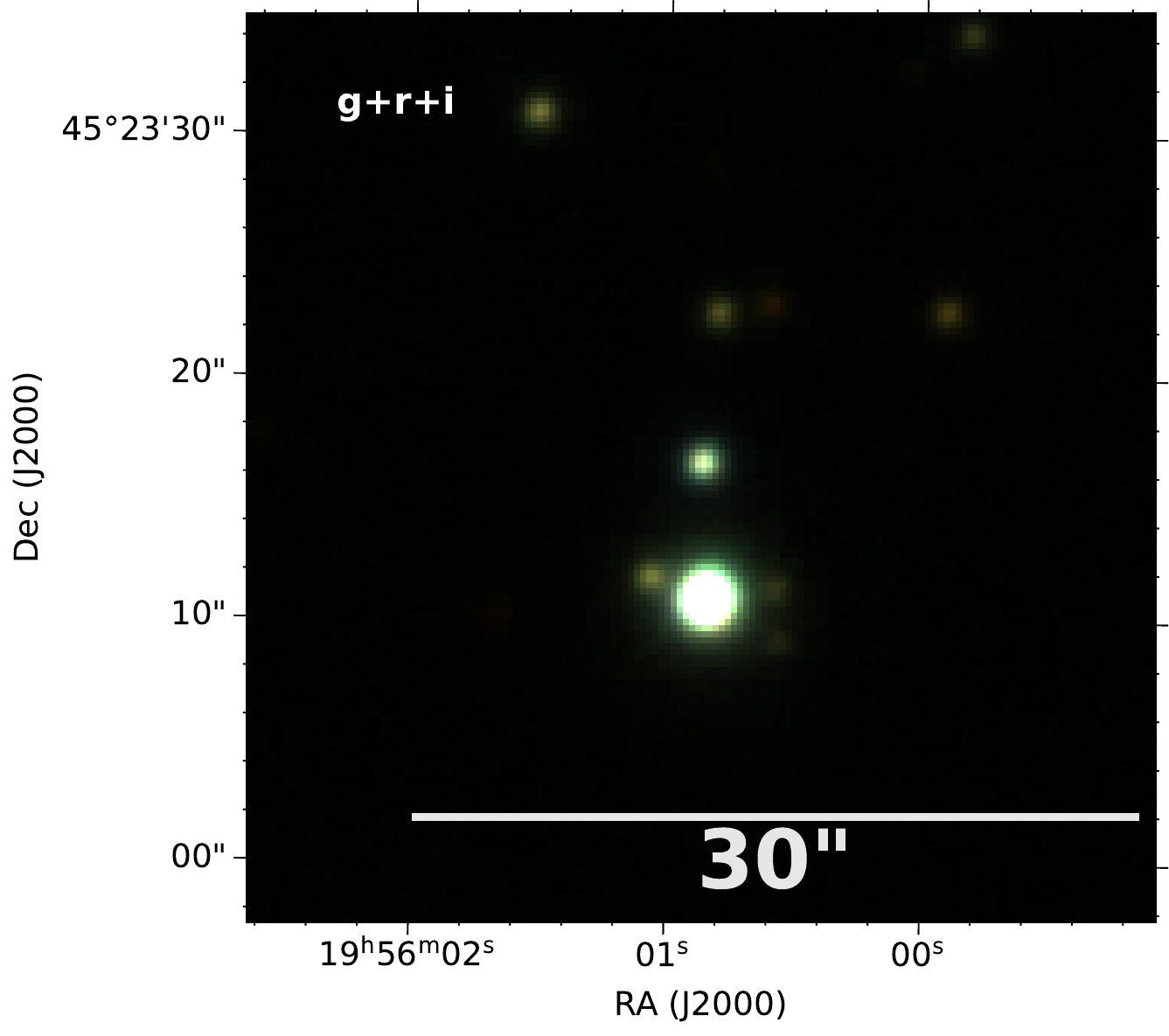}\\
        \rule{0ex}{1.3cm}%
  }
    \rule{0.25cm}{0ex}}
  \end{tabular}
  \caption{Same as Figure~\ref{fig:spectra-image-trurPN-better} but for the probable PN that are located outside of both zones in the color-color diagram.}
  \label{fig:spectra-image-LikelyPN-outside}
\end{figure*}

As in \ref{sec:spectra-tPNe}, Figure~\ref{fig:spectra-image-LikelyPN-medium} displays the spectra and images of the probable and likely PNe situated in the very extended PN zone according to the color-color diagrams. The only probable PNe located outside our established selection criteria is presented in Figure~\ref{fig:spectra-image-LikelyPN-outside}. Further details about these PNe, including additional parameters, can be found in Table~\ref{tab:LikelyPN-inf}.\\

\begin{table*}
	\centering
	\caption{List of probable and likely PNe in our study with their respective names, coordinates in Right Ascension (RA) and Declination (Dec), PN status (P for probable, L for likely), angular sizes, \((G - g)\), \((G - r)\), and \((G_{BP} - G_{RP})\) colors.  Additionally, Comments describe the region classification.}
	\label{tab:LikelyPN-inf}
	\begin{tabular}{lcccccccc} 
                \hline
		\hline
		Name     &           \(\mathrm{RA (J2000)}\) & \(\mathrm{Dec (J2000)}\) &  PN status & PN diameter (HASH) & $G - g$ & $G - r$ & $BP-RP$ & Comments$^{a}$  \\
          \hline
Dr 27 & 00:11:03.44 & 57:10:36.1 & L & 540.0 & 0.09 & -0.21 & -0.23 & Epn(g), Epn(r) \\
Fr 2-22 & 01:50:54.28 & 31:07:46.7 & P & 60.0 & 0.14 & -0.24 & -0.34 & Epn(g), Epn(r)\\
Fr 2-23 & 03:14:45.91 & 48:12:05.8 & L & 1650.0 & 0.13 & -0.23 & -0.30 & Epn(g), Epn(r) \\
Alves 3 & 03:43:14.37 & 10:29:38.1 & L & 468.0 & 0.16 & -0.23 & -0.40 & Epn(g), Epn(r) \\
Ber 1 & 06:06:50.63 & 54:58:23.9 & P & 1800.0 & 0.11 & -0.17 & -0.17 & Epn(g), Epn(r)\\
TK 1 & 08:27:05.52 & 31:30:08.1 & L & 2360.0 & 0.20 & -0.29 & -0.53 & Epn(g), Epn(r)\\
Abell 28 & 08:41:35.56 & 58:13:48.3 & L & 330.0 & 0.17 & -0.26 & -0.44 & Epn(g), Epn(r) \\
WPS 54 & 09:51:25.99 & 53:09:30.7 & P & 3600.0 & 0.15 & -0.33 & -0.55 & Epn(g), Epn(r)\\
WPS 4 & 15:50:54.73 & 14:59:33.7 & P & 120.0 & 0.12 & -0.18 & -0.17 & Epn(g), Epn(r) \\
Kn 27 & 19:56:00.87 & 45:23:16.3 & P & -- & -0.38 & 0.06 & 0.89 & Other\\
Fr 2-16 & 21:18:18.73 & 12:01:32.0 & P & 1800.0 & 0.18 & -0.29 & -0.52 & Epn(g), Epn(r)\\
Fr 2-18 & 23:11:28.11 & 29:29:33.8 & P & 1800.0 & 0.16 & -0.30 & -0.50 & Epn(g), Epn(r)\\
Fr 1-6 & 23:27:15.95 & 40:01:23.6 & P & 260.0 & 0.10 & -0.22 & -0.30 & Epn(g), Epn(r)\\
		
        \hline
	\end{tabular}
  \begin{tablenotes}
            \item \textbf{Note.} $^{a}$ See Table~\ref{tab:TruePN-inf} for abbreviations.
\end{tablenotes}
\end{table*}


During the preparation of this work the author(s) used ChatGPT3.5 in order to refine the English wording of certain phrases. After using this tool/service, the author reviewed and edited the content as needed and take(s) full responsibility for the content of the publication.

\bibliographystyle{elsarticle-harv} 
\bibliography{ref-pne}






\end{document}